\documentclass[aps,twocolumn,superscriptaddress,showpacs]{revtex4-1}
\usepackage{longtable}
\usepackage{verbatim}
\usepackage{multirow}
\usepackage{bm,graphicx}
\begin{document}
% You should use BibTeX and apsrev.bst for references
\bibliographystyle{apsrev}

%\preprint{}

\title{Multichannel parametrization of $\pi N$ scattering amplitudes\\
        and extraction of resonance parameters}
% Optional argument for running titles on pages
%\title[]{}


% other informatio

\author{M. Shrestha}
\author{D.~M.~Manley}
\affiliation{Department of Physics, Kent State University, Kent, OH 44242-0001}

%\thanks{}


\date{\today}

\begin{abstract}
We present results of a new multichannel partial-wave analysis for 
$\pi N$ scattering in the c.m.\ energy range 1080 to 2100 MeV.
This  work explicitly includes $\eta N$ and $K \Lambda$ channels and the single pion photoproduction channel.
Resonance parameters were extracted by fitting partial-wave amplitudes
from all considered channels using a multichannel parametrization that
is consistent with $S$-matrix unitarity. 
The resonance parameters so obtained  are compared to predictions of quark models.
\end{abstract}
%\keywords{}

\maketitle

\section{INTRODUCTION}

According to quark models, nucleons (baryons) are bound states of three constituent quarks. The excited states of these quarks give rise to the baryon resonance spectrum.  There are many models \cite{isgur-karl, capstick-isgur, glozman88, loring01, capstick-roberts} that describe the interactions between the quarks in baryons. 
%Isgur and karl model[1] uses the effective single gluon exchange interaction between the quarks. Similar assumption is made in the model by Capstick and Isgur[2]. The model by Plessas and collaborators[3] uses multi Goldstone bosons exchanges between the quarks. L\"{o}ring, Metsch and Petry[4] introduced the concept of instanton induced interactions. There are other theoretical analyses  describing the quarks interaction in baryons. 
In spite of all these different approaches, they pose the same common scenario: a greater number of predicted states as compared to the experimentally verified states. As an example, only nine $N^*$ resonances have been confirmed by experimental analyses (as listed by the Particle Data Group (PDG) \cite{pdg12}) whereas at least 21 states are predicted in the same energy range  \cite{capstick-roberts}. The possible reason for the discrepancy is either the models are based on wrong assumptions or the analyses for the extraction of resonance information are incomplete. In this work we focus on the second possibility.

Almost all analyses \cite{arndt91, cutkosky80} use the study of $\pi N$ elastic scattering as the main source for the extraction of $N^*$ resonance parameters. Information on the coupling of resonances to other channels has come mainly from analyses of $\pi N$ inelastic scattering. This might be the reason for the missing states as the $\pi N$ channel is predicted to decouple from many resonances with masses above 1.7 GeV \cite{capstick94}. It is thus logical to look for resonances in other inelastic channels including photoproduction channels at  the higher energies. An analysis that incorporates all possible channels simultaneously is highly desirable. 

 Currently, other groups working on analyses of $\pi N$ scattering are the EBAC (JLab) group \cite{julia-diaz07}, the Bonn-Gatchina Group \cite{sarantsev09}, and the GWU/SAID \cite{arndt06} group. The EBAC group \ uses a dynamical coupled-channels approach.
 %(DCC) approach that is based on the Sato-Lee model \cite{matsuyama07}. 
 The channels included are $\pi N$, $\pi \pi N$, $\eta N$, $K\Lambda$, and $\gamma N$. 
 %The J\"ulich DCC approach is based on J\"ulich $\pi N$ model \cite{schutz98, krehl00, gasparyan03} which is in turn based on the time-ordered perturbation theory \cite{schweber05}. The J\"ulich model includes $\pi N$, $\eta N$, $\sigma N$, $\rho N$, and $\pi \Delta$ channels and recently, the J\"ulich group have modified the model to include pion-photoproduction channels \cite{huang12}. 
 The group of Huang {\it et al.\ }\cite{huang12} has also used a dynamical coupled-channels model to investigate pion photoproduction. The Bonn-Gatchina Group uses a multichannel Breit-Wigner and $K$-matrix approach for the amplitude parametrization; channels included are $\pi N$, $\pi \pi N$, $\eta N$ and photoproduction channels. The SAID group maintains up-to-date partial-wave analyses (PWAs) of several reactions including $\pi N\rightarrow \pi N$ and $\gamma N\rightarrow \pi N$. Currently, they have started to use a multichannel PWA giving equal importance to the inelastic channels. 
 
 Our approach is different and unique in the sense that it uses a generalized energy-dependent Breit-Wigner parametrization of amplitudes treating all the channels on an equal footing and taking full account of non-resonant backgrounds. The channels included in this analysis are $\pi N$, $\pi \pi N$, $\eta N$, $K \Lambda$, and $\gamma N$. We begin with an energy-dependent model for fitting of the $\pi N$ partial-wave data. Our detailed partial-wave analyses of reactions $\pi^-p\rightarrow \eta n$ and $\pi^-p\rightarrow K^0\Lambda$ are presented elsewhere \cite{manoj12}. The reliability of the energy-dependent amplitudes extracted from this work is tested by using the fitted amplitudes to compare with various observables \cite{manoj012}. Our solution is in good agreement with available data for $\pi^-p\rightarrow \eta n$ and $\pi^-p\rightarrow K^0 \Lambda$. 

\section{Theoretical Aspects}
The KSU model, developed by Manley \cite{manley03},  employs a unitary multichannel parameterization to extract resonance parameters. T.-S. H. Lee has reviewed the KSU model as one of the available models for analyzing data from meson production reactions \cite{lee06}. He has shown that it can be derived starting from very general coupled-channel equations. The KSU model developed as a variation of the parameterization used in the analysis by Manley and Saleski \cite{manley92} for fitting $\pi N\rightarrow \pi N$ and $\pi N\rightarrow \pi\pi N$ amplitudes. The results presented in this paper supercede those of Ref.\ \cite{manley92}. The KSU model in its present form has been used to extract $N^*$ and $\Delta^*$ parameters from a combined fit of $\pi N\rightarrow \pi N$, $\pi N\rightarrow \pi\pi N$, and $\gamma N\rightarrow \pi N$ amplitudes \cite{niboh97}. It has also been successfully applied to extract $\Lambda^*$ and $\Sigma^*$ parameters from multichannel fits of $\bar KN$ scattering amplitudes \cite{john07, hongyu08}.

 In the KSU model,  the partial-wave $S$-matrix is defined as
\begin{equation}
   S= B^TRB=I+2{\rm i}T~,
 \end{equation}
where $T$ is the corresponding partial-wave $T$-matrix. Here $R$ is a unitary, symmetric, and generalized multichannel Breit-Wigner matrix while $B$ and its transpose $B^T$ are unitary matrices describing non-resonant background. The background matrix $B$ is constructed from a product of unitary matrices: $B = B_1B_2\cdots B_n$, where $n$ is a very small interger. Further details about the background parameterization can be found in Ref. \cite{manley03}. The pure resonant and background matrices $T_R$ and $T_B$ can be constructed as
\begin{equation}
T_R = (R-I)/2{\rm i}; T_B = (B^TB - I)/2{\rm i}~.
\end{equation}
With these, the total $T$-matrix takes the form
\begin{equation}
T = B^TT_RB + T_B~.
\end{equation}
To obtain the elements of $T_R$, a resonant $K$-matrix is constructed such that 
\begin{equation}
T_R = K(I - \rm iK)^{-1}~.
\end{equation}
The elements of the resonant $K$-matrix  are of the form

\begin{equation}
K_{ij}=\sum _{\alpha=1}^N X_{i\alpha}X_{j\alpha}\tan\delta_\alpha~, 
\end{equation}
where $\alpha$ denotes a specific resonance and $N$ is the number of resonances in the energy range of the fit; $\delta_\alpha$ is an energy-dependent phase and $X_{i\alpha}$ is related to the branching ratio for resonance $\alpha$ to decay into channel $i$, and for each resonance, $\sum_i(X_{i\alpha})^2 = 1$. 
%With this form for the $K$-matrix, the matrix elements of $T_R$ can be written in closed form as:
  If $N$=1, then $\tan \delta = ({\Gamma/2})/{(M - W)}$  and 
  $X_i^2 = {\Gamma_i}/{\Gamma}$, where $\Gamma_i$ is the partial width for the $i^{th}$ channel.
% A $K$-matrix element is $K_{ij} = x_{i}\cdot x_{j}(\Gamma/2)/(M-W)$.
   The corresponding $K$- and $T_R$-matrix elements are
  $K_{ij} = X_{i}\cdot X_{j}(\Gamma/2)/(M-W)$ and
   \begin{equation}
   [T_R]_{ij} = X_{i}\cdot X_{j}\frac{\Gamma /2}{M - W - \rm i\Gamma/2}~,
   \end{equation}
   respectively.
   For the special case of two resonances, $K_{ij} = X_{i1}\cdot X_{j1}\tan\alpha_1 + X_{i2}\cdot X_{j2}\tan\alpha_2$, so that \\
  \begin{eqnarray}
   [T_R]_{ij} &=& (X_{i1}\cdot X_{j1}) C_{11} + (X_{i1}\cdot X_{j2}) C_{12}   \\
                     &+&~(X_{i2}\cdot X_{j1}) C_{21} + (X_{i2}\cdot X_{j2}) C_{22}~,\nonumber
  \end{eqnarray}
where the energy-dependent coefficients $C_{ij}$ can be calculated analytically \cite{manley03}.
   Generalizing to $N$ resonances, we can write
\begin{equation}
[T_R]{_{ij}} =\sum_{\alpha=1}^N\sum_{\beta=1}^N~ X_{i\alpha}~[D^{-1}]_{\alpha\beta}~X_{j\beta}~.
\end{equation}
The energy dependence of the phases $\delta_\alpha$ is determined in a nontrivial and 
novel way such that 
\begin{equation}
[D^{-1}]_{\alpha\beta}\propto~\prod_{\gamma=1}^N[M_\gamma-W-\rm i(\Gamma_\gamma/2)]^{-1}~,
\end{equation} 
where $M_\gamma$ is a constant and $W$ is the total c.m.\ energy. $M_\gamma$ and $\Gamma_\gamma$  evaluated at $W = M_\gamma$ represent conventional Breit-Wigner parameters. Each of the resonances corresponds to a pole in $T_R$ and, therefore, 
also in the total $S$-matrix. The poles occur at complex energies $W = W_\gamma$ where $M_\gamma - W_\gamma - \rm i(\Gamma_\gamma/2) = 0$.

\section{Fitting Procedure}
The amplitudes for the multichannel energy-dependent fit were obtained from various single-energy analyses.  Our energy-dependent fit included the SAID SP06 solution for $\pi N\rightarrow\pi N$ \cite{arndt06}, the SAID FA07 solution $\gamma N\rightarrow\pi N$ \cite{arndt07}, and the solution of Manley {\it et al.} \cite{manley84} for $\pi N\rightarrow \pi\pi N$. In some of the photoproduction amplitudes there were no single-energy solutions above 1.8 GeV. In such cases we used an average of the SAID current solution and the SAID SM95 solution \cite{gwu95}. In addition, we included our single-energy amplitudes for $\pi N\rightarrow \eta N$ and $\pi N\rightarrow K\Lambda$ \cite{manoj12}. Previous single-channel analyses \cite{abaev96, vrana00, saxon80} of $\pi N\rightarrow \eta N$ and $\pi N \rightarrow K\Lambda$ were simplistic energy-dependent PWAs that failed to satisfy $S$-matrix unitarity. A multichannel energy-dependent fit was performed in the c.m.\ energy range from 1080 to 2100 MeV. Initially some approximately known fitting parameters were held fixed to yield a good fit. In some partial waves, $\omega N$ and $\rho\Delta$ channels were included as dummy channels (channels without data) to satisfy unitarity. In our final fits, uncertainties in resonance parameters were calculated with all fitting parameters free to vary.

 \section{Discussion of Resonance Parameters}
The hadronic resonance parameters for states with $I=1/2$ and $I=3/2$ are listed in Tables I and II, respectively. The first column lists the resonance name together with the fitted resonance mass and its fitted total width in MeV. The second column lists the fitted  hadronic decay channels, starting with $\pi N$ elastic channel. Quasi-two-body $\pi \pi N$ channels are tabulated as $\pi\Delta$, $\rho N$, $\sigma N$, or $\pi N^*$, where $\sigma$ denotes the $s$-wave $\pi\pi$ system with $J^P=0^+$ and $I_{\pi\pi}=0$, and $N^*$ denotes the $P_{11} (1440)$ resonance. Sometimes a subscript appears with a channel notation (e.g. $(\pi\Delta)_D$); here the subscript denotes the orbital angular momentum of the channel. Also a subscript after the meson symbol in a reaction channel (e.g.\ $\rho_3N$) refers to twice the sum of the intrinsic spins (2$S$) of the meson and baryon. The third column in Table I or II lists the partial decay widths ($\Gamma_i$) associated with corresponding channels. The symbol ${\cal B}_i$ in the fourth column denotes the branching ratio for a given channel. Finally, the $x$ and $x_i$ represent the ratio of elastic partial width and partial width for the $i^{th}$ channel respectively to the total width.

In Tables III and IV we compare our results on resonance parameters (resonance mass, width, and elasticity) for  $I=1/2$ and $I=3/2$ states with prior analyses.
Any resonance included above 2.1 GeV had its mass parameter initially fixed and resonance parameters for these states are generally not listed. For most resonances, the PDG star rating \cite{pdg12} is included in column 1.

In Tables V and VI we list the complex pole positions of resonances and compare our results with prior analyses. Here, the first column lists the name of the resonance, the second column lists the real part of the pole position (pole mass), and the third column lists the pole width, which is given by the negative of twice the imaginary part of the pole position.

Figure 1 shows  representative Argand diagrams for the elastic and two inelastic ($\pi N\rightarrow \eta N$ and $\pi N\rightarrow K\Lambda$) amplitudes for $I=1/2$ partial waves (for $D_{13}$ only elastic amplitude is shown) and representative Argand diagrams for the elastic $I=3/2$ amplitudes. %the $S_{11}$ and $P_{11}$ partial waves.
To discuss the resonance parameters we follow a logical sequence of partial waves.\\

%%\newpage
\begin{figure*}[htpb]
%\caption{Argand diagrams for two body or quasi-two body amplitudes}
\vspace{-10mm}
\scalebox{0.35}{\includegraphics*{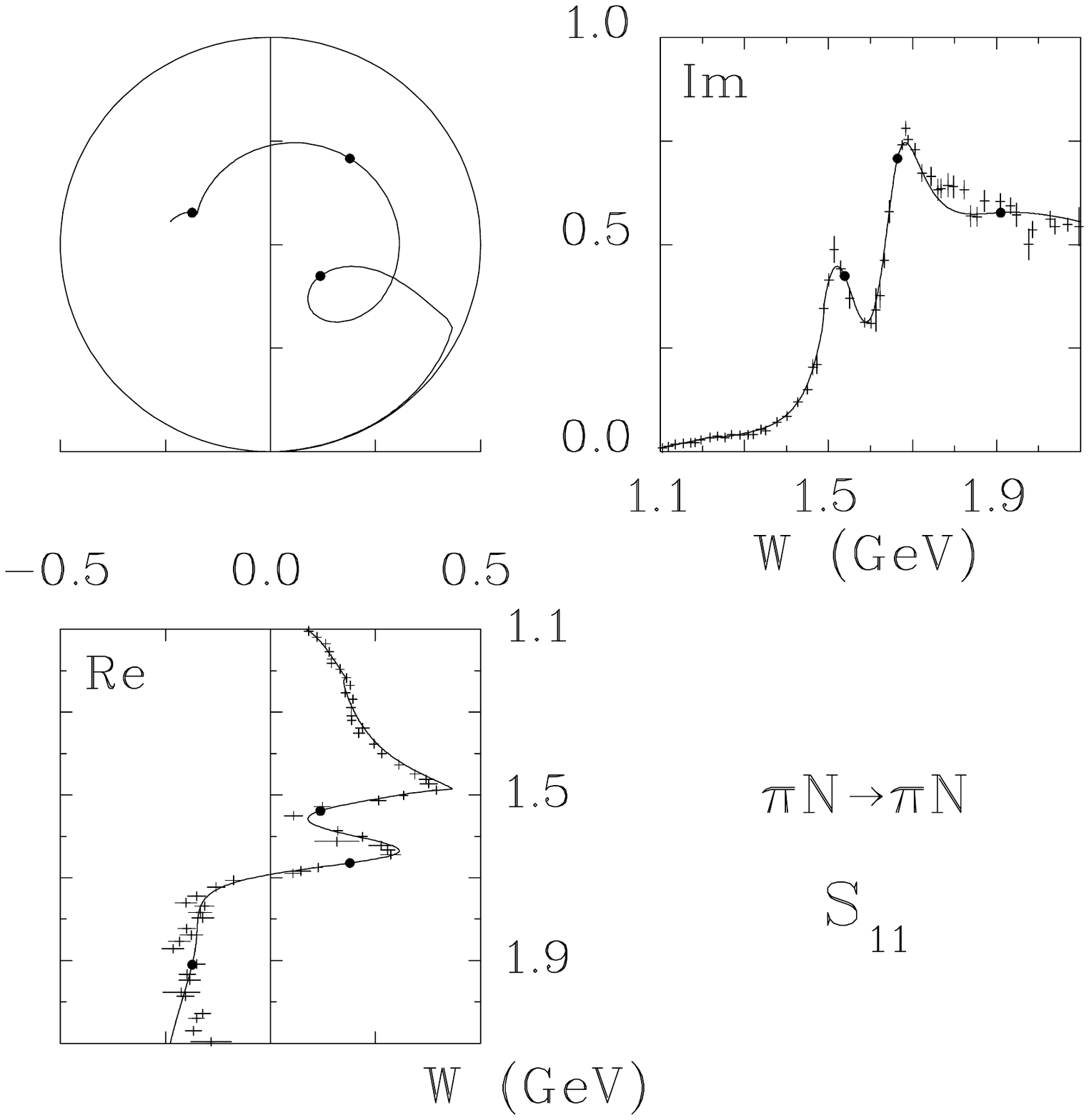}}
\vspace{-25mm}
\scalebox{0.35}{\includegraphics*{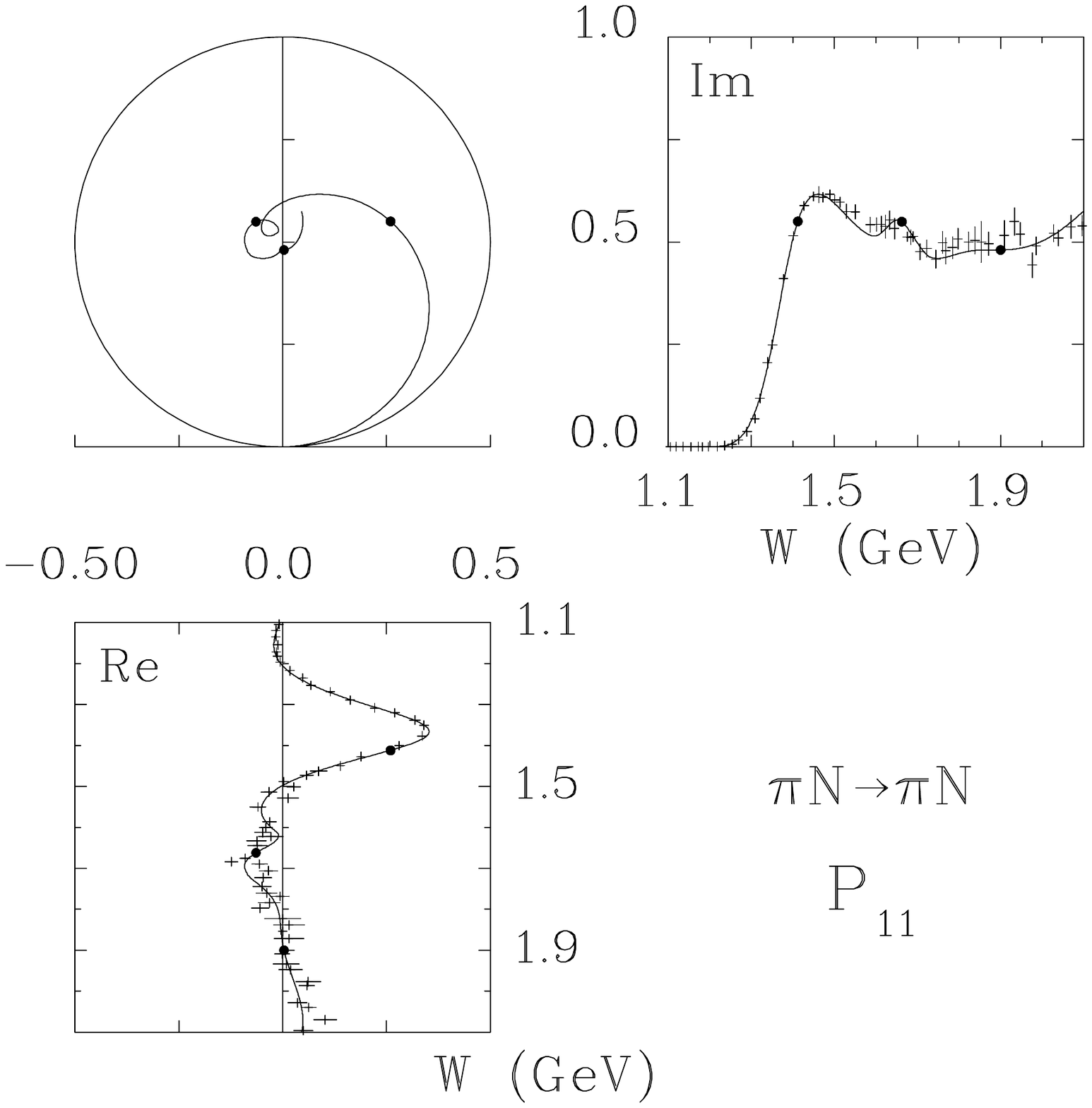}}
\vspace{-25mm}
\scalebox{0.35}{\includegraphics*{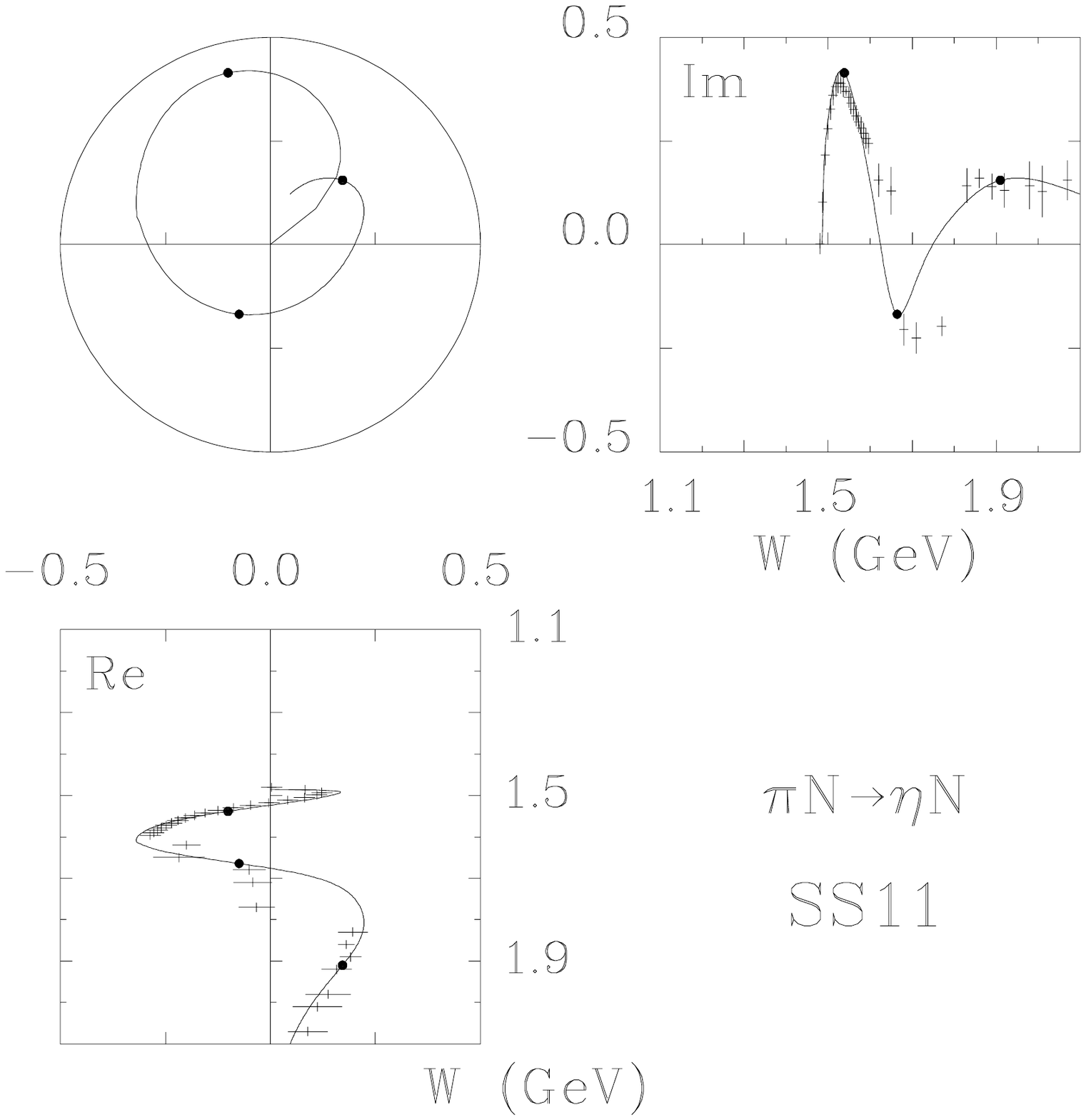}}
%\vspace{-25mm}
\scalebox{0.35}{\includegraphics*{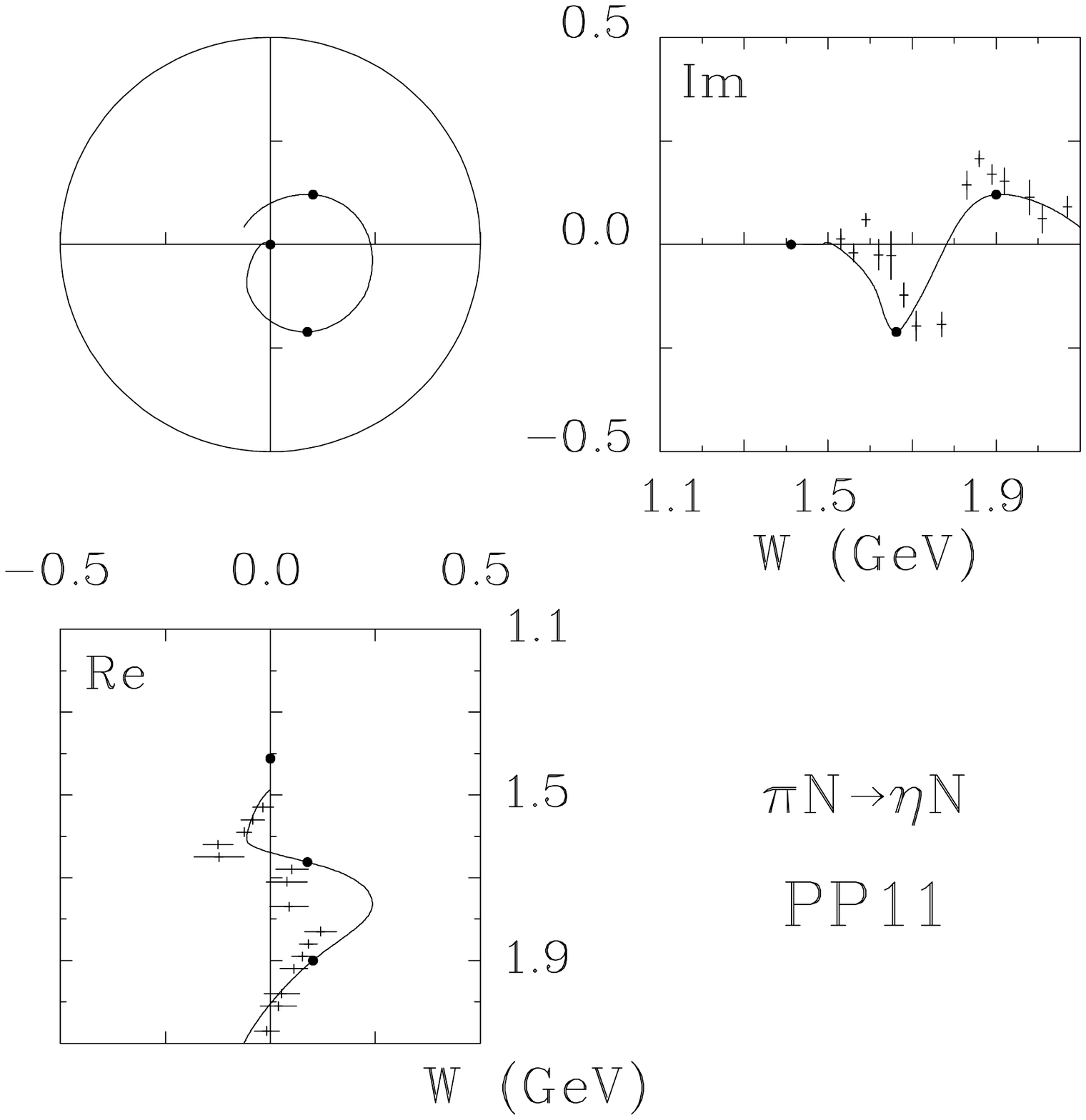}}
\vspace{-15mm}
\scalebox{0.35}{\includegraphics*{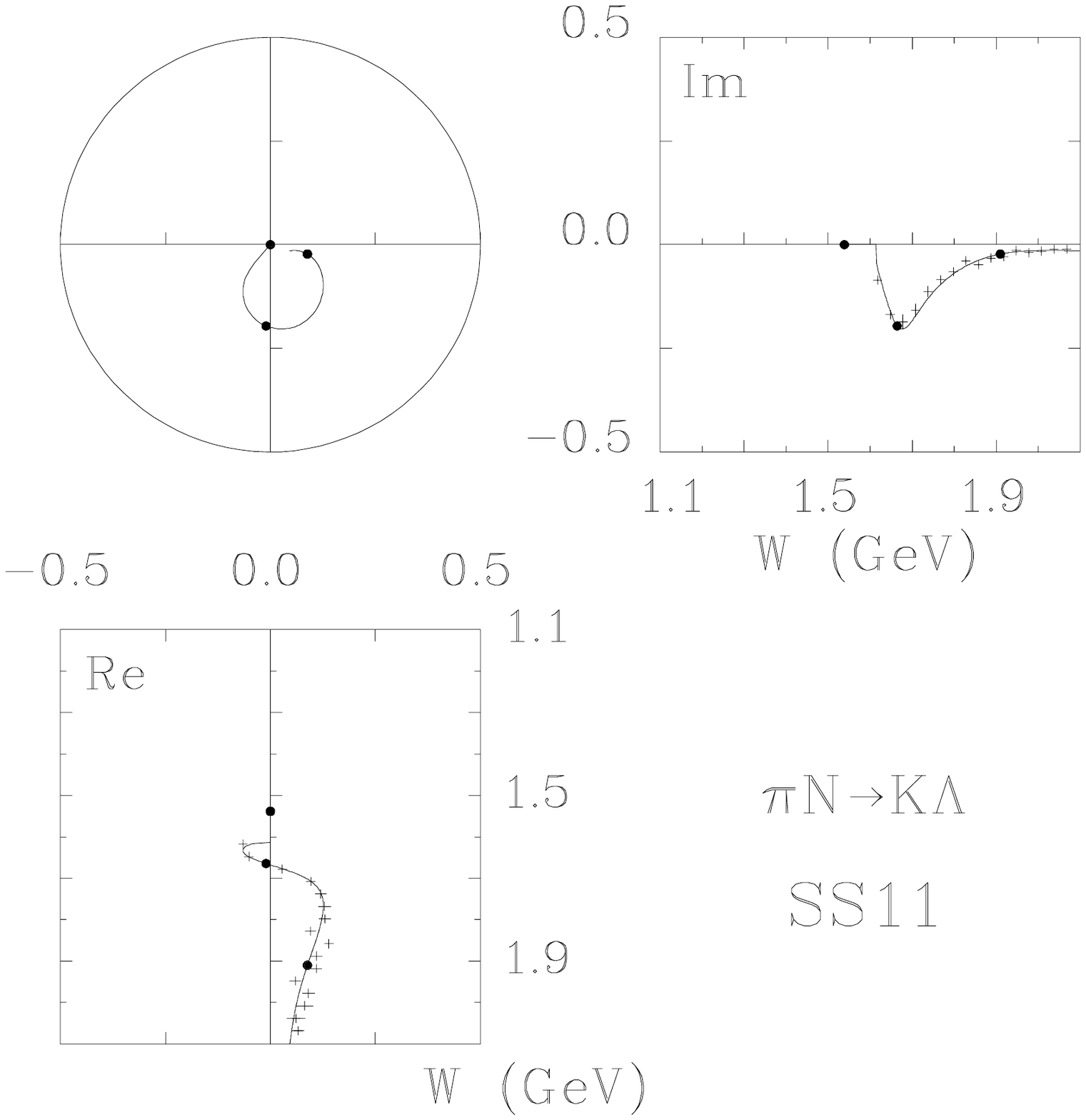}}
\scalebox{0.35}{\includegraphics*{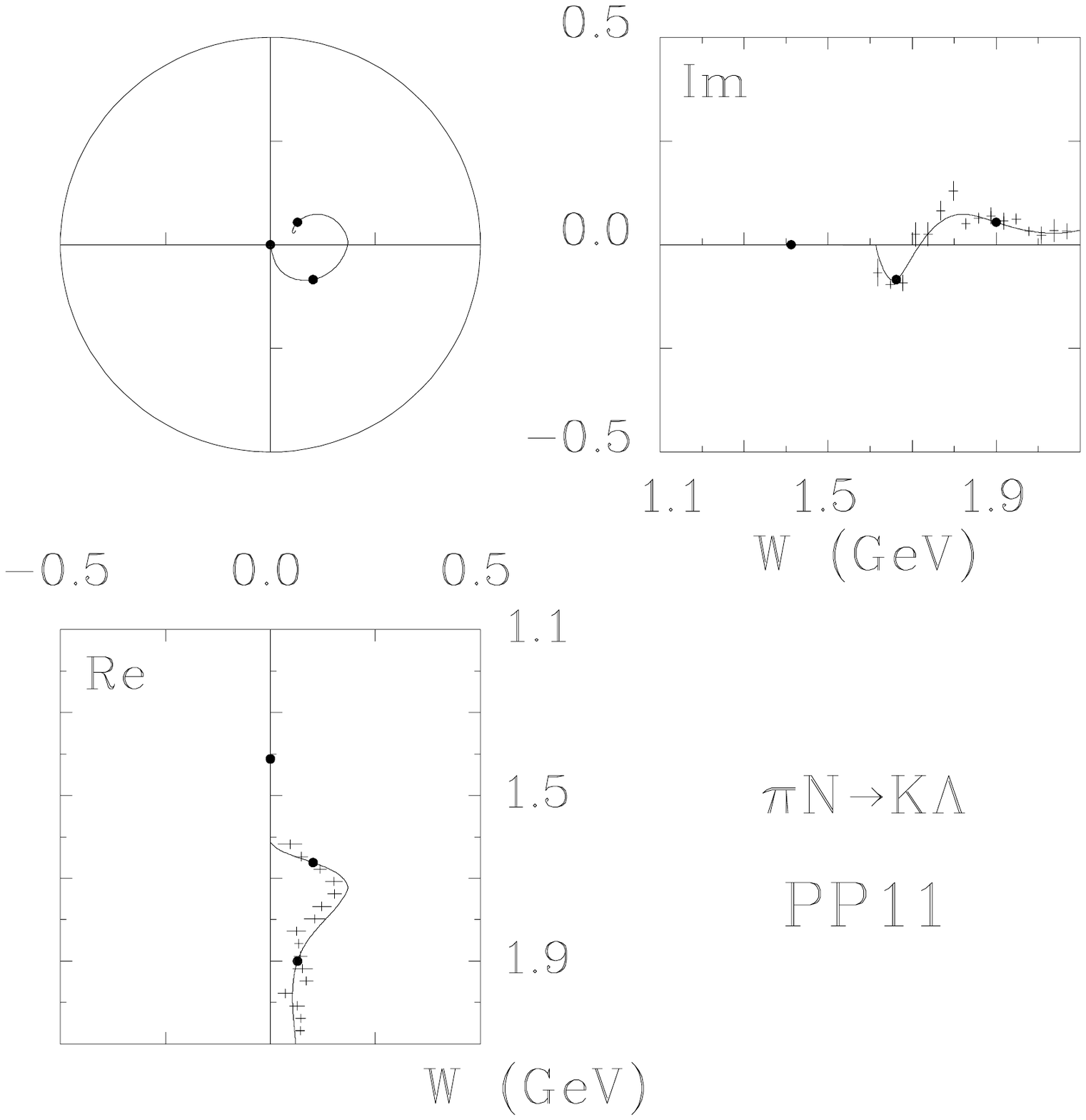}}
\caption{Argand diagrams for two-body amplitudes.}
\end{figure*}

\begin{figure*}[htpb]
\addtocounter{figure}{-1}
%\caption{Cont'd.}
\vspace{-10mm}
\scalebox{0.35}{\includegraphics*{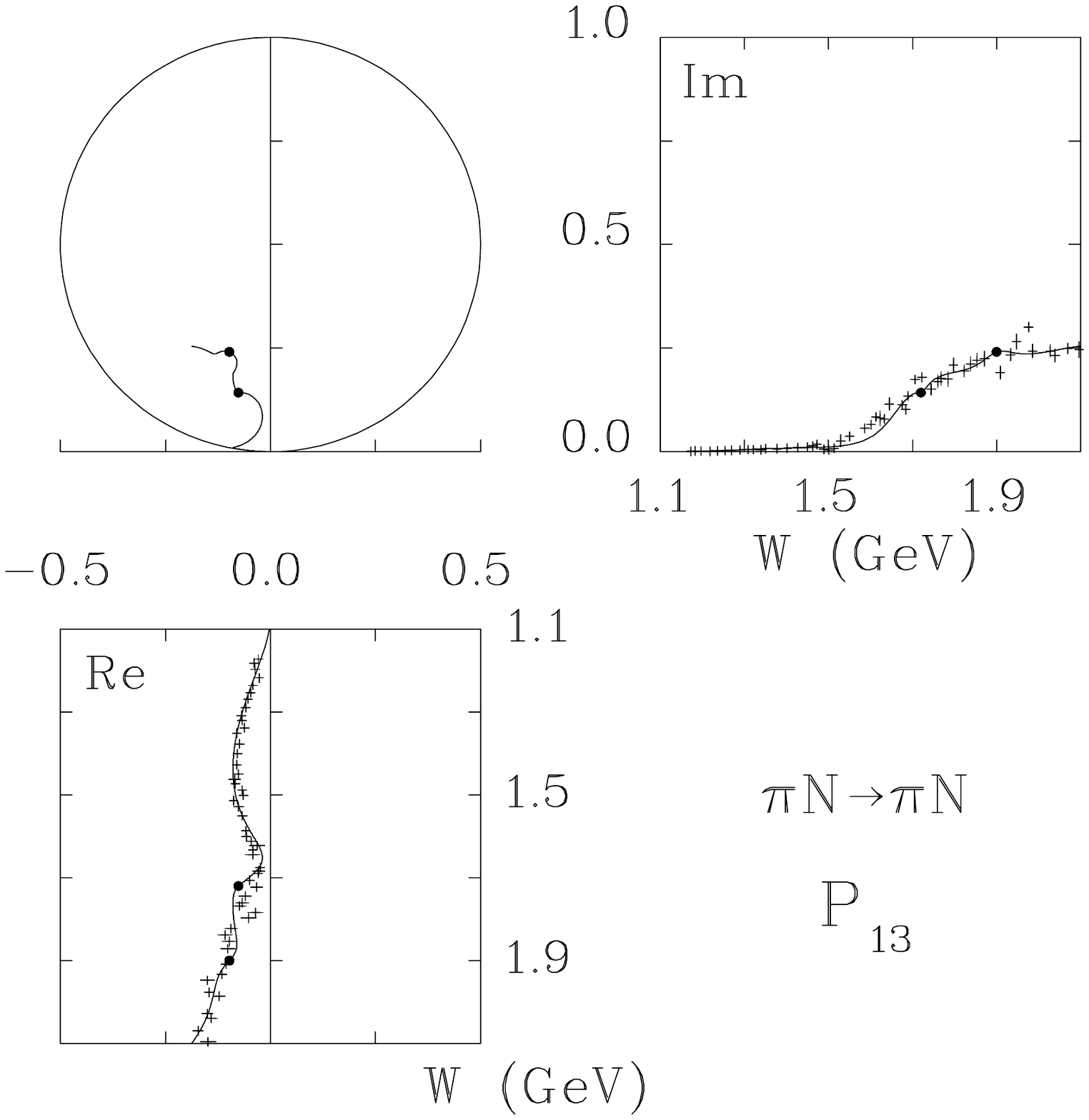}}
\vspace{-25mm}
\scalebox{0.35}{\includegraphics*{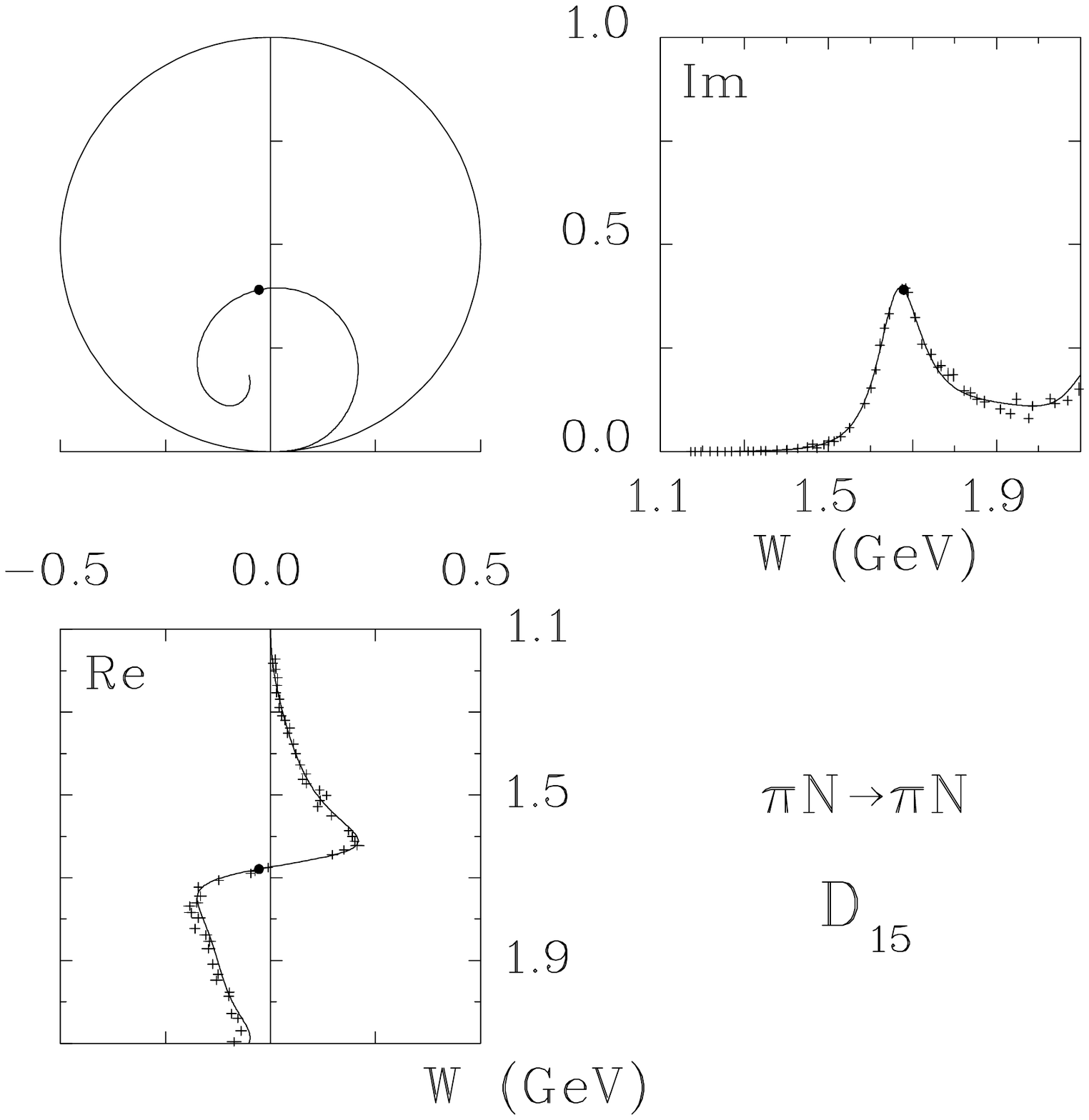}}
\vspace{-25mm}
\scalebox{0.35}{\includegraphics*{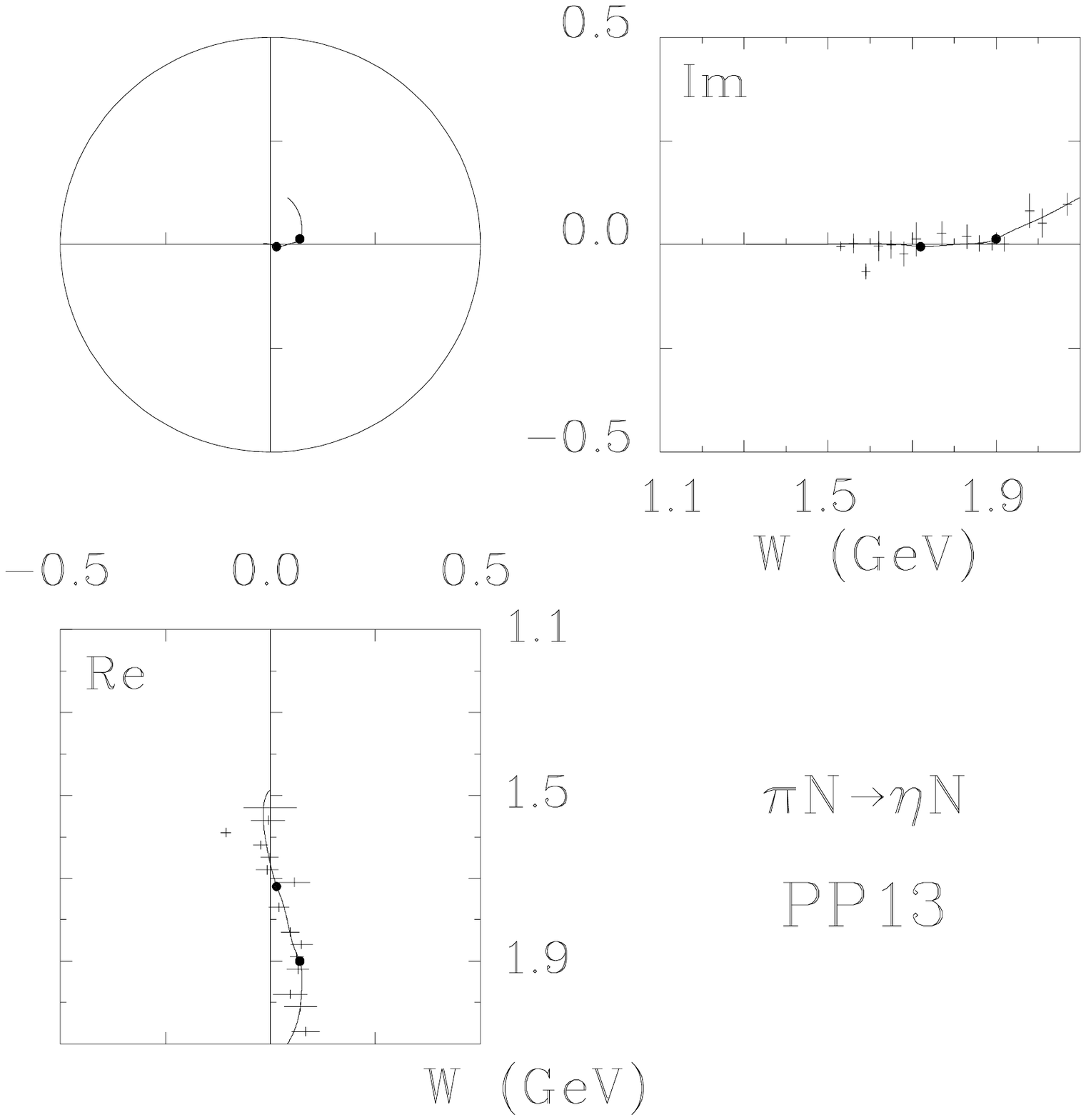}}
%\vspace{15mm}
\scalebox{0.35}{\includegraphics*{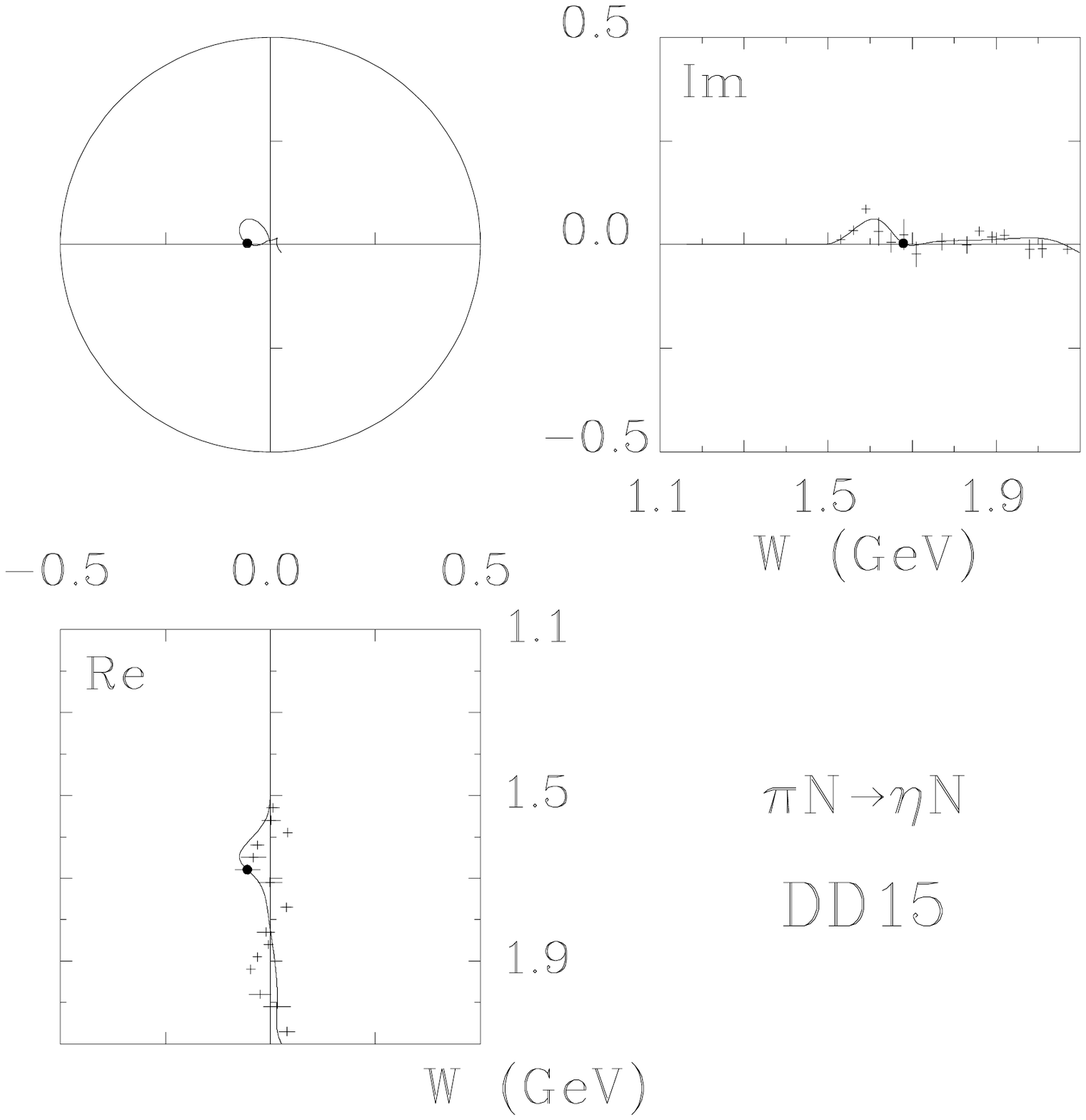}}
\vspace{-15mm}
\scalebox{0.35}{\includegraphics*{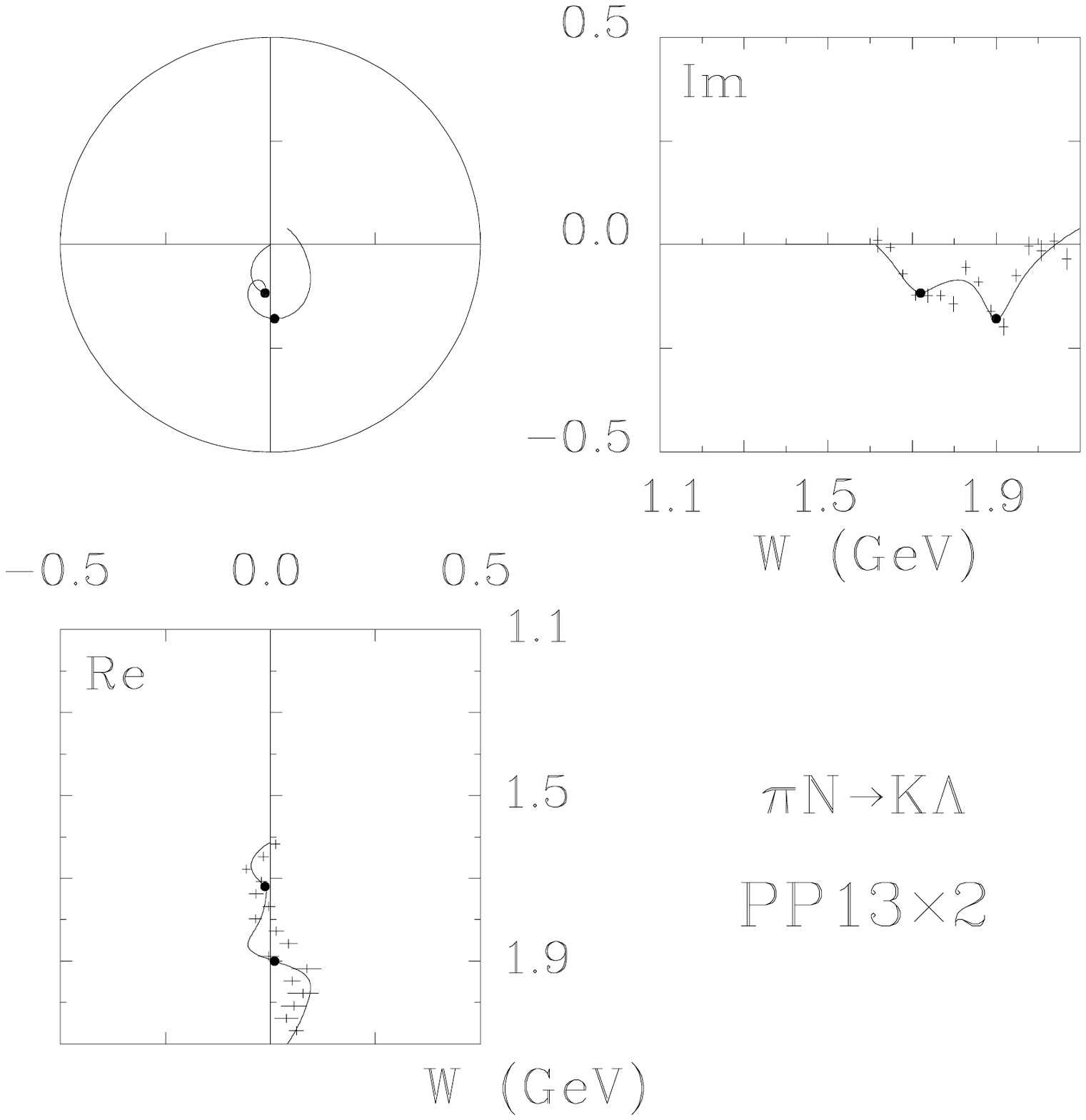}}
\scalebox{0.35}{\includegraphics*{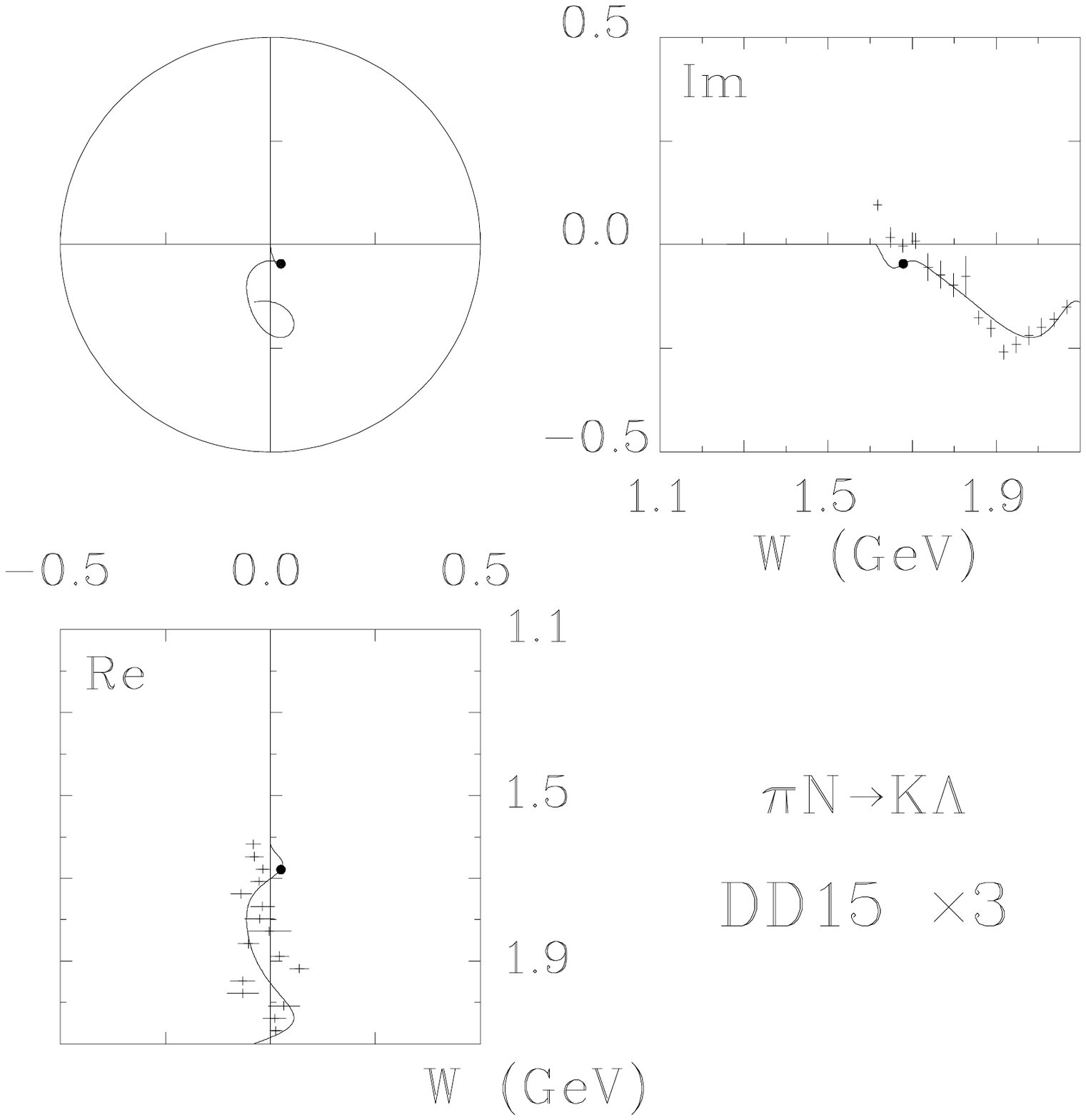}}
\caption{Cont'd.}
\end{figure*}

\begin{figure*}[htpb]
\addtocounter{figure}{-1}
\vspace{-10mm}
\scalebox{0.35}{\includegraphics*{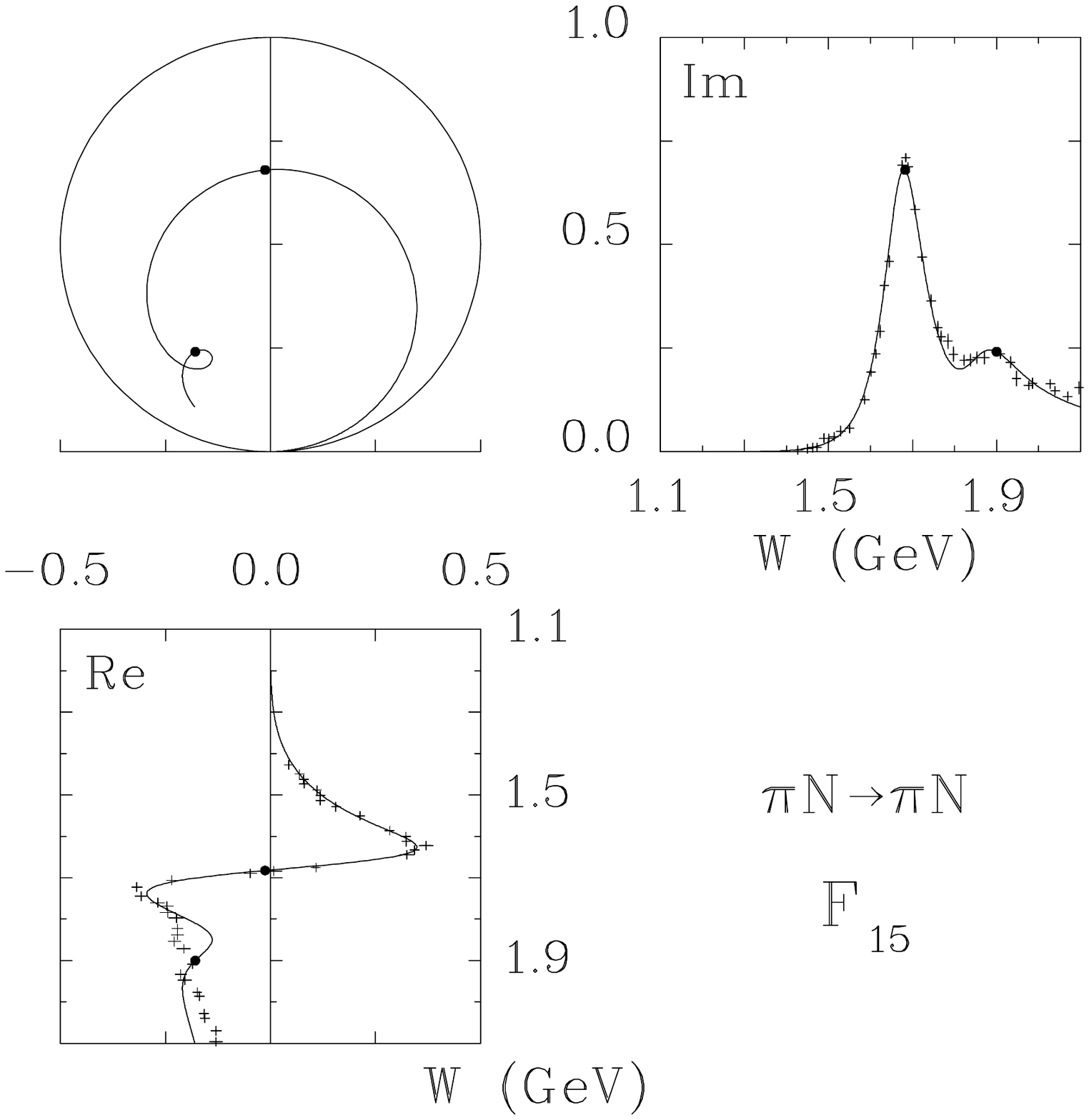}}
\vspace{-25mm}
\scalebox{0.35}{\includegraphics*{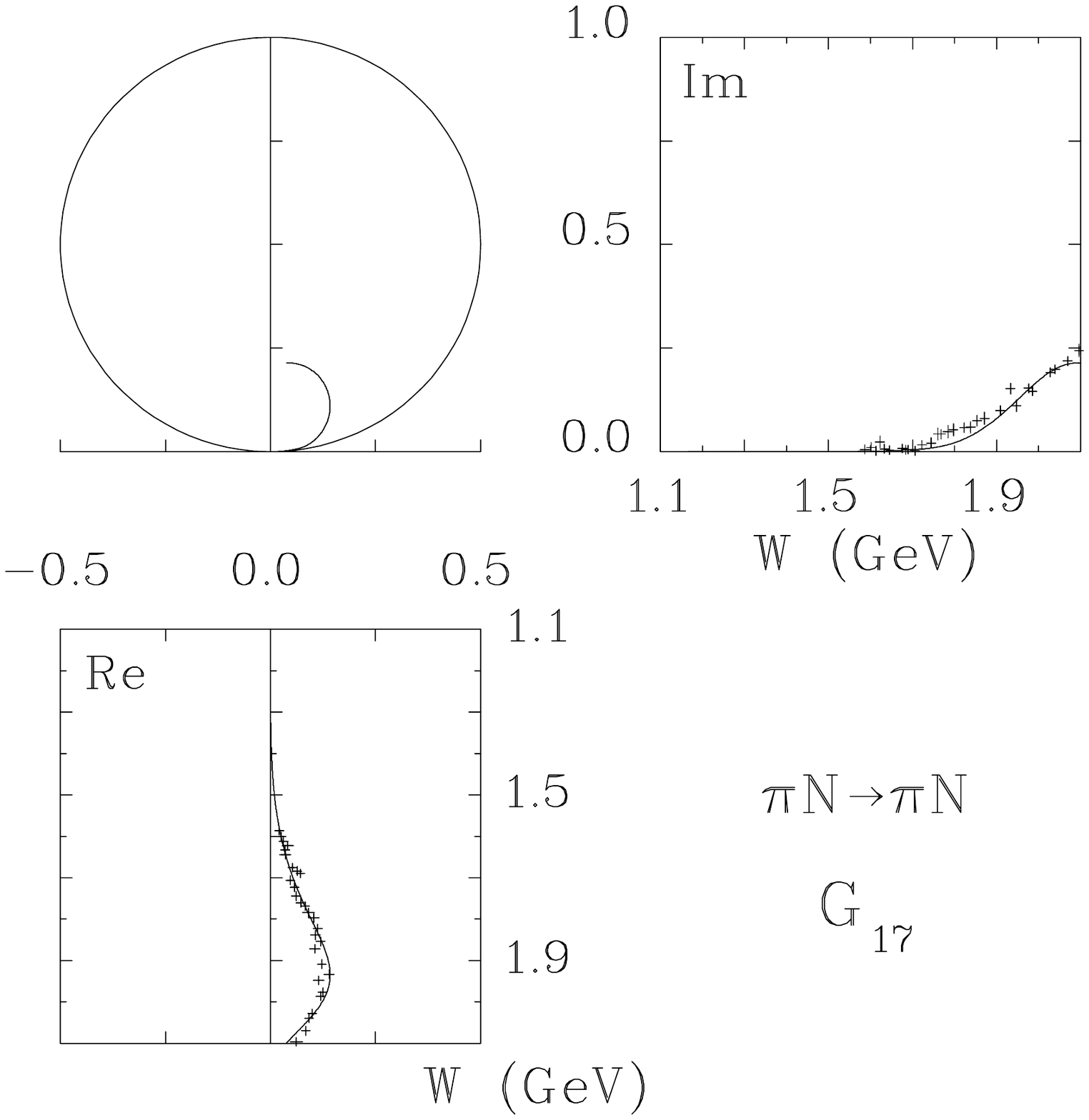}}
%\vspace{-25mm}
\scalebox{0.35}{\includegraphics*{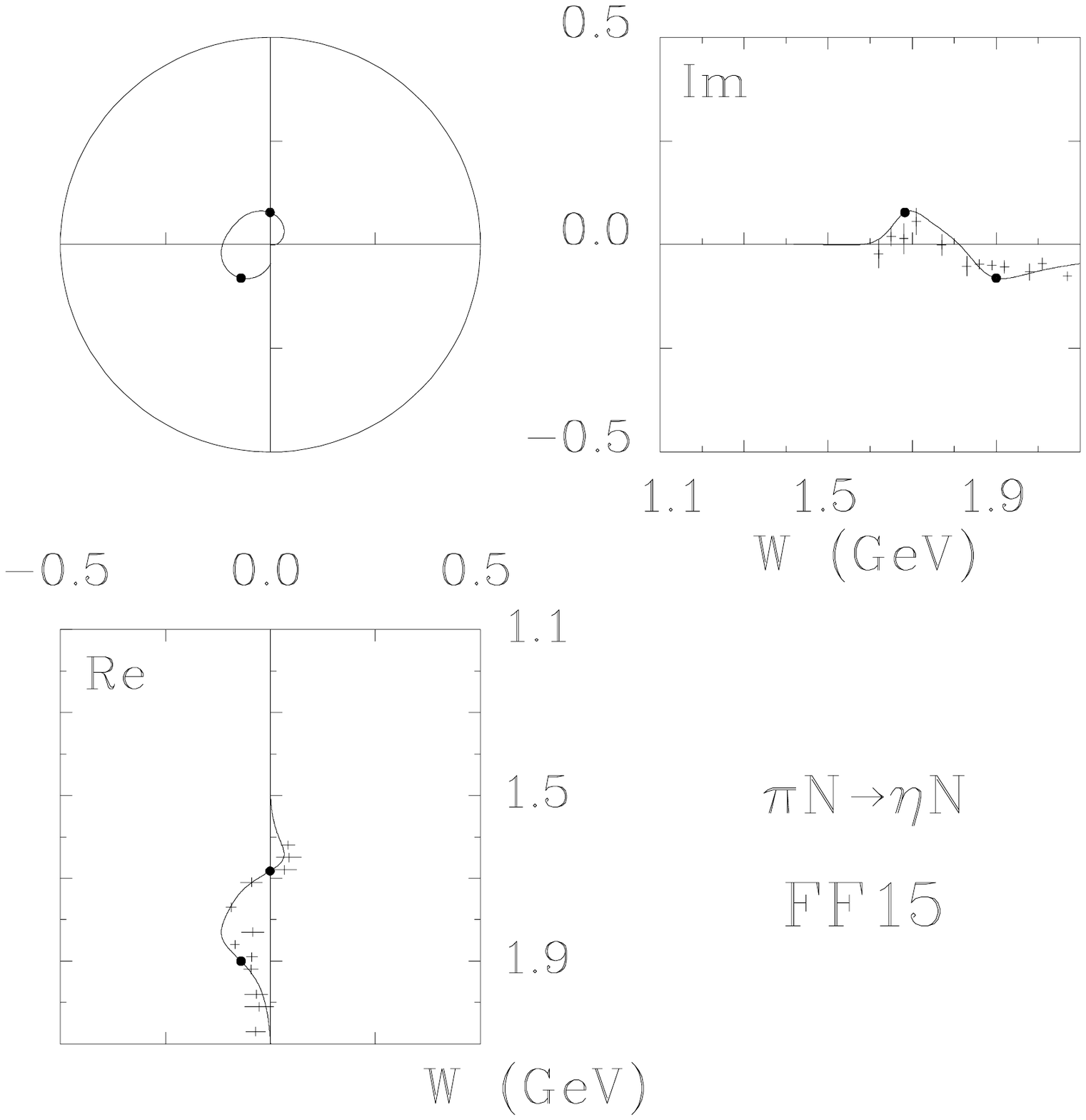}}
\vspace{-25mm}
\scalebox{0.35}{\includegraphics*{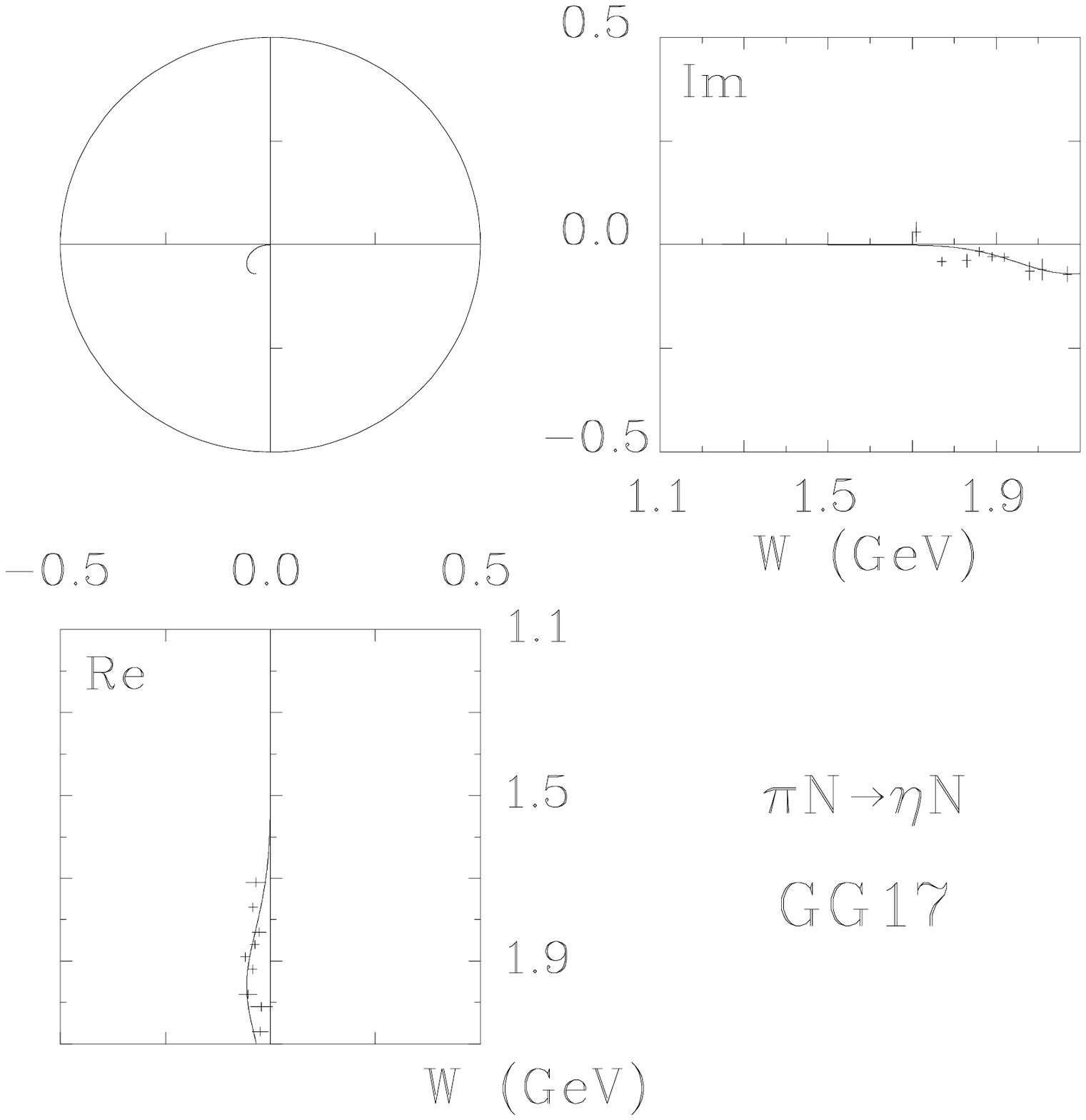}}
\vspace{-15mm}
\scalebox{0.35}{\includegraphics*{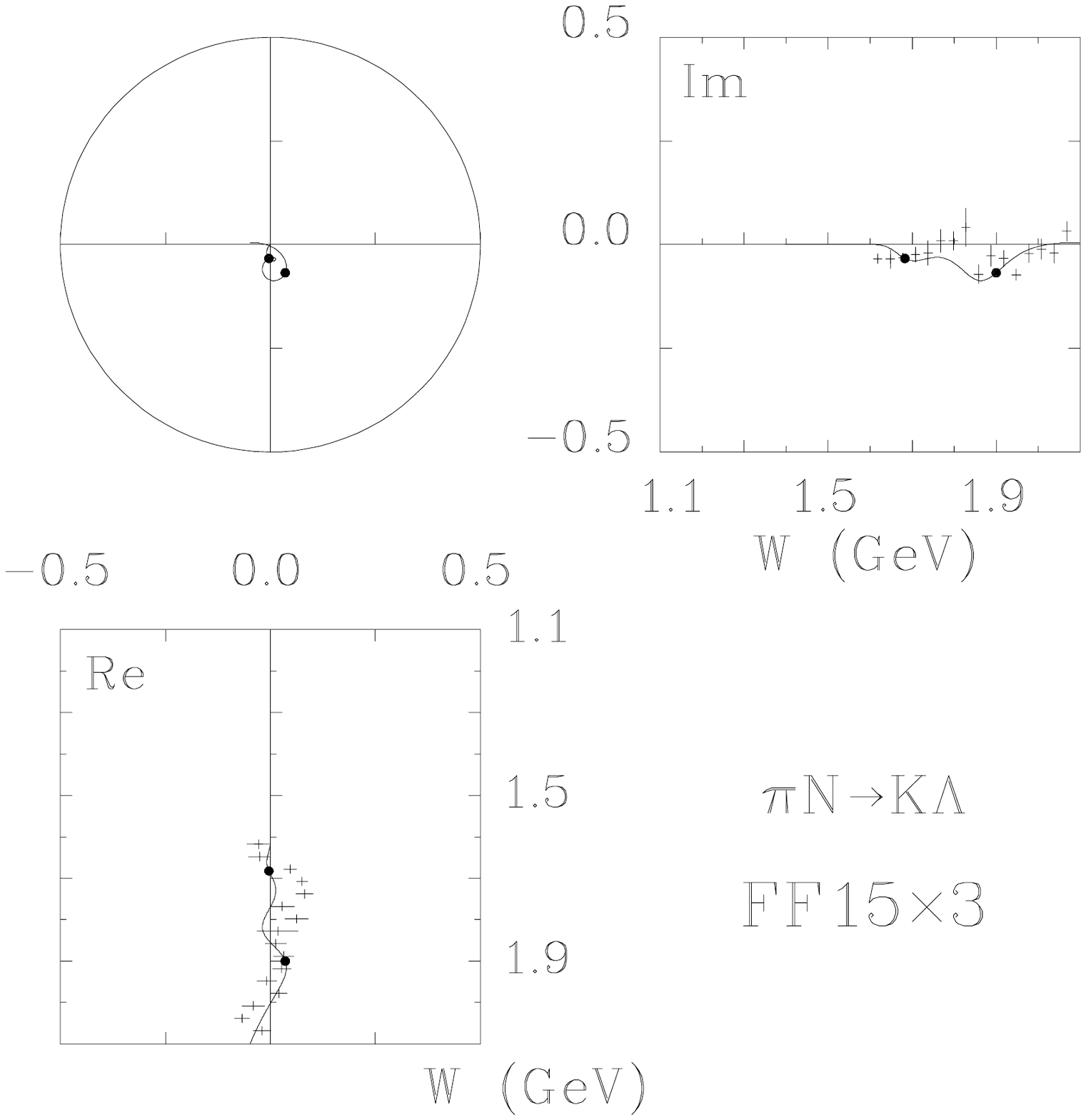}}
\scalebox{0.35}{\includegraphics*{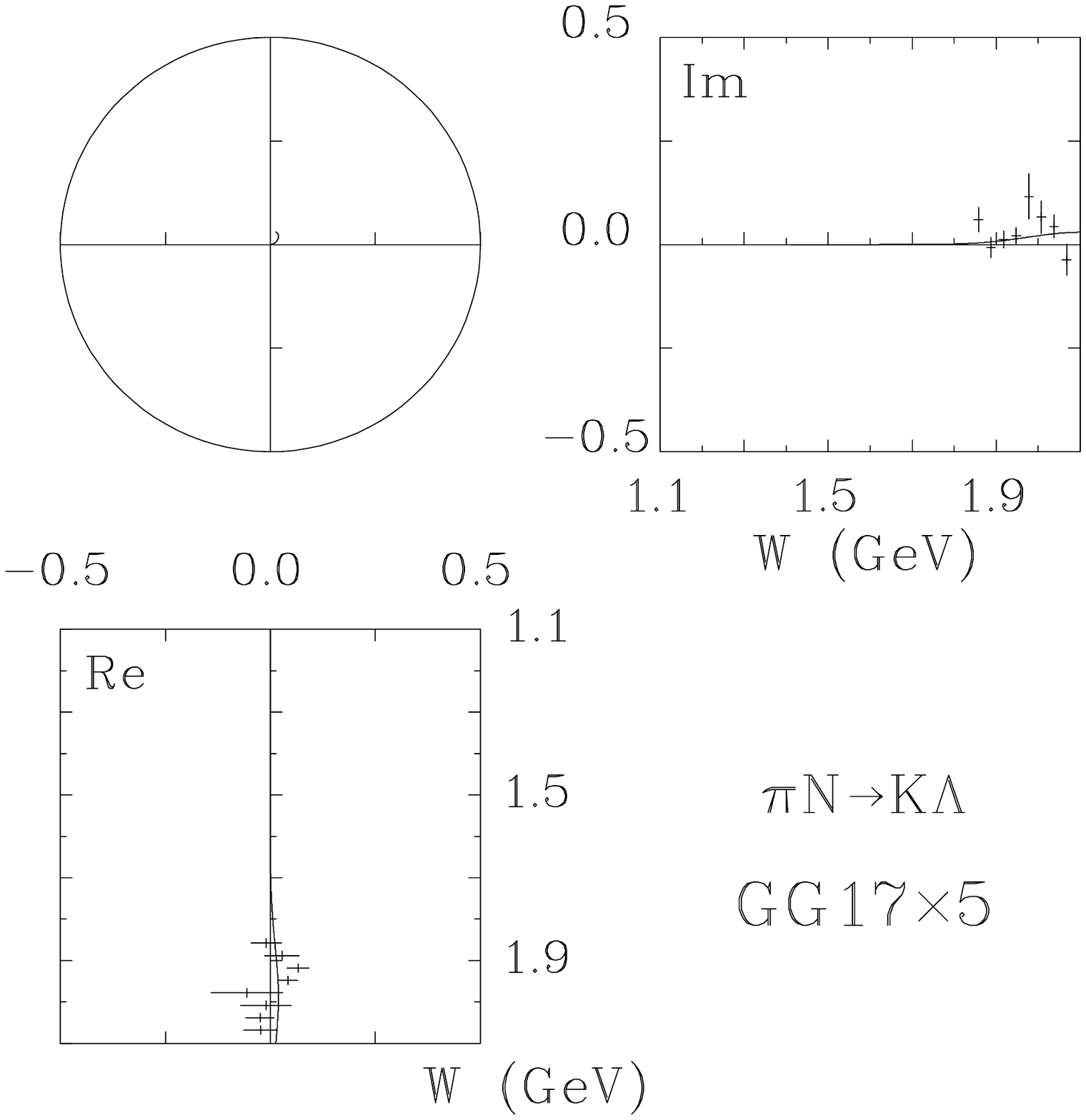}}
\caption{Cont'd.}
\end{figure*}

%\begin{comment}
%%\newpage
\begin{figure*}[htpb]
%\caption{Argand diagrams for two body or quasi-two body amplitudes}
%\vspace{-15mm}
\addtocounter{figure}{-1}
\vspace{-10mm}
\scalebox{0.35}{\includegraphics*{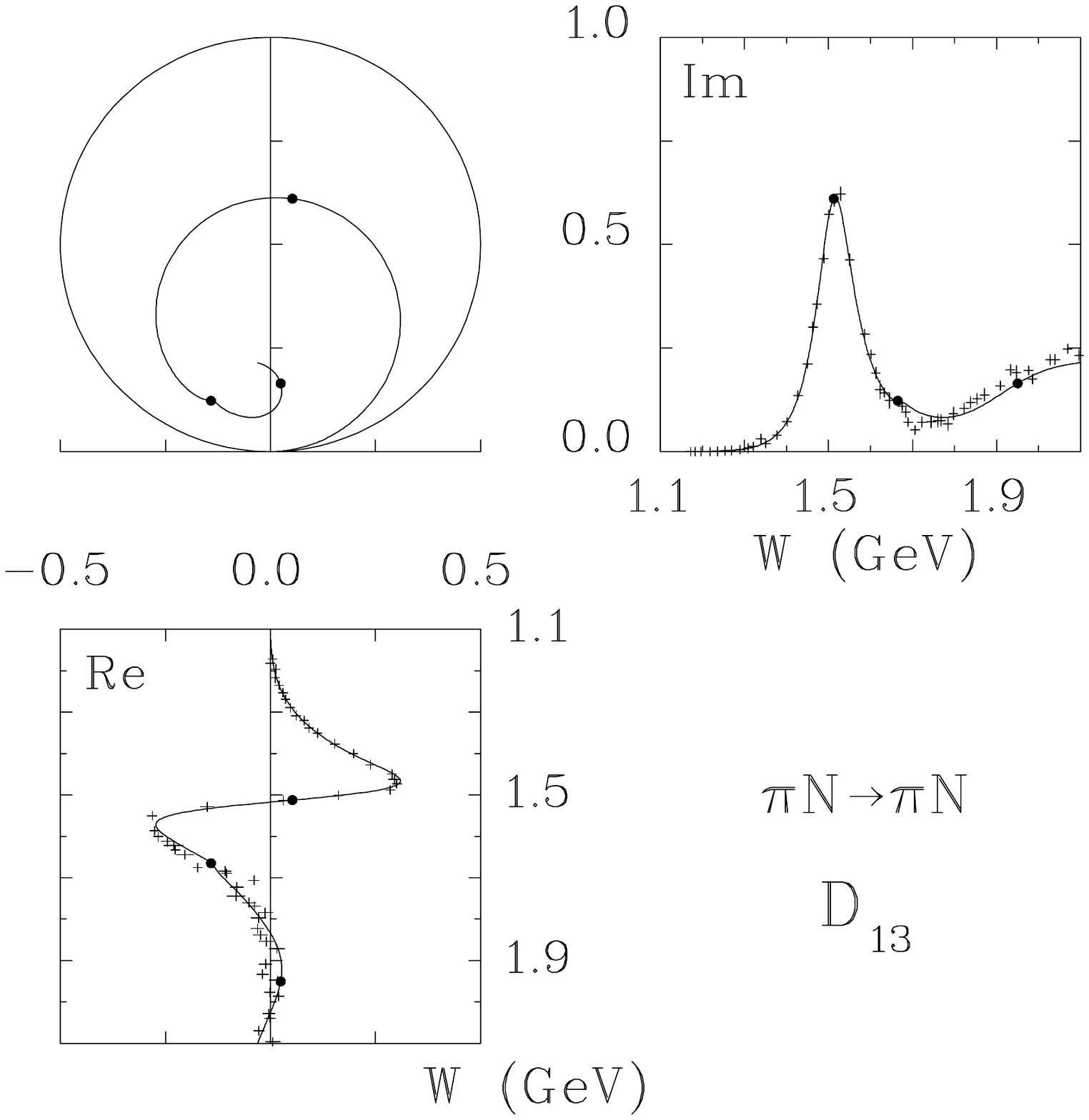}}
\vspace{-25mm}
\scalebox{0.35}{\includegraphics*{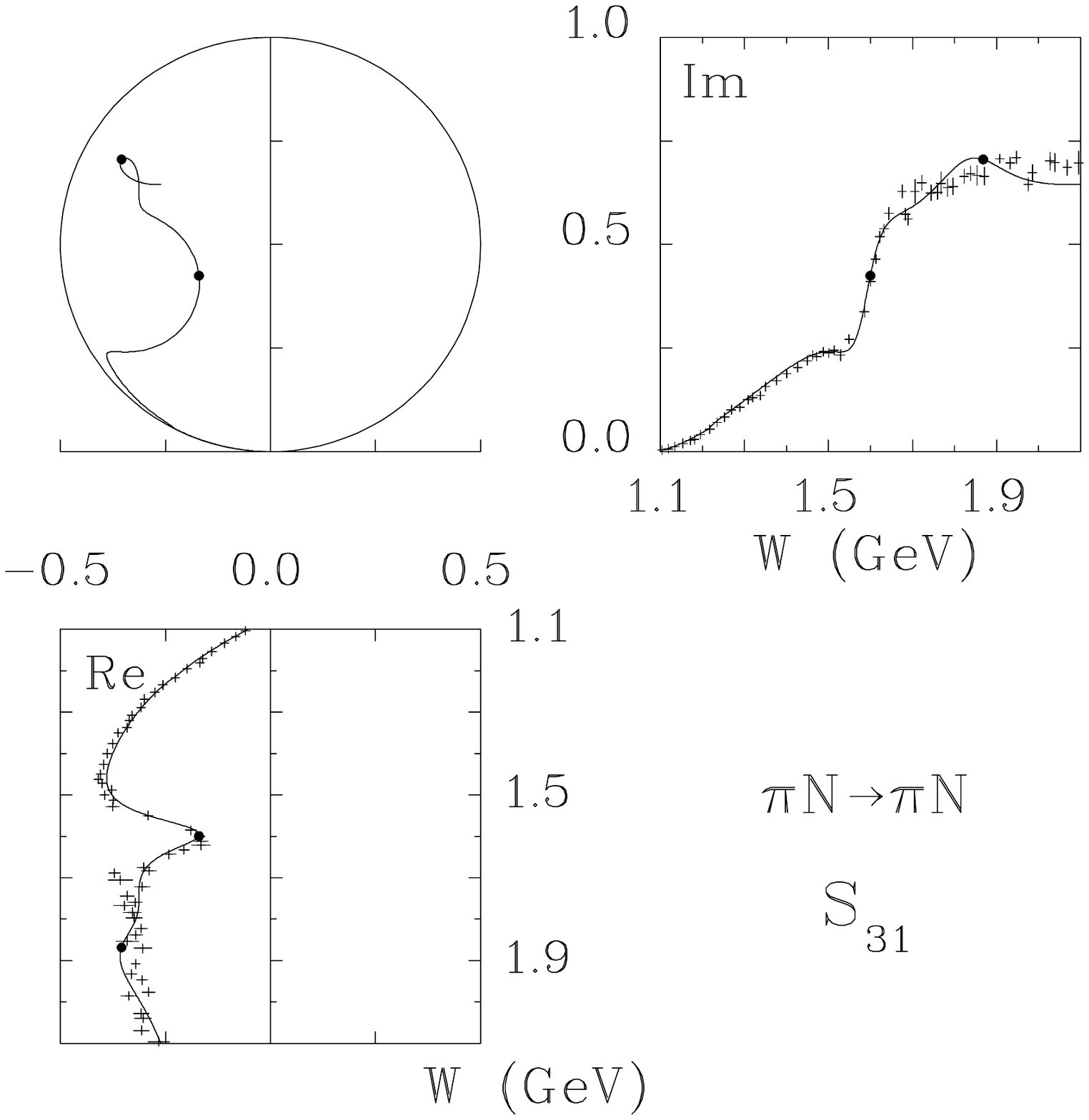}}
\vspace{-25mm}
%\vspace{15mm}
\scalebox{0.35}{\includegraphics*{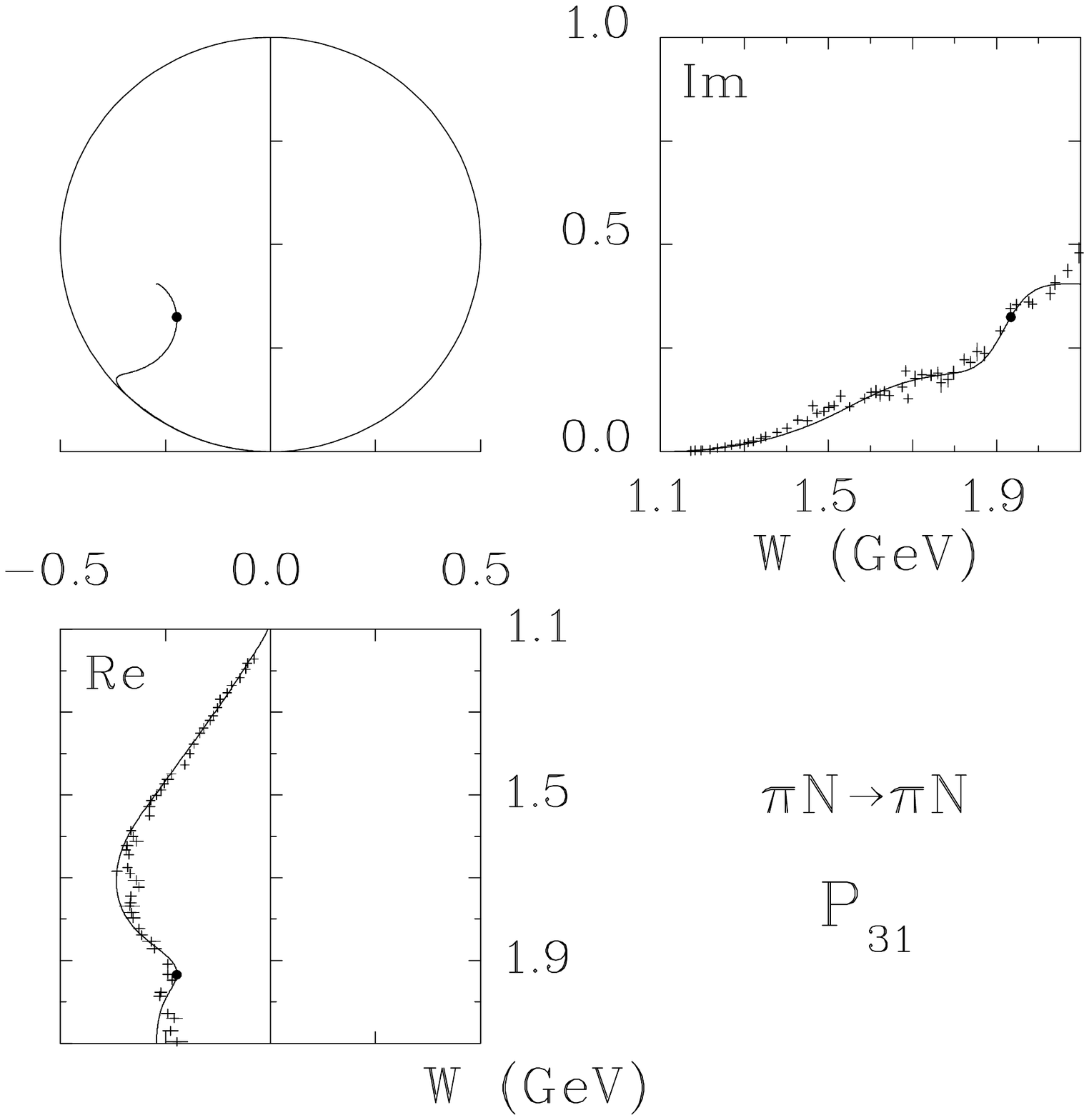}}
\scalebox{0.35}{\includegraphics*{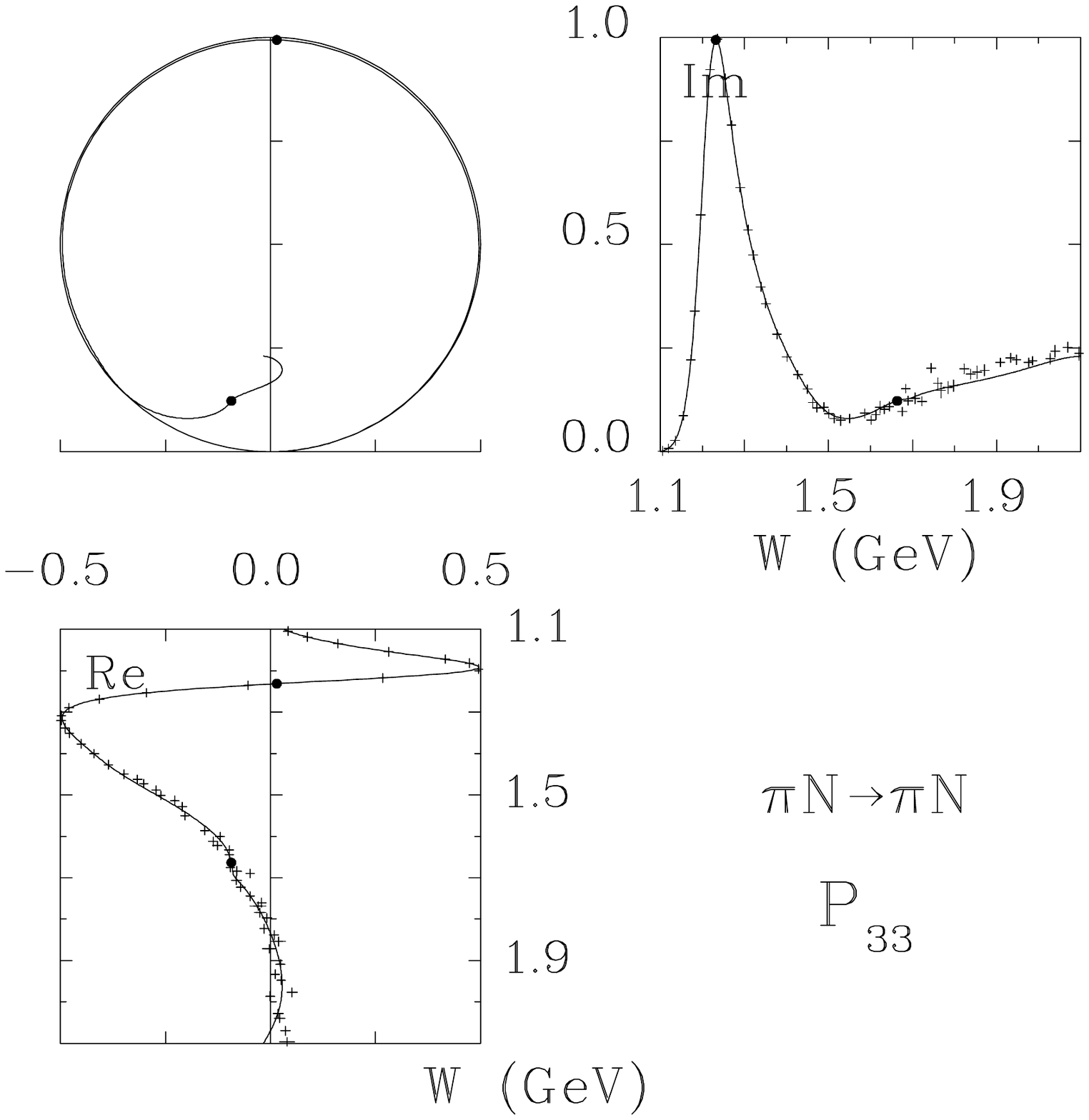}}
\vspace{-15mm}
\scalebox{0.35}{\includegraphics*{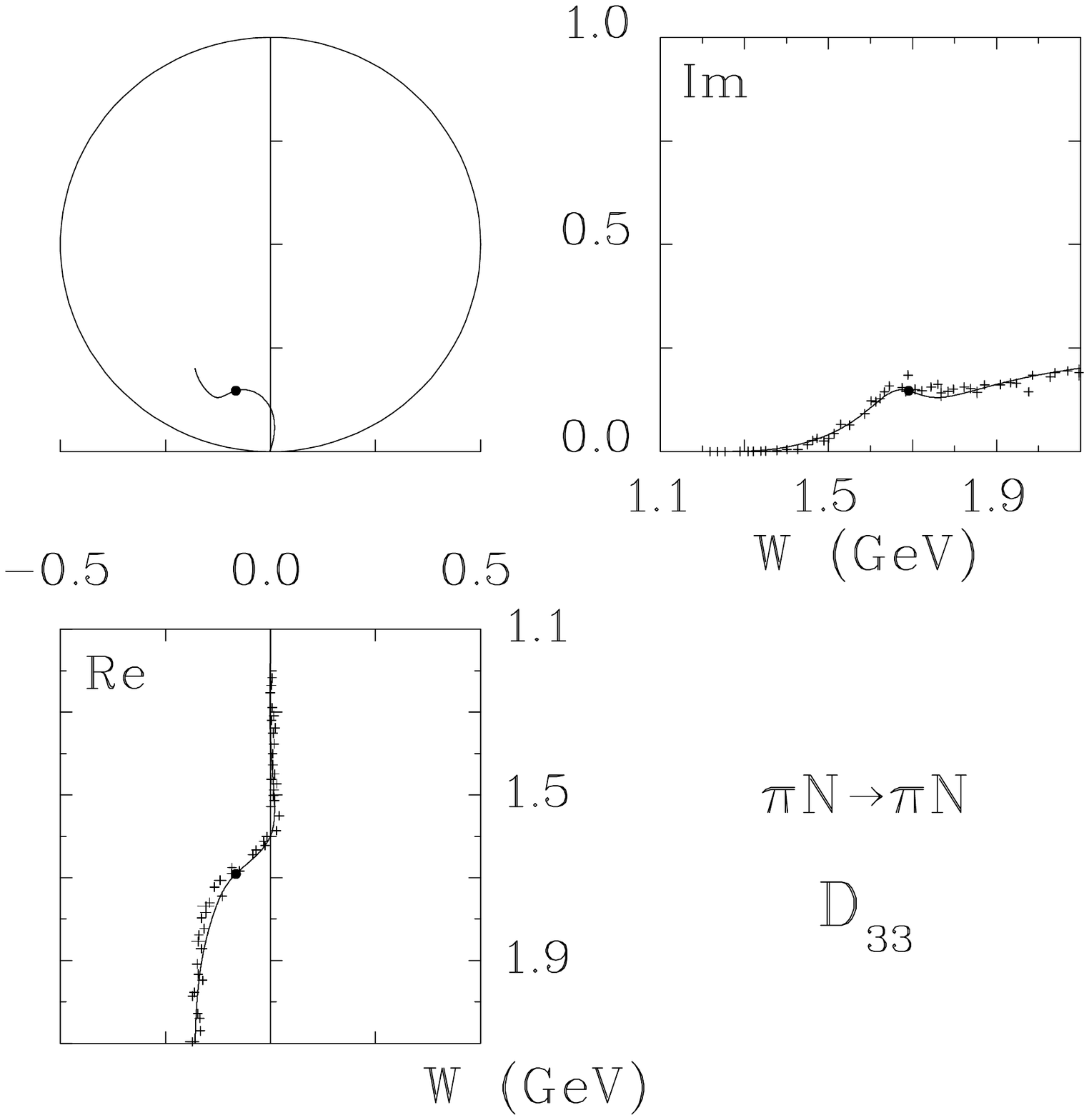}}
\scalebox{0.35}{\includegraphics*{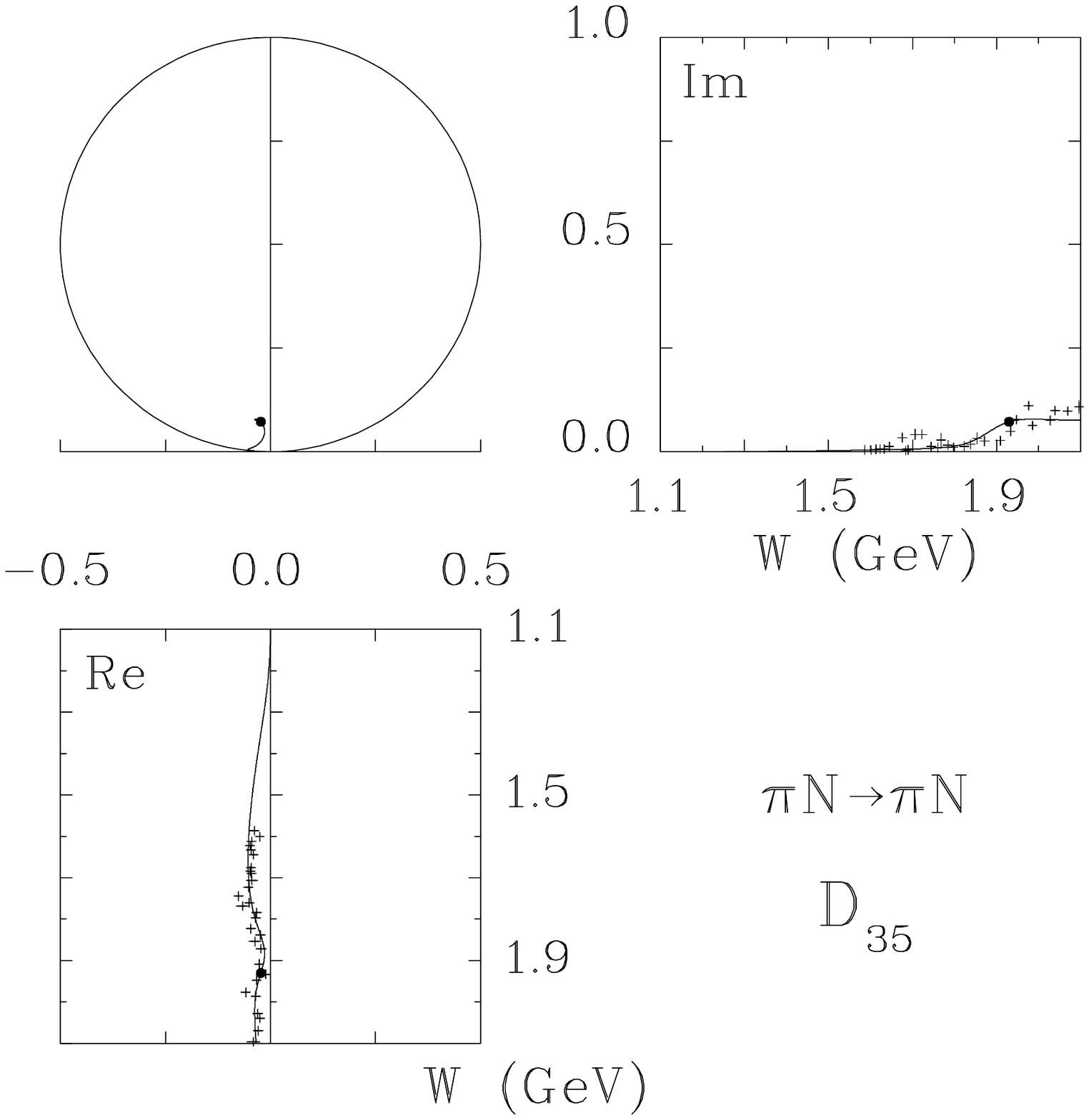}}

\caption{Cont'd.}
\end{figure*}

%\begin{comment}
\begin{figure*}[htpb]
\addtocounter{figure}{-1}
%\caption{Cont'd.}
%\vspace{-10mm}
\scalebox{0.35}{\includegraphics*{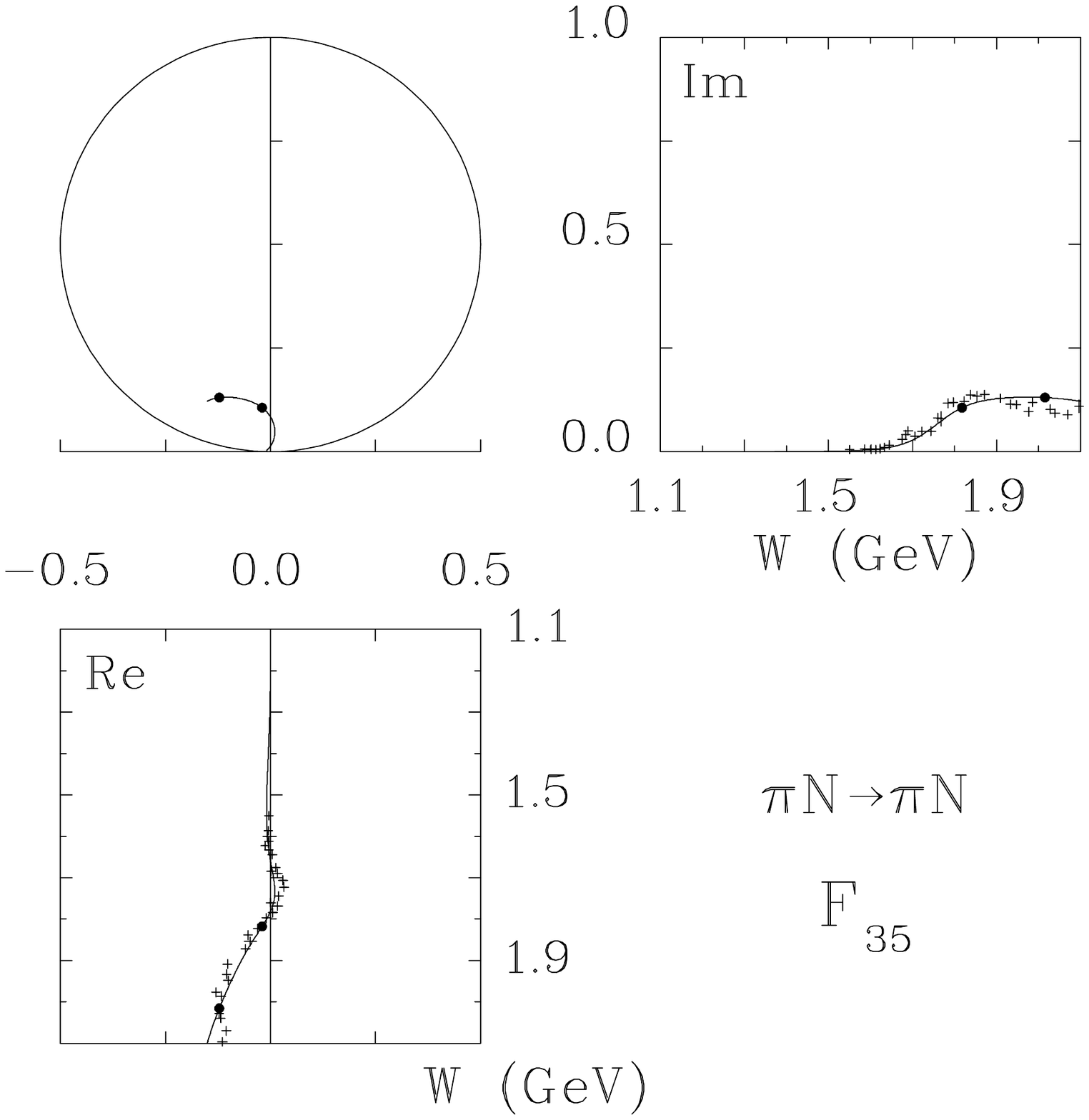}}
\scalebox{0.35}{\includegraphics*{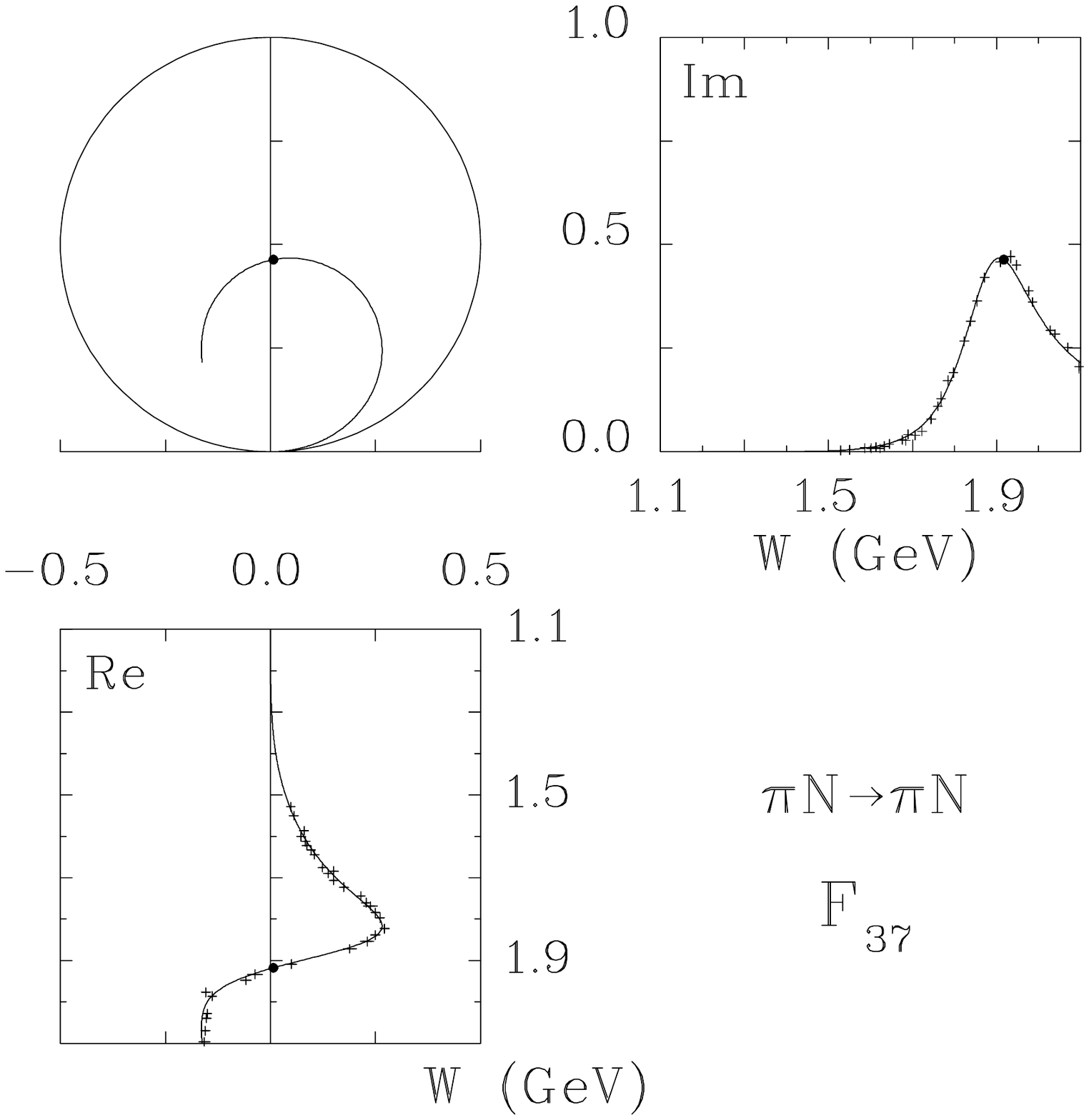}}
\vspace{-15mm}
%\scalebox{0.35}{\includegraphics*{../d35/plot1.pdf}}
%\vspace{15mm}
%\scalebox{0.35}{\includegraphics*{../p33/plot5.pdf}}
%\vspace{15mm}
%\scalebox{0.35}{\includegraphics*{../d35/plot6.pdf}}
%\vspace{15mm}
%\scalebox{0.35}{\includegraphics*{../p33/plot6.pdf}}
%\scalebox{0.35}{\includegraphics*{../d35/plot7.pdf}}
\caption{Cont'd.}
\end{figure*}
%\end{comment}
%The hadronic resonance parameters for states with I = $1/2$ and I = $3/2$ are listed in Tables I and II, respectively. %The helicity amplitudes from this work as a result of inclusion of pion-photoproduction channels are compiled in Table III.
%The rest of the tables are dedicated for comparison of the resonance parameters ,the helicity amplitudes and pole positions with other analyses.  
%\begin{longtable*}{c|c|c|c|c}

$\bf{S_{11}}$:\\
This partial wave was fitted with three resonances. The first resonance occurred with a mass M = $1538\pm 1$ MeV and width $\Gamma$ = $141\pm4$ MeV and corresponds to the 4* $S_{11}(1535)$. The strength of this resonance divides more or less equally to $\pi N$ and $\eta N$ at 37\% and 41\%, respectively, with the remainder going to $\pi\pi N$ channels. Our results for this state agree quite well with those from previous analyses, especially Ref. \cite{hoehler79}. The second resonance was seen to have M =  $1664\pm2$ MeV and $\Gamma$ = $126\pm3$ MeV corresponding to the 4* $S_{11}(1650)$. Decay modes for this state are primarily $\pi N$, $\eta N$, $K\Lambda$, $\pi\Delta$, and $\rho_1N$. We found the third resonance with M = $1910\pm15$ MeV and $\Gamma = 502\pm47$ MeV corresponding to the 2* $S_{11}(1895)$. There is striking resemblance of our $\eta N$ amplitude for the partial wave $S_{11}$ with one solution presented by Batini\'c {\it et al.\ }\cite{batanic95}.

For $S_{11} (1535)$, our pole mass $M_p = 1515$ MeV and pole width $\Gamma_p = 123$ MeV are in good agreement with previous analyses, especially that by Arndt {\it et. al.\ }\cite{arndt06} and the same is true with the second resonance $S_{11} (1650)$ with $M_p = 1655$ MeV and $\Gamma_p = 123$ MeV. For the $S_{11} (1895)$, $M_p = 1858$ MeV and $\Gamma_p = 479$ MeV.

$\bf{P_{11}}$:\\
This partial wave was fitted with four resonances. The first resonance occurred at $M = 1412\pm2$ MeV with $\Gamma = 248\pm 5$ MeV and corresponds to the 4* $P_{11} (1440)$ . These results agree quite well with those from prior analyses, especially Ref.\ \cite{cutkosky80}. The decay modes are $\pi N$ and $\pi \pi N$ channels.
Our analysis confirms the existence of the state $P_{11}$(1710), which is refuted or marked uncertain by the GWU analysis \cite{arndt06}. This resonance occurred at $M=1662\pm 7$ MeV with $\Gamma = 116\pm17$ MeV agreeing with the previous analysis by Cutkosky {\it et al.\ }\cite{cutkosky80}. Its elasticity is about 17\%. The branching ratios for the $\eta N$ and K$\Lambda$ channels are about 11\% and 8\%, respectively. The third resonance occurred at $M=1900\pm36$ MeV with $\Gamma = 485\pm142$ MeV. This resonance corresponds to the 2* $P_{11} (1880)$. The major decay modes are $\pi N$, $\eta N$, and $K\Lambda$. A fourth resonance was included at $M=2250\pm116$ MeV and $\Gamma=600\pm394$ MeV. During the fits, the mass of this resonance was held fixed but in the final zero-iteration fit we treated it as a free parameter.

For $P_{11}(1440)$, our pole mass $M_p = 1370$ MeV and pole width $\Gamma_p = 214$ MeV are in good agreement with prior analyses, especially that by Cutkosky {\it et al.\ }\cite{cutkosky80}. For $P_{11}(1710)$ our pole mass of $M_p = 1644$ MeV is slightly smaller than those given in previous analyses but the pole width $\Gamma_p = 104$ MeV is comparable with the results of Cutkosky and Wang \cite{cutkosky90} and Cutkosky {\it et al.\ }\cite{cutkosky80}.

$\bf{P_{13}}$:\\
This partial wave was fitted with two resonances. The first resonance occurred at $M=1720\pm5$ MeV with $\Gamma = 200\pm20$ MeV and can be identified with the 4* $P_{13}(1720)$. Our results for this state agree very well with those from prior analyses, especially Ref. \cite{hoehler79}. The major decay modes were found to be $\pi N$ (14\%), $\rho_1N (1\%)$, and $K\Lambda$ (3\%). We found no coupling to the $\eta N$ channel. Most of the flux was seen to be carried by dummy $\rho\Delta$ and $\omega N$ channels. The second resonance corresponds to the 2* $P_{13}(1900)$. The decay channels for this resonance were found to be $\pi N$ (7\%), $\rho_1 N$ (64\%), and $K\Lambda$ (14\%). Again no $\eta N$ coupling was seen. 

For $P_{13}(1720)$, the pole mass $M_p = 1687$ MeV and the pole width $\Gamma_p = 175$ MeV are in good agreement with previous analyses, especially that by H\"ohler \cite{hoehler93}.

$\bf{D_{13}}$:\\
This partial wave was fitted with four resonances. The first resonance occurred at $M=1512.6\pm0.5$ MeV with $\Gamma= 117\pm1$ MeV, which corresponds to the 4* $D_{13} (1520)$. Our results for this state agree very well with those from prior analyses. This state was seen to be highly elastic (63\%) and other major decay modes were found to be $(\pi\Delta)_S$ (9\%), $(\pi\Delta)_D$ (6\%), and $\rho_3 N$ (21\%). The second resonance occurred at $M=1665\pm3$ MeV with $\Gamma=56\pm8$ MeV. This state was found to be highly inelastic with major decay channels $(\pi\Delta)_S$ (31\%), $\rho_3 N$ (38\%), and $\sigma N$ (24\%). The third resonance occurred at $M=1951\pm27$ MeV with $\Gamma=500\pm45$ MeV. Its major decay channel was $(\pi\Delta)_S$ (87\%) and its elasticity was found to be about 7\%. We didn't find any coupling to $\eta N$ and $K\Lambda$ channels with any of these excited states. The fourth resonance was included at $M=2200\pm39$ MeV and $\Gamma=750\pm101$ MeV.

For $D_{13}(1520)$, the pole mass $M_p = 1501$ MeV and the pole width $\Gamma_p = 112$ MeV are in very good agreement with previous analyses. For $D_{13}(1700)$, our pole mass $M_p = 1662$ MeV agrees quite well with the analysis by Cutkosky {\it et al.\ }\cite{cutkosky80} and our pole width of $\Gamma_p = 55$ MeV as well. For $D_{13}(1875)$ the pole mass of $M_p = 1975$ MeV agrees well with the previous analyses within uncertainties but our pole width $\Gamma_p = 495$ MeV is greater than those given in the prior analyses.   
      
$\bf{D_{15}}$:\\
Two resonances were required to fit this partial wave. The first resonance occurred at $M=1679\pm1$ MeV with $\Gamma = 145\pm4$ MeV. It corresponds to the 4* $D_{15} (1675)$. Our results for this state agree very well with those from previous analyses. The major decay channels were $\pi N$ (39\%), and $\pi\Delta$ (46\%) with tiny couplings for $\eta N$ and $K\Lambda$. The second resonance was included at $M=2116\pm21$ MeV with $\Gamma = 307\pm112$ MeV.

For $D_{15}(1675)$, the pole mass $ M_p = 1656$ MeV and the pole width $\Gamma = 128$ MeV agree very well with previous analyses.

$\bf{F_{15}}$:\\
This partial wave was fitted with two resonances. The first resonance occurred at  $M= 1682.7\pm0.5$ MeV with $\Gamma=126\pm1$ MeV and can be identified with the 4* $F_{15} (1680)$. This resonance was found to be highly elastic with an elasticity of 68\%. The elastic amplitude for this state exhibited classic Breit-Wigner behavior. The other hadronic channels were $(\pi\Delta)_P$ (11\%), $(\pi\Delta)_F$ (1\%), $(\rho_3N)_P$ (7\%), $(\rho_3N)_F$ (2\%), $\sigma N$ (9\%), $\eta N$ (about 1\%), and $K\Lambda$ ($< 1\%$). The second resonance occurred at $M=1900\pm7$ MeV with $\Gamma=219\pm23$ MeV and corresponds to the 2* $F_{15} (1860)$. In our initial fits we kept this mass fixed. This resulted in a small wiggle in the real part of the elastic amplitude near 1900 MeV, thereby disagreeing slightly with the GWU single-energy solution. The major decay modes were $\pi N$ (17\%), $(\rho_3 N)_F$ (34\%), and $\sigma N$ (41\%).%, and with greater uncertainties for $\eta N$, $K\Lambda$, and other $\pi \pi N$ channels.

For $F_{15}(1680)$, the pole mass $M_p = 1669$ MeV and the pole width $\Gamma_p = 119$ MeV agree very well with previous analyses. For $F_{15}(1860)$, the pole mass $M_p = 1863$ MeV and pole width $\Gamma_p = 189$ MeV are slightly larger than those presented in the analysis by Arndt {\it et al.\ }\cite{arndt06}.

$\bf{F_{17}}$:\\
This partial wave was fitted with a single resonance at $M=1990\pm45$ MeV and $\Gamma=203\pm161$ MeV. This resonance is highly inelastic with an elasticity of only 2\%. The coupling to the $K\Lambda$ channel is also small with a branching ratio of $< 1\%$.

For $F_{17}(1990)$ the pole mass $M_p = 1941$ MeV agrees quite well with that by Cutkosky {\it et al.\ }\cite{cutkosky80} but our pole width $\Gamma_p = 130$ MeV is about half the value from that analysis.
\\
\\
$\bf{G_{17}}$:\\
This partial wave was fitted with one resonance at $M=2150\pm26$ MeV with $\Gamma=500\pm74$ MeV. This resonance corresponds to the 4* $G_{17} (2190)$. The decay channels were $\pi N$ (20\%), $(\rho_3 N)_D$ (9\%), $\eta N$ (2\%), and $K\Lambda$ ($< 1\%$), with most of the strength carried by a dummy $\omega N$ channel.

For $G_{17}(2190)$ the pole mass $M_p = 2062$ MeV and the pole width $\Gamma_p = 428$ MeV agree more or less with previous analyses.

$\bf{S_{31}}$:\\
Two resonances were required to fit this partial wave. The first resonance occurred at $M=1600\pm1$ MeV with $\Gamma=112\pm2$ MeV and can be identified with the 4* $S_{31} (1620)$. The primary decay modes for this state were found to be $\pi N$ (33\%), $(\pi\Delta)_D$ (32\%), $\rho_1N$ (26\%), and $\pi N^*$ (9\%). The second resonance occurred at $M=1868\pm 12$ MeV with $\Gamma = 234\pm27$ MeV and corresponds to the 2* $S_{31} (1900)$. The major decay modes were found to be $\pi N$ (8\%), $\pi\Delta$ (56\%), $\rho_3 N$ (23\%), and $\pi N^*$ (12\%).

For $S_{31}(1620)$ the pole mass and width were found to be $M_p = 1587$ MeV and $\Gamma_p = 107$ MeV, respectively. Our pole width is comparable with that in previous analyses but our pole mass is slightly smaller than those in other analyses. For $S_{31}(1900)$, both our pole mass $M_p = 1844$ MeV and the pole width $\Gamma_p = 223$ MeV are comparable to those in the analysis by Cutkosky {\it et al.\ }\cite{cutkosky80}.

$\bf{P_{31}}$:\\
This partial wave was fitted with a single resonance at $M=1934\pm5$ MeV with $\Gamma=211\pm 11$ MeV. The hadronic decay channels were found to be $\pi N$ (17\%), $\pi N^*$ (47\%), with most of the strength going into a dummy $\rho\Delta$ channel. This resonance can be identified with the 4* $P_{31}(1910)$.

The pole mass and the pole width were found to be $M_p = 1910$ MeV and $\Gamma_p = 199$ MeV, respectively. These results agree well with the analysis by Cutkosky {\it et al. }\cite{cutkosky80}.

$\bf{P_{33}}$:\\
This partial wave was fitted with three resonances. The first resonance occurred at $M=1231.1\pm0.2$ MeV with $\Gamma=113.0\pm0.5$ MeV and  corresponds to the 4* $P_{33}$(1232). This state has an elasticity of 99.4\%. The second resonance occurred at $M=1626\pm8$ MeV with $\Gamma=225\pm18$ MeV and corresponds to the 3*  $P_{33}(1600)$. The major decay modes were found to be $\pi N$ (8\%), $\pi \Delta$ (70\%), and $\pi N^*$ (22\%). The third resonance occurred at $M=2146\pm32$ MeV with $\Gamma = 400\pm80$ MeV.

For $P_{33}(1232)$, the pole mass $M_p = 1212$ MeV and the pole width $\Gamma_p = 98$ MeV agree very well with previous analyses. For $P_{33}(1600)$ our pole mass $M_p = 1599$ MeV and pole width $\Gamma_p = 211$ MeV agree well with the analysis by Cutkosky {\it et al.\ }\cite{cutkosky80}.

$\bf{D_{33}}$:\\
This partial wave was fitted with one resonance at $M=1691\pm4$ MeV and $\Gamma=248\pm9$ MeV. This state can be identified with the 4* $D_{33} (1700)$. The main hadronic decay modes were found to be $\pi N$ (14\%), $(\pi\Delta)_S$ (54\%), and $\rho_3N$ (30\%). 

The pole mass and pole width for this resonance were found to be $M_p = 1656$ MeV and $\Gamma_p = 226$ MeV, respectively. These results agree very well with previous analyses.

$\bf{D_{35}}$:\\
This partial wave was fitted with one resonance at $M=1930\pm14$ MeV with $\Gamma=235\pm39$ MeV. This resonance can be identified with the 3* $D_{35} (1930)$. This state was found to have an elasticity of only 8\% with all the inelasticity carried by the dummy $\rho\Delta$ channel.

The pole mass and the pole width for this resonance were found to be $M_p = 1882$ MeV and $\Gamma_p = 187$ MeV, respectively. These values agree very well with analyses by H\"ohler \cite{hoehler93} and Cutkosky {\it et al.\ }\cite{cutkosky80} but are smaller than those in the analysis by Arndt {\it et al.\ }\cite{arndt06}.

$\bf{F_{35}}$:\\
This partial wave was fitted with two resonances. The first resonance occurred at $M=1818\pm8$ MeV with $\Gamma=278\pm18$ MeV. This resonance can be identified with the 4* $F_{35} (1905)$. Its major hadronic decay channels were found to be $\pi N$ (6\%), $(\pi\Delta)_P$ (28\%), and $(\pi\Delta)_F$ (64\%). The second resonance occurred at $M=2015\pm24$ MeV with $\Gamma=500\pm52$ MeV. Its major decay modes were found to be $\pi N$ (7\%), $(\pi\Delta)_P$ (3\%), and $\rho_3N$ (90\%). This state corresponds to the 2* $F_{35}(2000)$.

The pole mass and the pole width for $F_{35} (1905)$ were found to be $M_p = 1769$ MeV and $\Gamma_p = 239$ MeV, respectively. This pole mass is somewhat smaller than that in previous analyses but the pole width is comparable with that by Arndt {\it et al.\ }\cite{arndt06}.
The pole mass and the pole width for $F_{35}(2000)$ were found to be $M_p = 1976$ MeV and $\Gamma_p = 488$ MeV, respectively. These results are comparable with those by Cutkosky {\it et al.\ }\cite{cutkosky80} but larger than the values presented in the analysis by Vrana {\it et al.\ }\cite{vrana00}.

$\bf{F_{37}}$:\\
This partial wave was fitted with one resonance at $M=1918\pm1$ MeV with $\Gamma=259\pm4$ MeV. This state can be identified with the 4* $F_{37} (1950)$. This state was found to have an elasticity of 46\%. Most of the inelasticity was carried by a dummy $\rho\Delta$ channel  while about 8\% was found to be associated with the $\pi\Delta$ channel.

The pole mass and the pole width were found to be $M_p = 1871$ MeV and $\Gamma_p = 220$ MeV, respectively. Theses results agree very well with previous analyses.
%\begin{comment}
% xxxxx&xxxxxxxx&xxxxxxxxxxxxxxxx&xxxxxxxxxxxxxxxx&xxxxxxxxxxxxxxxx \kill
\begin{table*}[htpb]
% \caption{Resonance parameters for states with isospin $I = {1}/{2}$. \label{Resonances}   Column 1 lists the resonance name followed by its \\ fitted mass (in MeV) and fitted total width (in MeV)}\\ \hline
\caption {Resonance parameters for states with isospin $I = {1}/{2}$.  Column 1 lists the resonance name
followed by its fitted mass (in MeV) and fitted total width (in MeV). Column 2 lists the decay channel (see text for explanation). Column 3 lists the partial width in MeV and column 4 lists the corresponding branching fraction. Column 5 lists the resonant amplitude (see text). \label{Resonances} }
\begin{center}
\begin{ruledtabular}
\begin{tabular}{cccccccccc}

{Resonance}    & {Channel} & {$\Gamma_i$ (MeV)} & {${\cal B}_i$ ($\%$)}  & {$\sqrt{xx_i}$ } &{Resonance}    & {Channel} & {$\Gamma_i$ (MeV)} & {${\cal B}_i$ ($\%$)}  & {$\sqrt{xx_i}$ }\\ \hline
       $P_{11}$(1440)  &  $\pi N$   & $161(3)$  & $64.8(9)$  & $+0.648(9)$    &       $P_{11}$(1710)  &  $\pi N$   & $17(5)$  & $15(4)$  & $+0.15(4)$\\
      $1412(2)$  & $(\pi\Delta)_P $   &  $16(2)$  &  $6.5(8)$ &  $+0.21(1)$      &        $1662(7)$  & $\eta N$   &  $13(7)$  &  $11(7)$ &  $-0.13(4)$\\
      $248(5)$  & $\rho_1 N$ & $3(1)$ & $1.3(4)$ & $+0.09(1)$                        &      $116(17)$  &  $K \Lambda$ & $10(6)$ & $8(4)$ & $-0.11(3)$\\
      & $\sigma N$   &  $68(4)$  &  $27(1)$  &  $+0.42(1)$                                  &       & $(\pi \Delta)_P$  &  $7(4)$  &  $6(3)$  &  $-0.09(2)$\\
      &&&&                                                                                                                     &       & $\rho_1 N$   &  $20(7)$  & $17(6)$ &   $-0.16(2)$\\
      &&&&                                                                                                                     &       & $\sigma N$   &  $<1$  & $<1$ &  $-0.00(2)$\\
      \hline
%      & $\rho\Delta $   &  $0(0)$  & $0(0)$ &   $0(0)$\\
      $D_{13}$(1520)   & $\pi N$     & $73(1)$      & $62.7(5)$         &   $+0.627(5)$   &    $P_{13}$(1720)  &  $\pi N$   & $27(3)$  & $13.6(6)$  & $+0.136(6)$\\
      $1512.6(5)$ & $(\pi\Delta)_S$   & $11(1)$      & $9.3(7)$        &   $-0.24(1)$      &           $1720(5)$  & $\eta N$   &  $<1$  &  $<1$ &  $+0.00(2)$\\
      $117(1)$      & $(\pi\Delta)_D$   & $7(1)$        &$6.3(5)$          &   $-0.20(1)$      &             $200(20)$  & $K \Lambda$ & $6(1)$ & $2.8(4)$ & $-0.061(5)$\\
                             & $(\rho_3 N)_S$           &  $24(1)$    & $20.9(7)$      & $-0.36(1)$ &                            & $\rho_1 N$  &  $3(1)$  &  $1.4(5)$  &  $+0.04(1)$\\
                             & $\sigma N$          & $<1$          & $<1$               & $+0.04(1)$&             &&&&\\
        \hline
      $S_{11}$(1535)  &  $\pi N$   & $52(1)$    & $37(1)$         &  $+0.37(1)$&              $F_{15}$(1860)  &  $\pi N$   & $37(3)$  & $17(1)$  & $+0.17(1)$\\
      $1538(1)$  & $\eta N$   &  $58(4)$           & $41(2)$          &  $+0.39(1)$&                $1900(7)$  & $\eta N$   &  $9(4)$  &  $4(2)$ &  $-0.08(2)$\\
      $141(4)$  & $(\pi \Delta)_D$     & $3(1)$       & $1.8(8)$            &  $+0.08(2)$&      $219(23)$  &  $K \Lambda$ & $0.8(4)$ & $<1$ & $-0.02(1)$\\
       & $(\rho_3 N)_D$  &  $12(2)$    & $8(1)$       & $-0.18(1)$&                                    & $(\pi \Delta)_P$  &  $<4$  &  $<2$  &  $-0.03(3)$\\
       & $\rho_1 N$  &  $14(1)$    & $10(1)$     & $-0.19(1)$&                                            & $(\pi \Delta)_F$  &  $<1$  &  $<1$  &  $+0.00(2)$\\
       & $\sigma N$    & $2(1)$   & $1.5(5)$       & $+0.07(1)$&                                          & $(\rho_3 N)_P$  &  $7(5)$  &  $3(2)$  &  $-0.07(3)$\\
       & $\pi N^*$ & $<1$ & $<1$ &  $+0.01(2)$&                                                                  & $\sigma N$   & $89(15)$  & $41(6)$  & $+0.26(2)$\\
%       & $(\gamma p)_E$ & $0.4(1)$& $0.20(5)$ & $-0.030(3)$\\
 %      & $(\gamma n)_E$ & $0.3(1)$ & $0.10(5)$ & $+0.022(4)$\\
        \hline
    $S_{11}$(1650)  &  $\pi N$   & $71(3)$  & $57(2)$  & $+0.57(2)$&                      $D_{13}$(1875)  &  $\pi N$   & $36(12)$  & $7(2)$  & $+0.07(2)$\\
      $1664(2)$  & $\eta N$   &  $26(3)$    &  $21(2)$ &  $-0.34(1)$&                          $1951(27)$  & $(\pi \Delta)_S$  &  $434(39)$  & $87(3)$  &  $-0.25(4)$\\
      $126(3)$  &  $K \Lambda$   & $11(1)$    &   $8(1)$ & $-0.22(1)$&                       $500(45)$    & $(\pi \Delta)_D$  &  $<28$  & $<6$  &  $-0.04(2)$\\
      & $(\pi \Delta)_D$  &  $9(2)$       &  $7(2)$   &    $+0.20(2)$&                                 & $(\rho_3 N)_S$  &  $<24$  & $<5$ &   $+0.04(2)$\\
      & $(\rho_3 N)_D$   &  $<1$         &  $<1$      &      $-0.04(2)$&                                & $\sigma N$    & $<19$  & $<4$  & $+0.03(3)$\\
      & $\rho_1 N$   &  $8(2)$      &  $6(1)$      &      $-0.19(2)$&                                     &&&&\\
      & $\sigma N$   & $<1$       &  $<1$     &     $+0.04(2)$&                                           &&&&\\                                   
      & $\pi N^*$ & $<1$ & $<1$ & $+0.02(3)$         &                                                      &&&&\\
  %    & $(\gamma p)_E$ & $0.07(3)$ & $0.05(2) $ & $-0.010(3)$\\
   %   & $(\gamma n)_E$ & $0.005(9)$ & $0.00(1)$ & $ -0.004(3)$\\  
   \hline
      $D_{15}$(1675)  &  $\pi N$   & $56(1)$  & $38.6(6)$  & $+0.386(6)$&    $P_{11}$(1880)  &  $\pi N$   & $74(25)$  & $15(5)$  & $+0.15(5)$\\
      $1679(1)$  & $\eta N$   &  $<1$  &  $<1$ &  $+0.03(1)$&             $1900(36)$  & $\eta N$   &  $80(42)$  &  $16(7)$ &  $+0.16(4)$\\ 
      $145(4)$  &  $K \Lambda$ & $<1$ & $<1$ & $-0.03(1)$&             $485(142)$  &  $K \Lambda$ & $157(61)$ & $32(10)$ & $+0.22(3)$\\
       & $(\pi \Delta)_D$  &  $66(3)$  &  $46(1)$  &  $+0.42(1)$&           & $(\pi \Delta)_P$  &  $<9$  &  $<2$  &  $-0.03(3)$\\
      & $(\rho_3 N)_D$   &  $<1$  & $<1$ &   $-0.03(1)$&                         & $\rho_1 N$   &  $<2$  & $<1$ &  $+0.01(4)$\\
      & $\rho_1 N$   &  $<1$  & $<1$ &  $+0.02(1)$&                                 & $\sigma N$   & $40(24)$  & $8(5)$  & $+0.11(3)$\\
      \hline
      $F_{15}$(1680)  &  $\pi N$   & $85.6(7)$  & $68.0(5)$  & $+0.680(5)$&       $S_{11}$(1895)  &  $\pi N$   & $85(11)$  & $17(2)$  & $+0.17(2)$\\
      $1682.7(5)$  & $\eta N$   &  $1.2(4)$  &  $1.0(3)$ &  $+0.08(1)$&    $1910(15)$      &  $\eta N$   &  $203(29)$  &  $40(4)$ &  $+0.26(2)$\\
      $126(1)$  &  $K \Lambda$ & $<1$   & $<1$ & $-0.01$&                       $502(47)$      &  $K \Lambda$ & $9(5)$ & $1.8(8)$ & $+0.06(1)$\\
      & $(\pi \Delta)_P$      &  $13(1)$       &  $10.5(9)$    &  $-0.27(1)$&& $(\pi \Delta)_D$   &  $37(15)$  &  $7(3)$  &     $+0.11(2)$\\
      & $(\pi \Delta)_F$      &  $1.2(2)$      &  $1.0(1)$   &  $+0.08(1)$&      & $(\rho_3 N)_D$    &  $43(12)$  & $9(3)$ &    $-0.12(2)$\\
      & $(\rho_3N)_P $      &  $9.3(9)$         & $7.4(7)$        &   $-0.22(1)$&& $\rho_1 N$    &  $<9$  &  $<2$ &    $+0.04(2)$\\
      & $(\rho_3N)_F $      &  $3.0(3)$      & $2.4(3)$        &  $-0.13(1)$&    & $\sigma N$ &  $<6$  &  $<2$  &   $+0.03(2)$\\
      & $\sigma N$   & $12(1)$    & $9.4(8)$        & $+0.25(1)$& & $\pi N^*$       &  $118(25)$ & $24(4)$ &  $-0.20(2)$\\
         \hline
      $D_{13}$(1700)  &  $\pi N$   & $1.5(3)$  & $2.8(5)$  & $+0.028(5)$&    $P_{13}$(1900)  &  $\pi N$   & $7(4)$  & $7(4)$  & $+0.07(4)$\\
      $1665(3)$& $(\pi \Delta)_S$  &  $17(6)$  &  $31(9)$  &  $-0.09(2)$&     $1900(8)$  & $\eta N$   &  $<1$  &  $<1$ &  $+0.00(2)$\\
      $56(8)$ & $(\pi\Delta)_D $   &  $2(1)$  & $3(2)$ &   $+0.03(1)$&            $101(15)$  &  $K \Lambda$ & $14(5)$ & $14(5)$ & $-0.10(4)$\\
      & $(\rho_3 N)_S$    &  $21(4)$  & $38(6)$ &  $-0.10(1)$&                         & $\rho_1 N$   &  $64(9)$  & $64(7)$ &  $+0.21(8)$\\
      & $\sigma N$    & $13(4)$  & $24(6)$  & $+0.08(1)$&       &&&&\\

    % %\newpage
    
    \end{tabular}
    \end{ruledtabular}
    \end{center}
    \end{table*}
   
    %\newpage
    \begin{table*}[htbp]
    \addtocounter{table}{-1}
    \caption{Cont'd.}
%       \hline
       \begin{center}
      \begin{ruledtabular}
\begin{tabular}{cccccccccc}
{Resonance}    & {Channel} & {$\Gamma_i$ (MeV)} & {${\cal B}_i$ ($\%$)}  & {$\sqrt{xx_i}$ } &{Resonance}    & {Channel} & {$\Gamma_i$ (MeV)} & {${\cal B}_i$ ($\%$)}  & {$\sqrt{xx_i}$ }\\  \hline
     $F_{17}$(1990)  &  $\pi N$   & $4(2)$  & $2(1)$  & $+0.02(1)$         &               $G_{17}$(2190)  &  $\pi N$   & $101(18)$  & $20(1)$  & $+0.20(1)$\\
      $1990(45)$  & $K\Lambda$   &  $1(1)$  &  $<1$ &  $-0.010(3)$     &              $2150(26)$  & $\eta N$   &  $9(7)$  &  $2(1)$ &  $-0.06(2)$\\
     $203(161)$  &   &   &   &                                                                            &             $500(74)$  &  $K \Lambda$ & $<1$ & $<1$ & $+0.01(1)$\\
                            &&&&                                                                                     &              & $(\rho_3 N)_D$  &  $45(33)$  &  $9(6)$  &  $-0.13(5)$\\
     \hline
          $D_{15}$(2060)  &  $\pi N$   & $26(10)$  & $9(2)$  & $+0.09(2)$ &&&&&\\
      $2116(21)$      &  $\eta N$   &  $<3$  &  $<1$ &  $+0.01(2)$               &&&&&\\
      $307(112)$      &  $K \Lambda$ & $<2$ & $<1$ & $+0.00(3)$            &&&&&\\
      & $(\pi \Delta)_D$   &  $123(39)$  &  $40(13)$  &     $+0.19(4)$         &&&&&\\
      & $(\rho_3 N)_D$    &  $<25$  & $<9$ &    $-0.06(3)$                           &&&&&\\
      & $\rho_1 N$    &  $63(40)$  &  $21(15)$ &    $-0.13(5)$                      &&&&&\\
          
    \end{tabular}
    \end{ruledtabular}
    \end{center}
    \end{table*}

\begin{table*}[htpb]
 \caption {Resonance parameters for states with isospin $I = {3}/{2}$. (See caption to Table I for details.)}
% \begin{ruledtabular}
\begin{center}
\begin{ruledtabular}
 \begin{tabular}{cccccccccc}

  Resonance     &Channel & $\Gamma_i$ (MeV) & ${\cal B}_i$ ($\%$) & $\sqrt{xx_i}$ &{Resonance}    & {Channel} & {$\Gamma_i$ (MeV)} & {${\cal B}_i$ ($\%$)}  & {$\sqrt{xx_i}$ }\\ \hline
 
      $P_{33}$(1232)  &  $\pi N$   & $112.4(5)$  & $99.4$  & $+0.994$ &    $P_{31}$(1910)  &  $\pi N$   & $36(2)$  & $17(1)$  & $+0.17(1)$\\           
      $1231.1(2)$  & $(\pi\Delta)_P $   &  $0$  &  $0$ &  $+0.00$ &               $1934(5)$   & $\pi N^*$  &  $99(14)$  &  $47(6)$  &  $-0.28(2)$\\
      $113.0(5)$  & $\pi N^*$ & $0$ & $0$ & $+0.00$   &                                 $ 211(11)$&    &   &   \\
      &&&&                                                                                                                  & &&&\\
    \hline
      $P_{33}$(1600)  &  $\pi N$   & $18(4)$  & $8(2)$  & $+0.08(2)$  &    $P_{33}$(1920)  &  $\pi N$   & $63(19)$  & $16(4)$  & $+0.16(4)$\\
        $1626(8)$ & $(\pi \Delta)_P$  &  $157(11)$  &  $70(3)$  &  $+0.24(2)$  &    $2146(32)$ & $(\pi \Delta)_P$  &  $27(21)$  &  $7(5)$  &  $-0.10(4)$\\
         $225(18)$  & $\pi N^*$   &  $49(9)$  & $22(3)$ &   $+0.13(1)$                 &    $400(80)$ & $\pi N^*$   &  $<83$  & $<20$ &   $+0.12(7)$\\
  %                         & $\rho \Delta$   &  $<36$  & $<14$ &  $+0.08(6)$\\
                             \hline
      $S_{31}$(1620)  &  $\pi N$   & $37(2)$  & $33(2)$  & $+0.33(2)$               &    $D_3$$_5$(1930)  &  $\pi N$   & $19(5)$  & $7.9(4)$  & $+0.079(4)$\\
      $1600(1)$  & $(\pi \Delta)_D$  &  $35(2)$  & $32(2)$  &  $-0.32(1)$          &    $1930(12)$  &    &              & \\
       $112(2)$   & $(\rho_3 N)_D$  &  $<1$  & $<1$ &   $-0.03(1)$                     &    $235(39)$&   &   & &  \\ 
       & $\rho_1 N$    &  $29(2)$  & $26(2)$ &  $+0.29(1)$                                     &    &&&&\\
       & $\pi N^*$ & $10(1)$ & $9(1)$ &  $-0.17(1)$                                                 &     &&&&\\
    \hline  
        $D_{33}$(1700)  &  $\pi N$   & $36(2)$  & $14(1)$  & $+0.14(1)$             &        $F_{37}$(1950)  &  $\pi N$   & $118(2)$  & $45.6(4)$  & $+0.456(4)$\\
        $1691(4)$ & $(\pi \Delta)_S$  &  $134(6)$  &  $54(3)$  &  $+0.28(1)$      &        $1918(1)$           & $(\pi\Delta)_F$&  $22(3)$   & $8(1)$     & $+0.20(1)$ \\
        $248(9)$   & $(\pi \Delta)_D$  &  $4(2)$  &  $1(1)$  &  $+0.05(2)$             &      $259(4)$&   &   & &  \\       
                           & $(\rho_3N)_S$   &  $75(7)$  & $30(3)$ &   $+0.21(1)$           &  &&&\\
     \hline
       $S_{31}$(1900)  &  $\pi N$   & $20(4)$  & $8(1)$  & $+0.08(1)$                  &     $F_{35}$(2000)  &  $\pi N$   & $34(6)$  & $7(1)$  & $+0.07(1)$\\ 
       $1868(12)$& $(\pi \Delta)_D$  &  $132(18)$  &  $56(6)$  &  $-0.22(2)$     &    $2015(24)$ & $(\pi \Delta)_P$  &  $13(13)$  &  $3(3)$  &  $-0.04(2)$\\
       $234(27)$& $(\rho_3 N)_D$   &  $53(15)$  & $23(5)$ &   $-0.14(2)$          &    $500(52)$   & $(\pi  \Delta)_F$  &  $<13$  & $<3$ &   $+0.02(3)$\\
       & $\rho_1 N$   &  $29(10)$  & $12(4)$ &  $-0.10(2)$                                      & &$(\rho_3 N)_P$   &  $449(49)$  & $90(3)$ &  $+0.25(1)$ \\
       & $\pi N^*$ & $<2$ & $<1$ &  $+0.01(2)$                                                         & &&&\\
      \hline
        $F_{35}(1905)$  &  $\pi N$   & $16(2)$  & $6(1)$  & $+0.06(1)$                &    &&&\\
         $1818(8)$ & $(\pi \Delta)_P$  &  $77(19)$  &  $28(7)$  &  $+0.13(2)$     & &&&\\
         $278(18)$ & $(\pi \Delta)_F$  &  $177(27)$  & $64(8)$ &   $+0.19(2)$    &&&&\\
         & $(\rho_3 N)_P$   &  $<14$  & $<6$ &  $-0.04(2)$         &                          &&&\\                                                                                                           
      \end{tabular}
      \end{ruledtabular}
      \end{center}
      \end{table*}

 \begin{table*}[htpb]
 \caption {Comparison of resonance parameters for $I=1/2$ states with other analyses.}
 \begin{center}
  \begin{ruledtabular}
  \begin{tabular}{ccclcccl}
   Mass (MeV)& Width (MeV)& Elasticity & ~~~~~Analysis &     Mass (MeV)& Width (MeV)& Elasticity & ~~~~~Analysis\\  
   \hline
 %  $S_{11}(1535)$ **** &&&     &  $D_{13}$(1700) ***         &&&\\
     \multicolumn{4}{c}    {$S_{11}(1535)$ **** }   & \multicolumn{4}{c}        { $D_{13}$(1700) ***} \\  \cline{1-4}     \cline{5-8}
 %  *****                        &&&         ****    &   &&&\\
         $1538(1)$        & $141(4)$        & $0.37(1)$         & This~Work&                                              $1665(3)$  & $56(8)$          & $0.028(5)$     & This~Work\\
                                   $1519(5)$    &$128(14)$          & $0.54(5)$      & Anisovich 12 \cite{anisovich12}                                      &  $1790(40)$    &$390(140)$          & $0.12(5)$      & Anisovich 12 \cite{anisovich12}\\
                                             $1547.0(7)$  &$188.4(38)$      & $0.355(2)$      & Arndt 06 \cite{arndt06}                                  &  $1737(44)$  &  $250(220)$     &$0.01(2)$           & Manley 92 \cite{manley92}\\%MANLEY 92\\
                                             $1534(7)$     &$151(27)$         &$0.51(5)$          & Manley 92 \cite{manley92}                           &  $1675(25)$  & $90(40)$        & $0.11(5)$          & Cutkosky 80 \cite{cutkosky80}\\%CUTKO. 80\\
                                             $1550(40)$   &$240(80)$         & $0.50(10)$      & Cutkosky 80 \cite{cutkosky80}                      &  $1731(15)$  & $110(30)$      & $0.08(3)$          &H\"ohler 79 \cite{hoehler79}\\%HOEHL. 79\\
                                             $1526(7)$     &$120(20)$         & $0.38(4)$         & H\"ohler 79 \cite{hoehler79}  &&&&\\
       \hline     
%         $S_{11}(1650)$ **** &&&&       $D_{13}$(1875)   **        &&&\\
           \multicolumn{4}{c}    { $S_{11}(1650)$ ****}   & \multicolumn{4}{c}        { $D_{13}$(1875)   **} \\  \cline{1-4}     \cline{5-8}
%   *****                        &&&         $**$    &   &&&\\                             
        $1664(2)$       & $126(3)$       & $0.57(2)$      & This~Work       &            $1951(27)$     & $500(45)$      & $0.07(2)$      & This~Work\\                        
                                 $1651(6)$    &$104(10)$          & $0.51(4)$      & Anisovich 12 \cite{anisovich12}                            &  $1880(20)$    &$200(25)$          & $0.03(2)$      & Anisovich 12 \cite{anisovich12}\\ 
                                           $1634.7(11)$  & $115.4(28)$   &$1.0$          & Arndt 06 \cite{arndt06}                &  $1804(55)$    &  $450(185)$   &$0.23(3)$       & Manley 92\cite{manley92}\\%MANLEY 92\\
                                           $1659(9)$       & $173(12)$       &$0.89(10)$     & Manley 92 \cite{manley92}         &  $1880(100)$ & $180(60)$       & $0.10(4)$      & Cutkosky 80 \cite{cutkosky80}\\%CUTKO. 80\\
                                           $1650(30)$     & $150(40)$       & $0.65(10)$     & Cutkosky 80 \cite{cutkosky80}   &  $2060(80)$ & $300(100)$       & $0.14(7)$      & Cutkosky 80 \cite{cutkosky80}\\%CUTKO. 80\\
                                           $1670(8)$       & $180(20)$       &$0.61(4)$        & H\"ohler 79 \cite{hoehler79}       &  $2081(20)$   & $265(40)$       & $0.06(2)$      & H\"ohler 79 \cite{hoehler79}\\%HOEHL. 79\\
        \hline
%          $S_1$$_1$[1900]  &  $1900(31)$       & $205(94)$       & $0.20(11)$      & This~Work\\
%        \\\hline
%  $S_{11}(1895)$ ** &&& &      $D_{15}$(1675)   ****        &&&\\
    \multicolumn{4}{c}    {$S_{11}(1895)$ ** }   & \multicolumn{4}{c}        { $D_{15}$(1675)   ****} \\  \cline{1-4}     \cline{5-8}
 %  $**$                        &&&         *****    &   &&&\\

           $1910(15)$       & $502(47)$    & $0.17(2)$      & This~Work                &  $1679(1)$      & $145(4)$       & $0.386(6)$          & This~Work\\
                                         $1895(15)$    &$90^{+30}_{-15}$          & $0.02(1)$      & Anisovich 12 \cite{anisovich12}                           &  $1664(5)$    &$152(7)$          & $0.40(3)$      & Anisovich 12 \cite{anisovich12}\\
                                               $1928(59)$       & $414(157)$     & $0.10(10)$      & Manley 92 \cite{manley92}      &  $1674.1(2)$  &$146.5(10)$  & $0.393(1)$        & Arndt 06 \cite{arndt06}\\%ARNDT 06\\
                                               $2180(80)$       & $350(100)$     & $0.18(8)$         & Cutkosky 80 \cite{cutkosky80}     &  $1676(2)$     &  $159(7)$      &$0.47(2)$            & Manley 92 \cite{manley92}\\%MANLEY 92\\
                                               $1880(20)$       & $95(30)$          & $0.09(5)$         &H\"ohler 79 \cite{hoehler79}      &  $1675(10)$   & $160(20)$     & $0.38(5)$          & Cutkosky 80 \cite{cutkosky80}\\%CUTKO. 80\\
                                              &&&                                                                                                                                       &  $1679(8)$     & $120(15)$     & $0.38(3)$          & H\"ohler 79 \cite{hoehler79}\\%HOEHL. 79\\
         \hline       
  %         $P_{11}(1440)$   ****&&&&       $D_{15}$(2060) **         &&&\\
             \multicolumn{4}{c}    {$P_{11}(1440)$   **** }   & \multicolumn{4}{c}        {$D_{15}$(2060) **} \\  \cline{1-4}     \cline{5-8}
 %  *****                        &&&         $\ast\ast$   &   &&&\\
                        
           $1412(2)$        & $248(5)$  & $0.648(9)$        & This~Work     &                                                 $2116(21)$     & $307(112)$      & $0.09(2 )$          & This~Work\\
                                  $1430(8)$    &$365(35)$          & $0.62(3)$      & Anisovich 12 \cite{anisovich12}                           &  $2060(15)$    &$375(25)$          & $0.08(2)$      & Anisovich 12 \cite{anisovich12}\\
                                              $1485.0(12)$   & $284(18)$  & $0.787(16)$ & Arndt 06 \cite{arndt06}                              &  $2180(80)$  & $400(100)$  & $0.10(3)$          & Cutkosky 80 \cite{cutkosky80}\\%CUTKO. 80\\
                                              $1462(10)$      & $391(34)$  &$0.69(3)$         & Manley 92 \cite{manley92}                    &  $2228(30)$   & $310(50)$    & $0.07(2)$          & H\"ohler 79 \cite{hoehler79}\\%HOEHL. 79\\
                                              $1440(30)$      & $340(70)$  & $0.68(4)$        &Cutkosky 80  \cite{cutkosky80}              &&&&\\
                                              $1410(12)$      & $135(10)$  & $0.51(5)$        & H\"ohler 79 \cite{hoehler79}                  &&&&\\
                              
       \hline
 %        $P_{11}(1710)$ *** &&&&       $F_{15}$(1680)  ****         &&&\\
           \multicolumn{4}{c}    {$P_{11}(1710)$ *** }   & \multicolumn{4}{c}        {$F_{15}$(1680)  ****} \\  \cline{1-4}     \cline{5-8}
%   ****                        &&&         *****    &   &&&\\

             $1662(7)$   & $116(17)$     & $0.15(4)$         & This~Work                                          &     $1682.7(5)$      & $126(1)$      & $0.680(5)$       & This~Work\\
          $1710(20)$    &$200(18)$          & $0.05(4)$      & Anisovich 12 \cite{anisovich12}                              &  $1689(6)$    &$118(6)$          & $0.64(5)$      & Anisovich 12 \cite{anisovich12}\\
                                              $1717(28)$   & $480(230)$   & $0.09(4)$         & Manley 92 \cite{manley92}                                                  &  $1680.1(2)$   &$128.0(11)$ & $0.701(1)$    & Arndt 06 \cite{arndt06}\\%ARNDT 06\\
                                              $1650(30)$   & $150(40)$     & $0.65(10)$       & Cutkosky 80 \cite{cutkosky80}                                            &  $1684(4)$      &  $139(8)$     &$0.70(3)$       & Manley 92 \cite{manley92}\\%MANLEY 92\\
                                              $1670(8)$   & $180(20)$     & $0.61(4)$          &H\"ohler 79 \cite{hoehler79}                                                  &  $1680(10)$    & $120(10)$   & $0.62(5)$      & Cutkosky 80 \cite{cutkosky80}\\%CUTKO. 80\\
                                            &&&                                                                                                                                                                                 &  $1684(3)$      & $128(8)$      & $0.65(2)$      & H\"ohler 79 \cite{hoehler79}\\%HOEHL.79\\                                    
         \hline
%           $P_{11}(1880)$ ** &&& &      $F_{15}$(1860) **         &&&\\
             \multicolumn{4}{c}    {$P_{11}(1880)$ ** }   & \multicolumn{4}{c}        { $F_{15}$(1860) **} \\  \cline{1-4}     \cline{5-8}
 %  $\ast\ast$                       &&&         $**$    &   &&&\\

                $1900(36)$   & $485(142)$       & $0.15(5)$           & This~Work                               &             $1900(7)$           & $219(23)$      & $0.17(1)$   & This~Work\\
                             $1870(35)$    &$235(65)$          & $0.05(3)$      & Anisovich 12 \cite{anisovich12}&           $1860_{-60}^{+120}$    &$270_{-50}^{+140}$   & $0.20(6)$      & Anisovich 12 \cite{anisovich12}\\  
                                               $1885(30)$     & $113(44)$      & $0.15(6)$           & Manley 92 \cite{manley92}     &  $1817.7$          &$117.6$           & $0.127$      & Arndt 06 \cite{arndt06}\\%ARNDT 06\\
                                             &&&                                                                                                                                      &  $1903(87)$     &  $490(310)$   &$0.08(5)$     & Manley 92 \cite{manley92}\\%MANLEY 92\\
                                             &&&                                                                                                                                     &  $1882(10)$     & $95(20)$        & $0.04(2)$     & H\"ohler 79 \cite{hoehler79}\\%HOEHL.79\\
       \hline
%         $P_{13}(1720)$  **** &&& &      $F_{17}$(1990)  **        &&&\\
           \multicolumn{4}{c}    {$P_{13}(1720)$  ****}   & \multicolumn{4}{c}        { $F_{17}$(1990)  ** } \\  \cline{1-4}     \cline{5-8}
  % *****                        &&&         $**$    &   &&&\\

           $1720(5)$       & $200(20)$   & $0.136(6)$         & This~Work                                   &      $1990(45)$       & $203(161)$   & $0.02(1)$     & This~Work\\
       $1690^{+70}_{-35}$    &$420(100)$          & $0.10(5)$      & Anisovich 12 \cite{anisovich12}     &  $2060(65)$    &$240(50)$          & $0.02(1)$      & Anisovich 12 \cite{anisovich12}\\
                                           $1763.8(46)$   & $210(22)$   &$0.094(5)$        & Arndt 06 \cite{arndt06}                                                       &  $2086(28)$        &  $535(120)$    &$0.06(2)$      & Manley 92 \cite{manley92}\\%MANLEY 92\\
                                           $1717(31)$    & $380(180)$ &$0.13(5)$          &Manley 92 \cite{manley92}                  &  $1970(50)$        & $350(120)$     & $0.06(2)$     & Cutkosky 80 \cite{cutkosky80}\\%CUTKO. 80\\
                                           $1700(50)$    & $125(70)$   & $0.10(5)$          & Cutkosky 80 \cite{cutkosky80}            &  $2005(150)$      & $350(100)$    & $0.04(2)$     &H\"ohler 79  \cite{hoehler79}\\%HOEHL. 79\\
                                           $1710(20)$    & $190(30)$   & $0.14(3)$          & H\"ohler 79 \cite{hoehler79}                &&&&\\
             \end{tabular}
        \end{ruledtabular}
       \end{center}
       \end{table*}

 \begin{table*}[htpb]
 \addtocounter{table}{-1}
 \caption {Cont'd.}
 \begin{center}
  \begin{ruledtabular}
  \begin{tabular}{ccclcccl}
   Mass (MeV)& Width (MeV)& Elasticity & ~~~~~Analysis &     Mass (MeV)& Width (MeV)& Elasticity & ~~~~~Analysis\\  
            \hline
%              $P_{13}(1900)$  ** &&& &      $G_{17}$(2190)   ****       &&&\\
                \multicolumn{4}{c}    { $P_{13}(1900)$  **}   & \multicolumn{4}{c}        {$G_{17}$(2190)   **** } \\  \cline{1-4}     \cline{5-8}
%   $\ast\ast$                      &&&           *****    &   &&&\\

          $1900(8)$       & $101(15)$   & $0.07(4)$         & This~Work   &           $2150(26)$       & $500(74)$   & $0.20(1)$     & This~Work\\

                         $1905(30)$    &$250^{+120}_{-50}$          & $0.03(2)$      & Anisovich 12 \cite{anisovich12}  &                                    $2180(20)$    &$335(40)$          & $0.16(2)$    & Anisovich 12 \cite{anisovich12}\\          
                                          $1879 (17) $    & $498 (78)$ &$0.26(6)$          &Manley 92 \cite{manley92}      &  $2152.4(14)$  &$484(13)$      & $0.22(1)$     & Arndt 06 \cite{arndt06}\\%ARNDT 06\\
                                        &&&                                                                                                                                      &  $2127(9)$        &  $550(50)$    &$0.70(3)$      & Manley 92 \cite{manley92}\\%MANLEY 92\\
                                        &&&                                                                                                                                      &  $2200(70)$      & $500(150)$  & $0.12(6)$     & Cutkosky 80 \cite{cutkosky80}\\%CUTKO. 80\\
                                        &&&                                                                                                                                      &  $2140(12)$      & $390(90)$    & $0.14(2)$     & H\"ohler 79 \cite{hoehler79}\\%HOEHL. 79\\
                                         
      \hline
%        $D_{13}(1520)$  ****&&&&                &&&\\
          \multicolumn{4}{c}    { $D_{13}(1520)$  ****}   & \multicolumn{4}{c}        { } \\  \cline{1-4}    
%   *****                        &&&            &   &&&\\

           $1512.6(5)$     & $117(1)$      & $0.627(5)$     & This~Work   &&&&\\
           $1517(3)$    &$114(5)$          & $0.62(3)$      & Anisovich 12 \cite{anisovich12}&&&&\\
                                               $1514.5(2)$   &$103.6(4)$    & $0.632(1)$   & Arndt 06 \cite{arndt06} &&&&\\
                                               $1524(4)$     &  $124(8)$      &$0.59(3)$      & Manley 92 \cite{manley92} &&&&\\
                                               $1525(10)$   & $120(15)$    & $0.58(3)$     &Cutkosky 80  \cite{cutkosky80}&&&&\\
                                               $1519(4)$     & $114(7)$       & $0.54(3)$     & H\"ohler 79 \cite{hoehler79}&&&&\\

       \end{tabular}
        \end{ruledtabular}
       \end{center}
       \end{table*}
         
%%%%%%%%%%  

 %    This analysis confirms the existence of the state $P_{11}$(1710) which is refuted or marked uncertain by the GWU analysis \cite{arndt06}. The elasticity is about 10\% agreeing to the previous analysis \cite{manley92} but the width is $149\pm 48$ which agrees better with the older analyses \cite{cutkosky80} and \cite{hoehler79} . The b.f.s for the $\eta N$ and K$\Lambda$ channels are comparable about 28\% and 18\% respectively.

 \begin{table*}[htpb]
 \caption {Comparison of resonance parameters for $I=3/2$ states with other analyses.}
 \begin{center}
 \begin{ruledtabular}
  \begin{tabular}{ccclcccl}
 Mass (MeV) & Width (MeV) & Elasticity  &~~~~~Analysis  & Mass (MeV) & Width (MeV) & Elasticity  &~~~~~Analysis\\
\hline
  \multicolumn{4}{c}    { $P_3$$_3$(1232) ****}   & \multicolumn{4}{c}        { $S_3$$_1$(1900) **} \\  \cline{1-4}     \cline{5-8}                             
                                           $1231.1(2)$       & $113.0(5)$   & $0.99$         & This~Work  &  $1868(12)$       & $234(27)$      & $0.08(1)$     & This~Work\\
                                            $1228(2)$    &$110(3)$          & $1.0$      & Anisovich 12 \cite{anisovich12}   &  $1840(30)$    &$300(45)$          & $0.07(3)$      & Anisovich 12 \cite{anisovich12}\\     
                                           $1233.4(4)$ & $118.7(6)$   &$1.0$            & Arndt 06 \cite{arndt06}                &  $1920(24)$       & $263(39)$       &$0.41(4)$     & Manley 92 \cite{manley92}\\%MANLEY 92\\
                                           $1231(1)$    & $118(4)$      &   $1.0$          &Manley 92 \cite{manley92}         &  $1890(50)$    &$170(50)$          & $0.10(3)$      & Cutkosky 80 \cite{cutkosky80}\\
                                           $1228(2)$    &$120(5)$          & $1.0$      & Cutkosky 80 \cite{cutkosky80}     &  $1908(30)$    &$140(40)$          & $0.08(4)$      & H\"ohler 79 \cite{hoehler79}\\
                                           $1228(2)$    &$116(5)$          & $1.0$      & H\"ohler 79 \cite{hoehler79}         &&&&\\
      \hline
%        $P_3$$_3$(1600)  *** &&&                                                                                               &  $F_3$$_5$(1905)****   &&&\\
          \multicolumn{4}{c}    { $P_3$$_3$(1600)  ***}   & \multicolumn{4}{c}        { $F_3$$_5$(1905) ****} \\  \cline{1-4}     \cline{5-8}                             
                                              $1510(20)$    &$220(45)$          & $0.12(5)$      & Anisovich 12 \cite{anisovich12}  &  $1861(6)$    &$335(18)$          & $0.13(2)$      & Anisovich 12 \cite{anisovich12}\\  
                                             $1706(10)$    & $430(73)$    &   $0.12(2)$          &Manley 92 \cite{manley92}          &  $1857.8(16)$  &$320.6(86)$  & $0.122(1)$  & Arndt 06 \cite{arndt06}\\%ARNDT 06\\
                                             $1600(50)$    &$300(100)$          & $0.18(4)$      & Anisovich 12 \cite{cutkosky80}  &  $1881(18)$     &  $327(51)$    &$0.12(3)$     & Manley 92 \cite{manley92}\\%MANLEY 92\\
                                             $1522(13)$    &$220(40)$          & $0.21(6)$      & Anisovich 12 \cite{hoehler79}      &  $1910(30)$    &$400(100)$          & $0.08(3)$      & Cutkosky 80 \cite{cutkosky80}\\
                                           &&&                                                                                                                                               &  $1905(20)$    &$260(20)$          & $0.15(2)$      & H\"ohler 79 \cite{hoehler79}\\
    \hline
%       $S_3$$_1$(1620)****  &&&                                                                                         &     $P_3$$_1$(1910)****&&&\\
         \multicolumn{4}{c}    { $S_3$$_1$(1620) ****}   & \multicolumn{4}{c}        { $P_3$$_1$(1910) ****} \\  \cline{1-4}     \cline{5-8}                             
                                           $1600(1)$   & $112(2)$        & $0.33(2)$          &  This~Work                                         &  $1934(5)$        & $211(11)$       & $0.17(1)$         & This~Work\\
                                           $1600(8)$    &$130(11)$          & $0.28(3)$      & Anisovich 12 \cite{anisovich12}    &  $1860(40)$     &$350(55)$          & $0.12(3)$      & Anisovich 12 \cite{anisovich12}\\    
                                            $1615.2(4)$&$146.9(1.9)$ & $0.315(1)$       &  Arndt 06 \cite{arndt06}                    &  $2067.9(17)$  & $543.0(101)$  & $0.239(1)$     & Arndt 06 \cite{arndt06}\\%ARNDT 06\\
                                            $1672(7)$  &  $154(37)$     &$0.09(2)$          &  Manley 92 \cite{manley92}            &  $1882(10)$     & $239(25)$         &$0.23(8)$          & Manley 92 \cite{manley92}\\%MANLEY 92\\
                                            $1620(20)$    &$140(20)$          & $0.25(3)$      & Cutkosky 80 \cite{cutkosky80}   &  $1910(40)$    &$225(50)$          & $0.19(3)$      & Cutkosky 80 \cite{cutkosky80}\\
                                            $1610(7)$    &$139(18)$          & $0.35(6)$      & H\"ohler 79 \cite{hoehler79}         &  $1888(20)$    &$280(50)$          & $0.24(6)$      & H\"ohler 79 \cite{hoehler79}\\
    \hline 
%       $D_3$$_3$(1700) ****&&&                                                                                                         &      $P_3$$_3$(1920) ***&&&\\    
         \multicolumn{4}{c}    {$D_3$$_3$(1700) **** }   & \multicolumn{4}{c}        { $P_3$$_3$(1920) ***} \\  \cline{1-4}     \cline{5-8}                                                       
                                              $1691(4)$     & $248(9)$   & $0.14(1)$     & This~Work                                                                      &  $2146(32)$       & $400(80)$    & $0.16(4)$         & This~Work\\                                                                                                           
                                               $1715^{+30}_{-15}$    &$310^{+40}_{-15}$          & $0.22(4)$      & Anisovich 12 \cite{anisovich12}       &  $1900(30)$    &$310(60)$          & $0.08(4)$      & Anisovich 12 \cite{anisovich12}\\  
                                              $1695.0(13)$ &$375.5(7)$    & $0.156(1)$  & Arndt 06 \cite{arndt06}                                                        &  $2014(16)$       & $152(55)$       &   $0.02(2)$         &Manley 92 \cite{manley92}\\%MANLEY 92\\
                                              $1762(44)$    &  $599(248)$   &$0.14(6)$   & Manley 92 \cite{manley92}                                                &  $1920(80)$    &$300(100)$          & $0.20(5)$      & Cutkosky 80 \cite{cutkosky80}\\
                                             $1710(30)$    &$280(80)$          & $0.12(3)$      & Cutkosky 80 \cite{cutkosky80}                                     &  $1868(80)$    &$220(80)$          & $0.14(4)$      & H\"ohler 79 \cite{hoehler79}\\
                                            $1680(70)$    &$230(80)$          & $0.20(3)$      & H\"ohler 79 \cite{hoehler79}                                            &&&&\\  
\end{tabular}
\end{ruledtabular}
\end{center}
\end{table*}   

%NOTE:\\
%          $\ast\ast\ast\ast$~~~      Existence is certain, and properties are fairly well explored.\\
%          $\ast\ast\ast$~~~~~            Existence is likely, but further confirmation is desirable.\\
%          $\ast\ast$ ~~~~~~~                 Evidence is only fair.\\
%          $\ast$   ~~~~~~~~~                    Evidence is poor.   \\

\begin{table*}[htpb]
 \addtocounter{table}{-1}
 \caption {Cont'd.}
 \begin{center}
  \begin{ruledtabular}
  \begin{tabular}{ccclcccl}
   Mass (MeV)& Width (MeV)& Elasticity & ~~~~~Analysis &     Mass (MeV)& Width (MeV)& Elasticity & ~~~~~Analysis\\  
            \hline
%$D_3$$_5$(1930)***&&&&                $F_{35} (2000)$ $\ast \ast$ &&&\\
  \multicolumn{4}{c}    { $D_3$$_5$(1930) ***}   & \multicolumn{4}{c}        {$F_{35} (2000)$ **} \\  \cline{1-4}     \cline{5-8}                             
    $1930(12)$     & $235(39)$     & $0.079(4)$      & This ~Work              &  $2015(24)$    &    $500(52)$    &   $0.07(1)$   &    This~Work\\
    $2233(53)$   &$773(187)$    & $0.081(12)$ & Arndt 06 \cite{arndt06}             &  $1752(32)$    &    $251(93)$    &   $0.02(1)$   &     Manley 92 \cite{manley92}\\
      $1956(22)$    &  $526(142)$  &$0.18(2)$       & Manley 92 \cite{manley92}   &  $2200(125)$   &    $400(125)$   &  $0.07(4)$   &    Cutkosky 80 \cite{cutkosky80}\\
      $1940(30)$    &$320(60)$          & $0.14(4)$      & Cutkosky 80 \cite{cutkosky80}          &  $1724(61)$    &    $138(68)$     &  $0.00(1)$   &    Vrana 00 \cite{vrana00}\\
    $1901(15)$    &$195(60)$          & $0.04(3)$      & H\"ohler 79 \cite{hoehler79} &&&&\\ \hline
 %       $F_3$$_7$(1950) ****  & &&&&&&\\
          \multicolumn{4}{c}    { $F_3$$_7$(1950) ****}   & \multicolumn{4}{c}        { } \\  \cline{1-4}                                
                                        $1918(1)$     & $259(4)$     & $0.456(4)$    & This~Work &&&&\\
                                       $1915(6)$    &$246(10)$          & $0.45(2)$      & Anisovich 12 \cite{anisovich12} &&&&\\          
                                              $1921.3(2)$  &$271.1(11)$  & $0.471(1)$  & Arndt 06 \cite{arndt06}   &&&&\\%ARNDT 06\\
                                              $1945(2)$     &  $300(7)$     &$0.38(1)$     & Manley 92 \cite{manley92}   &&&&\\%MANLEY 92\\
                                              $1950(15)$     &   $340(50)$    &$0.39(4)$     &  Cutkosky 80 \cite{cutkosky80}   &&&&\\
   %                                         &  $1950(15)$   &   $340(5)$      &$0.39(4)$     &  Cutkosky   80 \cite{???}
                                              $1913(8)$     &   $224(10)$    &$0.38(2)$     &  H\"ohler     79 \cite{hoehler79}   &&&&\\

\end{tabular}
\end{ruledtabular}
\end{center}
\end{table*}   
 
% \clearpage
 
    \begin{table*}[htpb]
 \caption {Comparison of pole positions (in MeV) for $I=1/2$ states with other analyses.}
 \begin{center}
 \begin{ruledtabular}
  \begin{tabular}{ccclcccl}
  Resonance             &   Real Part  &     $-2\times$Imaginary Part   &  ~~~~Analysis &     Resonance             &   Real Part  &     $-2\times$Imaginary Part   &  ~~~~Analysis\\
\hline
           $P_{11}$(1440)   &  $1370$     &  $214$    &  This~Work &                                                                 $F_{15}(1860)$   &$1863$             &$189$              &This~Work\\
              $\ast \ast$$ \ast \ast$ &$1370(4)$         &  $190(7)$        &  Anisovich 12 \cite{anisovich12} & $\ast \ast$ &$1830_{-60}^{+120}$         &  $250^{+150}_{-50}$        &  Anisovich 12 \cite{anisovich12}\\
           &$1359$               &  $162$        &  Arndt 06 \cite{arndt06}                       &                                     &$1807$               &$109$               & Arndt 06 \cite{arndt06}\\
           &$1385$               &  $164$        & H\"ohler 93 \cite{hoehler93} &&&&\\
           &$1375(30)$        &  $180(40)$&  Cutkosky 80 \cite{cutkosky80}&&&&\\
           \hline
           $D_{13}$(1520)   &$1501$       & $112$    &  This~Work                         &   $D_{13}$(1875)   &$1975$          &$495$     &  This~Work\\
            $\ast \ast$$ \ast \ast$ &$1507(3)$         &  $111(5)$        &  Anisovich 12 \cite{anisovich12} &  $\ast \ast$ &$1860(25)$         &  $200(20)$        &  Anisovich 12 \cite{anisovich12}\\
           &$1515$                &$113$         & Arndt 06 \cite{arndt06}                  &                 &$1880(100)$       &$160(80)$     & Cutkosky 80 \cite{cutkosky80}\\%C
           &$1510$                &$120$         & H\"ohler 93 \cite{hoehler93}    &&&&\\
           &$1510(5)$           &$114(10)$  & Cutkosky 80 \cite{cutkosky80}   &&&&\\
           \hline
           $S_{11}(1535)$    &$1515$       &$123$           &This~Work          &                                                                     $P_{11}$(1880)   &$1801$         & $383$    &  This~Work\\
            $\ast \ast $$\ast \ast$ &$1501(4)$         &  $134(11)$        &  Anisovich 12 \cite{anisovich12}     &   $\ast \ast$ &$1860(35)$         &  $250(70)$        &  Anisovich 12 \cite{anisovich12}\\
            &$1502$               &$95$            & Arndt 06 \cite{arndt06}      &&&&\\
            &$1487$               &--                   &H\"ohler 93  \cite{hoehler93}  &&&&\\
            &$1510(10)$       &$260(80)$   &Cutkosky 80 \cite{cutkosky80}   &&&&\\
            \hline
            $S_{11}(1650)$  &$1655$        & $123$          &This~Work                             &    $S_{11}$(1895)   &$1858$         & $479$    &  This~Work\\
             $\ast \ast $$\ast \ast$ &$1647(6)$         &  $103(8)$        &  Anisovich 12 \cite{anisovich12}   &   $\ast \ast$ &$1900(15)$         &  $90^{+30}_{-15}$        &  Anisovich 12 \cite{anisovich12}\\
            &$1648$               &$80$            & Arndt 06 \cite{arndt06}  &$2150(70)$       &   &$350(100)$  & Cutkosky 80 \cite{cutkosky80}\\
            &$1670$               &$163$          & H\"ohler 93 \cite{hoehler93}               & &$1937 ~\rm or~1949$  &$139 ~\rm or~ 131$& Longacre 78 \cite{longacre78}\\
            &$1640(20)$       &$150(30)$   &Cutkosky 80 \cite{cutkosky80}   &&&&\\
            \hline
            $D_{15}(1675)$  &$1656$          &$128$            &This~Work      &                 $P_{13}(1900)$      &$1895$         &$100$              &This~Work\\
             $\ast \ast$$ \ast \ast\ $ &$1654(4)$         &  $151(5)$        &  Anisovich 12 \cite{anisovich12}   &     $\ast \ast$ &$1900(30)$         &  $200^{+100}_{-60}$        &  Anisovich 12 \cite{anisovich12}\\  
            &$1657$               &$139$            & Arndt 06 \cite{arndt06} &&&&\\
            &$1656$               &$126$            & H\"ohler 93 \cite{hoehler93}   &&&&\\
            &$1660(10)$        &$140(10)$     &Cutkosky 80 \cite{cutkosky80}   &&&&\\
            \hline
                       $F_{15}(1680)$    &$1669$          &$119$      &This~Work   &    $F_{17}(1990)$   &$1941$             &$130$              &This~Work\\
            $\ast \ast $$\ast \ast$ &$1676(6)$         &  $113(4)$        &  Anisovich 12 \cite{anisovich12}   &   $\ast \ast$ &$2030(65)$         &  $240(60)$        &  Anisovich 12 \cite{anisovich12}\\
            &$1674$               &$115$            & Arndt 06 \cite{arndt06}   &   &   $1900(30)$               &$260(60)$               & Cutkosky 80 \cite{cutkosky80}\\
            &$1673$               &$135$            & H\"ohler 93 \cite{hoehler93}   &&&&\\
            &$1667(5)$          &$110(10)$     & Cutkosky 80 \cite{cutkosky80}   &&&&\\
            \hline
             $D_{13}(1700)$  &$1662$           &$55$         &This~Work    &    $D_{15}(2060)$  &$2064$          &$267$       &This~Work\\
             $\ast \ast$$ \ast$ &$1770(40)$         &  $420(180)$        &  Anisovich 12 \cite{anisovich12}   &    $\ast \ast$ &$2040(15)$         &  $390(25)$        &  Anisovich 12 \cite{anisovich12}\\
            &$1700$               &$120$             & H\"ohler 93 \cite{hoehler93}                            &                      &$2100(60)$        &$360(80)$  & Cutkosky 80 \cite{cutkosky80}\\
            &$1660(30)$        &$90(40)$        & Cutkosky 80 \cite{cutkosky80}   &&&&\\
            \hline
               $P_{11}(1710)$  &$1644$         &$104$            &This~Work                                              &   $G_{17}(2190)$  &$2062$          &$428$       &This~Work\\
            $\ast \ast $$\ast$ &$1687(17)$         &  $200(25)$        &  Anisovich 12 \cite{anisovich12}  &  $\ast \ast $$\ast \ast$ &$2150(25)$       &$330(30)$        &  Anisovich 12 \cite{anisovich12}\\
            &$1690$               &$200$              &H\"ohler 93 \cite{hoehler93}   & &$2070$               &$520$            & Arndt 06 \cite{arndt06}\\
            &$1698$               &$88$                & Cutkosky 90 \cite{cutkosky90} &  &$2042$               &$482$            &H\"ohler 93 \cite{hoehler93}\\
            &$1690(20)$       &$80(20)$         & Cutkosky 80 \cite{cutkosky80}    &&$2100(50)$        &$400(160)$  & Cutkosky 80 \cite{cutkosky80}\\
            \hline
            $P_{13}(1720)$      &$1687$         &$175$              &This~Work          &&&&\\
             $\ast \ast$$ \ast \ast$ &$1660(30)$         &  $450(100)$        &  Anisovich 12 \cite{anisovich12}   &&&&\\
            &$1666$               &$355$               & Arndt 06 \cite{arndt06}    &&&&\\
            &$1686$               &$187$               & H\"ohler 93 \cite{hoehler93}   &&&&\\
            &$1680(30)$        &$120(40)$       & Cutkosky 80 \cite{cutkosky80}   &&&&\\

            \end{tabular}
            \end{ruledtabular}
            \end{center}
            \end{table*}
 %\clearpage

%\clearpage 

     \begin{table*}[htpb]
 \caption {Comparison of pole positions (in MeV) for $I=3/2$ states with other analyses.}
 \begin{center}
 \begin{ruledtabular}
  \begin{tabular}{ccclcccl}
  Resonance            &   Real Part  &     $-2\times$Imaginary Part   &  ~~~~Analysis &   Resonance            &   Real Part  &     $-2\times$Imaginary Part   &  ~~~~Analysis\\ \hline
           $P_{33}$(1232)   &  $1212$     &  $98$    &  This~Work       &        $F_{35}(1905)$    &$1769$          &$239$      &This~Work\\
         $\ast \ast$$ \ast \ast$   &$1210.5(10)$         &  $99(2)$        &  Anisovich 12 \cite{anisovich12} &   $\ast \ast$$\ast \ast $ &$1805(10)$         &  $300(15)$        &  Anisovich 12 \cite{anisovich12}\\
           &$1211$               &  $99$         &  Arndt 06 \cite{arndt06}                                                 &               &$1819$               &$247$            & Arndt 06 \cite{arndt06}\\%ARNDT 06\\
           &$1209$               &  $100$      & H\"ohler 93 \cite{hoehler93}                                            &            &$1829$               &$303$            & H\"ohler 93 \cite{hoehler93}\\%HOEHLER 93\\
           &$1210(1)$        &  $100(2)$   &  Cutkosky 80 \cite{cutkosky80}                                          &          &$1830(5)$          &$280(60)$     & Cutkosky 80 \cite{cutkosky80}\\%CUTKOSKY 80\\
           &$1211(1)$       &  $100(2)$  & Anisovich 10 \cite{anisovich10}   &&&&\\
           \hline
           $P_{33}$(1600)   &$1599$       & $211$    &  This~Work     &     $P_{31}(1910)$  &$1910$           &$199$         &This~Work\\
           $\ast \ast$$\ast$ &$1498(25)$         &  $230(50)$        &  Anisovich 12 \cite{anisovich12}     &     $\ast \ast $$\ast\ast$ &$1850(40)$         &  $350(45)$        &  Anisovich 12 \cite{anisovich12}\\
           &$1457$                &$400$         & Arndt 06 \cite{arndt06}                                               &              &$1771$               &$479$            & Arndt 06 \cite{arndt06}\\
           &$1550$                &   --         & H\"ohler 93 \cite{hoehler93}                                             &            &$1874$               &$283$             & H\"ohler 93 \cite{hoehler93}\\%HOEHLER 93\\
           &$1550(40)$           &$200(60)$  & Cutkosky 80 \cite{cutkosky80}                                  &          &$1880(30)$        &$200(40)$        & Cutkosky 80 \cite{cutkosky80}\\
           \hline
           $S_{31}(1620)$    &$1587$       &$107$           &This~Work           &                                $P_{33}(1920)$  &$2110$         &$386$            &This~Work\\
            $\ast \ast$$\ast\ast$ &$1597(4)$         &  $130(9)$        &  Anisovich 12 \cite{anisovich12}   &    $\ast \ast$$ \ast$    &$1890(30)$         &  $300(60)$        &  Anisovich 12 \cite{anisovich12}\\
            &$1595$               &$135$            & Arndt 06 \cite{arndt06}                                    &                          &$1900$               &--              &H\"ohler 93 \cite{hoehler93}\\%HOEHLER 93\\
            &$1608$               &$116$          &H\"ohler 93  \cite{hoehler93}                               &                   &$1900(80)$       &$300(100)$         & Cutkosky 80 \cite{cutkosky80}\\%CUTKOSKY 80\\
            &$1600(15)$       &$120(20)$   &Cutkosky 80 \cite{cutkosky80}    &&&\\
            &$1596(7)$       &  $130(10)$  & Anisovich 10 \cite{anisovich10}  &&&\\
            \hline
            $D_{33}(1700)$  &$1656$        & $226$          &This~Work                                                            &   $D_{35}(1930)$      &$1882$         &$187$              &This~Work\\
           $\ast \ast$$\ast \ast$  &$1680(10)$         &  $305(15)$        &  Anisovich 12 \cite{anisovich12}  &    $\ast \ast$$\ast$  &$2001$                  &$387$               & Arndt 06 \cite{arndt06}\\%ARNDT 06\\
            &$1632$               &$253$            & Arndt 06 \cite{arndt06}                                 &                              &$1850$                  &$180$               & H\"ohler 93 \cite{hoehler93}\\%HOEHLER 93\\
            &$1651$               &$159$          & H\"ohler 93 \cite{hoehler93}                           &                                &$1890(50)$           &$260(60)$       & Cutkosky 80 \cite{cutkosky80}\\%CUTKOSKY 80\\
            &$1675(25)$       &$220(40)$   &Cutkosky 80 \cite{cutkosky80}                   &&&\\
            &$1650(30)$       &  $275(35)$  & Anisovich 10 \cite{anisovich10}              &&&\\
            \hline
            $S_{31}(1900)$  &$1844$          &$223$            &This~Work                                        &               $F_{37}(1950)$   &$1871$             &$220$              &This~Work\\
            $\ast \ast $ &$1845(25)$         &  $300(45)$        &  Anisovich 12 \cite{anisovich12} &    $\ast \ast $$\ast \ast$ &$1890(4)$         &  $243(8)$        &  Anisovich 12 \cite{anisovich12}\\
           &$1780$               &--            & H\"ohler 93 \cite{hoehler93}                             &                 &$1876$               &$227$               & Arndt 06 \cite{arndt06}\\
            &$1870(40)$        &$180(50)$     &Cutkosky 80 \cite{cutkosky80}                 &               &$1878$                  &$230$               & H\"ohler 93 \cite{hoehler93}\\%HOEHLER 93\\
            &&&                                                                                                                            &               &$1890(15)$           &$260(40)$       & Cutkosky 80 \cite{cutkosky80}\\%CUTKOSKY 80\\
          \hline
            &&&                                                                                                                                          & $F_{35}(2000)$   &$1976$             &$488$              &This~Work\\
            &&&                                                                                                                                             &$\ast \ast$ &$2150(100)$               &$350(100)$            & Cutkosky 80 \cite{cutkosky80}\\  
            &&&                                                                                                                               &               &$1697$                        &$112$                      & Vrana  00 \cite{vrana00}\\                                    
            
            \end{tabular}
            \end{ruledtabular}
            \end{center}
            \end{table*}

%\newpage
In Table VII we present our results on helicity amplitudes for $I = 1/2$ and $I = 3/2$ states, and compare these results with those from earlier analyses. The first column lists the resonance name. The second column and third columns list the helicity-1/2 amplitudes for proton and neutron targets, respectively, and the fourth and fifth columns list the helicity-3/2 amplitudes for proton and neutron targets, respectively.
 Our results on helicity amplitudes, in most cases, are comparable  both in magnitude and sign with previous analyses. For $P_{11}(1880)$, our result for proton target agrees quite well with the solution ``b" of the analysis by Anisovich {\it et al.\ }\cite{anisovich12} but disagrees in sign with the solution ``a". Resonances where we differ significantly with previous analyses are seen to be $S_{11}(1650)$, $S_{31}(1620)$, $P_{33} (1600)$, and $F_{35}(1905)$. For $S_{11}(1650)$, the helicity-1/2 amplitude for the neutron target was found to be +0.011 ${\rm GeV}^{-1/2}$, which differs in sign and is considerably smaller than the values in analyses by Anisovich {\it et al.\ }\cite{anisovich09} and Arndt {\it et al.\ }\cite{arndt06}. The same is true for $S_{31}(1620)$ and $P_{33} (1600)$ with our helicity amplitudes of $-0.003$ and $+0.006$ ${\rm GeV}^{-1/2}$, respectively. For $F_{35}(1905)$ our helicity-1/2 and helicity-3/2 amplitudes are considerably larger than in other analyses although we agree in sign.  

Figure 2 shows representative Argand diagrams for pion photoproduction amplitudes ($\gamma N\rightarrow \pi N$) for both $I=1/2$ and $I=3/2$ partial waves. These amplitudes are dimensionless unlike those from the Ref. \cite{arndt06}, which are expressed in milli-fermi (mfm). Details of the conversion from dimensioned amplitudes to our dimensionless amplitudes can be found elsewhere \cite{niboh97,piN98}.

 \begin{table*}[htpb]
 \caption {Comparison of helicity amplitudes (in $10^{-3}~{\rm GeV}^{-1/2}$) for $I=1/2$ states with other analyses. }
 \begin{center}
 \begin{ruledtabular}
 \begin{tabular}{cccclccccl}
 $A_{\frac12}^p$& $A_{\frac12}^n$ & $A_{\frac32}^p$  & $A_{\frac32}^n$&Analysis &  $A_{\frac12}^p$& $A_{\frac12}^n$ & $A_{\frac32}^p$  & $A_{\frac32}^n$&Analysis \\ \hline
%      $P_1$$_1$(1440) $\ast \ast $$\ast \ast$       &&&& &                                                                   $P_{11}$(1710)  $\ast \ast$$ \ast$      &&&&\\   
        \multicolumn{5}{c}    {  $P_1$$_1$(1440) $\ast \ast $$\ast \ast$}   & \multicolumn{5}{c}        {$P_{11}$(1710)  $\ast \ast$$ \ast$ } \\  \cline{1-5}     \cline{6-10}                                    
                                           $-84(3)$    &           $40(5)$   &          &   &  This ~Work  &                                $-8(3)$      & $17(3)$         &            &         & This~Work \\
                                           $-61(8)$         &   &     &   & Anisovich 12 \cite{anisovich12}   &                           $52(15)$         &   &     &   & Anisovich 12 \cite{anisovich12}   \\       
                                            $-51(2)$    &         &    &      & Dugger 07 \cite{dugger07}           &                         $7(15)$    &  $-2(15)$      &           &          & Arndt 96 \cite{arndt96}   \\          
                                            $-63(18)$  &         $45(15)$ &      &    & Arndt 96 \cite{arndt96}   &                         $6(18)$    &                        &          &           & Crawford 83 \cite{crawford83}   \\        
                                        &&&&                                                                                                                              &      $28(9)$    &  $0(18)$       &          &            &   Awaji 81 \cite{awaji81} \\      
 %                                        &&&&                                                                                                                                &         
      \hline
 %     $D_1$$_3$(1520) $\ast \ast$$ \ast \ast$  &    &  &&   &                   $P_1$$_3$(1720) $\ast \ast $$\ast \ast$   &&&&\\  
       \multicolumn{5}{c}    { $D_1$$_3$(1520) $\ast \ast$$ \ast \ast$ }   & \multicolumn{5}{c}        { $P_1$$_3$(1720) $\ast \ast $$\ast \ast$} \\  \cline{1-5}     \cline{6-10} 
                                         $-34(1)$   & $-38(3)$          & $127(3)$           & $-101(4)$         & This~Work   &                               $57(3)$      & $-2(1)$         & $-19(2)$            & $-1(2)$        & This~Work\\
                                             $-22(4)$         &    & $131(10)$    &   & Anisovich 12 \cite{anisovich12} &                                  $110(45)$         &    & $150(30)$    &   & Anisovich 12 \cite{anisovich12}\\
                                            $-28(2)$   &                          &$143(2)$            &                             &  Dugger 07 \cite{dugger07} &       $97(3)$       &                        & $-39(3)$           &                        & Dugger 07 \cite{dugger07}\\ 
                                            $-38(3)$   &                          &$147(10)$         &                             &  Ahrens 02 \cite{ahrehs02}  &   $-15(15)$  &  $7(15)$         & $7(10)$             & $-5(25)$       & Arndt 96 \cite{arndt96}\\
                                            $-20(7)$   &  $-48(8)$         &$167(5)$           &  $-140(10)$      &  Arndt 96 \cite{arndt96}            &  $44(66)$     &                       &  $-24(6)$         &                        & Crawford 83 \cite{crawford83}\\
                                            $-28(14)$ &                          &$156(22)$         &  $-124(9)$        &  Crawford 83 \cite{crawford83}& $-4(7)$        &  $2(5)$          & $-4(16)$         &  $-15(19)$  &  Awaji 81 \cite{awaji81}\\
                                            $-7(4)$     &  $-66(13)$      &$168(13)$          &                            &  Awaji 81 \cite{awaji81}            & &&&&\\
      \hline
%      $S_1$$_1$(1535) $\ast \ast$$ \ast \ast$  &    &&      &        &                                                          $F_{15}$(1860)$\ast \ast$  & &&&\\
       \multicolumn{5}{c}    { $S_1$$_1$(1535) $\ast \ast$$ \ast \ast$ }   & \multicolumn{5}{c}        { $F_{15}$(1860) $\ast \ast$} \\  \cline{1-5}     \cline{6-10} 
                                              $59(3)$      &        $-49(3)$                       &              &           & This~Work     &      $-17(3)$      & $10(5)$         & $29(4)$            & $-9(5)$        & This~Work\\
                                            $105(10)$    &         & &    & Anisovich 12 \cite{anisovich12}                      &                 $-19(11)$         &    & $48(18)$    &   & Anisovich 12 \cite{anisovich12}\\                                 
                                            $90(25)$    & $-80(20)$        & &    & Anisovich 09 \cite{anisovich09}                                      &&&&&\\
                                            $91(2)$      &           &   &   & Dugger 07 \cite{dugger07}                                               &&&&&\\                                         
                                            $120(13)$  &         &   &   & Krusche 97 \cite{krusche97}                                              & &&&&\\
                                            $60(15)$    &         $-20(35)$  & &   & Arndt 96 \cite{arndt96}                                       &&&&&\\
                                            $97(6)$      &          &  &  & Benmerrouch 95 \cite{bemner95}                                       &&&&&\\
                                          \hline
 %     $S_1$$_1$(1650) $\ast \ast$$ \ast \ast$ &            &  &&  &                               $D_{13}$(1875) $\ast \ast$ &&&&\\
       \multicolumn{5}{c}    { $S_1$$_1$(1650) $\ast \ast$$ \ast \ast$ }   & \multicolumn{5}{c}        { $D_{13}$(1875) $\ast \ast$ } \\  \cline{1-5}     \cline{6-10} 
                                         $30(3)$              & $11(2)$           &    &   &  This~Work       &                $7(8)$  & $55(21)$   & $43(22)$    & $-85(31)$         & This~Work\\           
                                            $33(7)$         &    &     &   & Anisovich 12 \cite{anisovich12}   &                                       $18(10)$         &    &$-9(5)$     &   & Anisovich 12 \cite{anisovich12}\\
                                          $100(35)$       & $-55(20)$   &     &   & Anisovich 09 \cite{anisovich09}                & $-20(8)$  & $7(13)$  & $17(11)$    &  $-53(34)$         &  Awaji 81 \cite{awaji81}\\%AWAJI 81\\                            
                                           $22(7)$    &       &                              &        & Dugger 07 \cite{dugger07}                      &&&&&\\ 
                                            $69(5)$             &  $-15(5)$    &   &   & Arndt 96 \cite{arndt96}                                       &&&&&\\
                                          \hline
 %              $D_1$$_5$(1675) $\ast \ast $$\ast \ast$ &              && &        &                    $P_{11}$(1880)$\ast \ast$  &  &&&\\  
                \multicolumn{5}{c}    {$D_1$$_5$(1675) $\ast \ast $$\ast \ast$  }   & \multicolumn{5}{c}        {$P_{11}$(1880) $\ast \ast$ } \\  \cline{1-5}     \cline{6-10} 
                                             $11(1)$            & $-40(4)$    & $20(1)$        &$-68(4)$        & This~Work                 &  $21(6)$      & $14(7)$      &                            &                               & This~Work\\
                                             $24(3)$         &    & $25(7)$    &   & Anisovich 12 \cite{anisovich12}     &                                              $-13(3)$         &  &  &   & Anisovich 12a \cite{anisovich12}\\      
                                            $18(2)$    &     &  $21(1)$                                  &                                & Dugger 07 \cite{dugger07}& $34(11)$      &  &  &   & Anisovich 12b \cite{anisovich12}\\
                                            $15(10)$  &  $-49(10)$    & $10(7)$      &$-51(10)$        & Arndt 96 \cite{arndt96}                           &&&&&\\
                                            $21(11)$  &    & $15(9)$                                   &                               & Crawford 83 \cite{crawford83} &&&&&\\
                                            $34(5)$     &  $-57(24)$    & $24(8)$      & $-77(18)$      &  Awaji 81 \cite{awaji81}                          &&&&&\\
                                                               & $-33(4)$        &  &$-69(4)$          & Fujii 81 \cite{fuji81}                                                     &&&&&\\
                                          
                            \hline

 %     $F_1$$_5$(1680)$\ast \ast$$ \ast \ast$  &    && &   &                                  $S_{11}$(1895) $\ast \ast$&  &&&\\
       \multicolumn{5}{c}    { $F_1$$_5$(1680) $\ast \ast$$ \ast \ast$ }   & \multicolumn{5}{c}        {$S_{11}$(1895) $\ast \ast$ } \\  \cline{1-5}     \cline{6-10} 
                                       $-17(1)$     & $29(2)$   & $136(1)$       &$-59(2)$   & This~Work        &  $12(6)$      & $3(7)$         &            &        & This~Work\\      
                                           $-13(3)$         &    & $135(6)$    &   & Anisovich 12 \cite{anisovich12}   &                  $-11(6)$         &    &     &   & Anisovich 12 \cite{anisovich12}\\
                                           $-17(1)$     &                   &  $134(2)$     &                            &  Dugger 07 \cite{dugger07}          &&&&&\\
                                           $-10(4)$     & $30(5)$   &$145(5)$        & $-40(15)$         & Arndt 96 \cite{arndt96}                   &&&&&\\
                                           $-17(18)$   &                  &$132(10)$     &                             & Crawford 83 \cite{crawford83}      &&&&&\\
                                           $-9(6)$       & $17(14)$ & $115(8)$      & $-33(13)$           &  Awaji 81 \cite{awaji81}                 &&&&&\\
                                                               & $32(3)$   &                        & $-23(5)$              & Fujii 81 \cite{fuji81}                         &&&&&\\                                           
                           \end{tabular}
                            \end{ruledtabular}
                            \end{center}
                            \end{table*}

 \begin{table*}[htpb]
 \addtocounter{table}{-1}
 \caption {Cont'd. }
 \begin{center}
 \begin{ruledtabular}
 \begin{tabular}{cccclccccl} 
 $A_{\frac12}^p$& $A_{\frac12}^n$ & $A_{\frac32}^p$  & $A_{\frac32}^n$&Analysis &  $A_{\frac12}^p$& $A_{\frac12}^n$ & $A_{\frac32}^p$  & $A_{\frac32}^n$&Analysis \\ \hline
%       $D_{13}$(1700)$\ast \ast$$\ast$  &      && &     &                  $P_1$$_3$(1900)$\ast  \ast$ &  &&&\\
        \multicolumn{5}{c}    {  $D_{13}$(1700) $\ast \ast$$\ast$ }   & \multicolumn{5}{c}        { $P_1$$_3$(1900) $\ast  \ast$} \\  \cline{1-5}     \cline{6-10} 
                                          $21(5)$    &  $-49(8)$ & $50(9)$     & $-92(14)$       & This~Work         &  $41(8)$     & $-10(4)$         & $-4(6)$            & $-11(7)$        & This~Work\\
                                              $41(17)$         &    &  $-34(13)$   &   & Anisovich 12 \cite{anisovich12}    &                 $26(15)$         &    &   $-65(30)$  &   & Anisovich 12 \cite{anisovich12}\\                                            
                                            $-16(14)$  &                        & $-9(12)$   &                            &  Crawford 83 \cite{crawford83}  &   &&&&\\
                                            $-2(13)$    &  $6(24)$         & $29(14)$ & $-33(17)$         &   Awaji 81 \cite{awaji81}              &   &&&&\\
\hline
%  $D_{15}$(2060)$\ast \ast$ &  &&&&&&&&\\
   \multicolumn{5}{c}    {$D_{15}$(2060) $\ast \ast$  }   & \multicolumn{5}{c}        { } \\  \cline{1-5}     
                                                    $18(4)$      & $-12(17)$         & $10(4)$            & $-23(23)$        & This~Work&&&&&\\
                                                     $67(15)$    &  & $55(20)$      &   & Anisovich 12 \cite{anisovich12}&&&&&\\        \hline                                    
 
%     $P_{33}(1232)$  $\ast \ast$$ \ast \ast$             &  &&& &      $F_{35}(1905)$ $\ast \ast$$\ast \ast$   &&&   &  \\
       \multicolumn{5}{c}    { $P_{33}(1232)$  $\ast \ast$$ \ast \ast$ }   & \multicolumn{5}{c}        {$F_{35}(1905)$ $\ast \ast$$\ast \ast$} \\  \cline{1-5}     \cline{6-10} 
%        $\ast \ast$$ \ast \ast$                              & $-131(4)$       & $-254(5)$   &  Anisovich 12 \cite{anisovich12}&&         $\ast \ast$$\ast \ast$                               & $25(5)$       & $-49(4)$   &  Anisovich 12 \cite{anisovich12}\\
                                        $-137(1)$     &      & $-251(1)$    &            &  This ~Work   &  $66(18)$   &  &$-223(29)$ &  & This~Work\\
                                           $-131(4)$  &     & $-254(5)$   & & Anisovich 12 \cite{anisovich12}  & $25(5)$   &    & $-49(4)$   &&  Anisovich 12 \cite{anisovich12}\\
                                            $-139(4)$    &&  $-258(5)$  && Dugger 07 \cite{dugger07}   &  $21(4)$     &  &$-46(5)$      &  & Dugger 07 \cite{dugger07}\\
                                            $-141(5)$   & & $-261(5)$   && Arndt 96 \cite{arndt96}          &  $22(5)$     &  &$-45(5)$      &  & Arndt 96 \cite{arndt96}\\
                          %               & $-136(5)$       & $-267(8)$   &  Anisovich 10 \cite{anisovich10}\\
                                          $-129(5)$      && $-243(1)$   & & Arndt 02 \cite{arndt02}  &   $21(10)$   & &$-56(28)$     & & Crawford 83 \cite{crawford83}\\
                                         $-145(15)$    && $-263(26)$ &  & Crawford 83 \cite{crawford83}  &$43(20)$   & &$-25(23)$     & &  Awaji 81 \cite{awaji81}\\
                                           $-138(4)$    & & $-259(6)$   &  &Awaji 81 \cite{awaji81} & & & & &\\
 %     \hline
 \hline
%      $P_{33}(1600)$  $\ast \ast$$\ast$     &  &&  &&    $P_{31}(1910)$ $\ast \ast$$\ast \ast$  &&&&   \\
       \multicolumn{5}{c}    { $P_{33}(1600)$  $\ast \ast$$\ast$  }   & \multicolumn{5}{c}        {$P_{31}(1910)$ $\ast \ast$$\ast \ast$} \\  \cline{1-5}     \cline{6-10} 
   %       $\ast \ast$$\ast$                                & $-50(9)$       & $-40(12)$   &  Anisovich 12 \cite{anisovich12}&&  $\ast \ast$$\ast \ast$                               & $22(9)$       &    &  Anisovich 12 \cite{anisovich12}\\
                                           $6(5)$       &  &$52(8)$          &                        & This~Work    &  $30(2)$  &&      &            & This~Work\\
                                          $-50(9)$       && $-40(12)$   & & Anisovich 12 \cite{anisovich12} & $22(9)$   &&    &    &  Anisovich 12 \cite{anisovich12}\\
                                            $-18(15)$   &&$-25(15)$          &            &  Arndt 96 \cite{arndt96}   &         $40(14)$   &       &$23(17)$      &     &  Awaji 81 \cite{awaji81}\\
                                            $-39(30)$  & &$-13(14)$            &         &  Crawford 83 \cite{crawford83} & &&&&\\
                                            $-46(13)$   &&$25(31)$                &       &  Awaji 81 \cite{awaji81} &&& && \\
%      \hline
\hline
%      $S_{31}(1620)$ $\ast \ast$$\ast\ast$ &&&&      &  $P_{33}(1920)$  $\ast \ast$$ \ast$  & &&& \\
       \multicolumn{5}{c}    {  $S_{31}(1620)$ $\ast \ast$$\ast\ast$}   & \multicolumn{5}{c}        {$P_{33}(1920)$  $\ast \ast$$ \ast$ } \\  \cline{1-5}     \cline{6-10} 
  %       $\ast \ast$$\ast\ast$                                 &  $52(5)$    &        & Anisovich 12 \cite{anisovich12} &&  $\ast \ast$$ \ast$                              & $130^{+30}_{-60}$       & $-115^{+25}_{-50}$   &  Anisovich 12 \cite{anisovich12}\\
                                           $-3(3)$        &  && &                        This~Work         &  $51(10)$   &      &$17(15)$          & & This~Work\\
                                            $50(2)$      &  &&&   Dugger 07 \cite{dugger07}  &          $40(14)$ &         &$23(17)$ &          &  Awaji 81 \cite{awaji81}\\
                                            $35(10)$   &  &&&   Crawford 83 \cite{crawford83}& &&&&\\
                                            $35(20)$   &   &&&   Arndt 96 \cite{arndt96}               & &&&&\\
                                            $10(15)$   &   &&&   Awaji 81 \cite{awaji81}             &&& &&\\
%          \hline
\hline
%      $D_{33}(1700)$ $\ast \ast$$\ast \ast$      &&&  &&     $D_{35}(1930)$ $\ast \ast$$\ast$ &&&&\\
       \multicolumn{5}{c}    {$D_{33}(1700)$ $\ast \ast$$\ast \ast$  }   & \multicolumn{5}{c}        {$D_{35}(1930)$ $\ast \ast$$\ast$ } \\  \cline{1-5}     \cline{6-10} 
 %         $\ast \ast$$\ast \ast$                                & $160(20)$       & $165(25)$   &  Anisovich 12 \cite{anisovich12}  &&  $\ast \ast$$\ast$       & $-7(10)$     &$5(10)$      & Arndt 96 \cite{arndt96}\\
                                            $58(10)$   &      &$97(8)$      &                            & This~Work   &  $11(3)$    &  & $2(2) $        & & This~Work\\
                                           $160(20)$   &    & $165(25)$   &&  Anisovich 12 \cite{anisovich12}      & $-7(10)$  &   &$5(10)$   &   & Arndt 96 \cite{arndt96}\\
                                             $125(3)$    &&  $105(3)$  && Dugger 07 \cite{dugger07} &    $9(9)$      &    &$-25(11)$  && Awaji 81 \cite{awaji81}\\
                                            $90(25)$   && $97(20)$          &            &  Arndt 96 \cite{arndt96} &&&  &&\\
                                            $111(17)$   && $107(15)$        &             &  Crawford 83 \cite{crawford83}&&& &&\\ 
                                            $89(33)$   && $60(15)$               &        &  Awaji 81 \cite{awaji81}&&& &&\\
                                          
   %                         \hline
   \hline
 %      $S_{31}(1900)$ $\ast \ast$       &&&   &&  $F_{37}(1950)$ $\ast \ast $$\ast \ast$  &&&&\\
        \multicolumn{5}{c}    { $S_{31}(1900)$ $\ast \ast$ }   & \multicolumn{5}{c}        {$F_{37}(1950)$ $\ast \ast $$\ast \ast$ } \\  \cline{1-5}     \cline{6-10} 
 %           $\ast \ast$            &  $59(16)$   &    & Anisovich 12 \cite{anisovich12}&& $\ast \ast $$\ast \ast$                                     & $-71(4)$       & $-94(5)$   &  Anisovich 12 \cite{anisovich12}\\
                                           $-82(9)$     & &&  &  This~Work    &$-65(1)$   &  & $-83(1)$  &      & This~Work\\
                                            $59(16)$   &   && & Anisovich 12 \cite{anisovich12}     & $-71(4)$    &   & $-94(5)$  & &  Anisovich 12 \cite{anisovich12}\\
                                            $-4(16)$      &  && & Crawford 83 \cite{crawford83} &   $-79(6)$     &&$-103(6)$    &  & Arndt 96 \cite{arndt96}\\
                                            $29(8)$       &   & && Awaji 81\cite{awaji81} & $-68(7)$    &  &$-94(16)$  &    & Awaji 81 \cite{awaji81}\\
                                          &&&&&       $-83(8)$   &   &$-92(8) $   &    & Anisovich 10 \cite{anisovich10}\\
                                          &&&&&&&\\ \hline
        \multicolumn{5}{c}    {  }   & \multicolumn{5}{c}        {$F_{35}(2000)$  $\ast\ast$ } \\       \cline{6-10}                                  
   %                                       &&&&&$F_{35}(2000)$  $\ast\ast$ & &&& \\
                                       &&&&&    $-61(18)$ &  &$158(32)$   &    & This~Work\\

                           \end{tabular}
                            \end{ruledtabular}
                            \end{center}
                            \end{table*}
\section{COMPARISONS WITH QUARK-MODEL PREDICTIONS}
 In Tables VIII and IX we present decay amplitudes for various channels and compare our results for $I=1/2$ and $I=3/2$ states, respectively, with quark models. The magnitude of the decay amplitude is equal to $\sqrt{\Gamma_i}$~, the square root of the partial width for the channel. Its sign is the phase relative to the $\pi N$ coupling (taken to be positive). The values in the first row are our results, while those in the second row are from Koniuk and Isgur \cite{koniuk80} and the third from Capstick and Roberts \cite{capstick94, capstick98}. The channels included are $\pi N$, $\eta N$, $K\Lambda$, $\pi \Delta$, and $\rho N$. The subscript `l' or `h' that appears with a channel represents the lower or higher orbital angular momentum of that channel. The states are listed in the first column in an ascending order in terms of their masses. 
 
 On comparing with predictions of quark models we find that our results over-all agree well with either one or both models. If we break down channel by channel, we have excellent agreements for the elastic decay amplitudes with at least one of the quark models for all states except $D_{13}(1700)$ and $F_{15}(1860)$. For $F_{15}(1860)$ our $\pi N$ amplitude of 6.1 is larger than the model values of 1.3. For $D_{13}(1700)$ our elastic coupling of 1.2 is smaller than the model values of 3.6 and 5.8.
 
% We did not agree with theories for all $D_{13}$ states on results for $\eta N$ and $K\Lambda$ channels because we did not find any couplings to these channels. 
 The predicted $\eta n$ and $K\Lambda$ decay amplitudes for $D_{13}(1520)$ and $D_{13}(1700)$ are small, in agreement with our results. 
 
 Our $\eta N$ results are in excellent agreement with the model predictions for $S_{11}(1535)$, $S_{11}(1650)$, and $F_{15}(1680)$. For $P_{11}(1710)$ and $G_{17}(2190)$, our $\eta N$ results agree very well in magnitude but not in sign with the predictions.  The states for which our $\eta N$ results agree in sign but not in magnitude with the predictions are $P_{13}(1720)$, $P_{11}(1880)$, $S_{11}(1895)$, and $F_{17}(1990)$. For $D_{15}(1675)$, our $\eta N$ amplitude (+0.6) disagrees both in magnitude and sign with model predictions ($-2.8$ and $-2.5$).

 Our $K\Lambda$ results are in excellent agreement with the predictions for $S_{11}(1650)$, $F_{15}(1680)$, $P_{11}(1710)$, $P_{13}(1720)$, $S_{11}(1895)$, and $F_{17}(1990)$. For $F_{15}(1860)$, our $K\Lambda$ results agree very well in magnitude but not in sign with model predictions. For $P_{11}(1880)$ and $P_{13}(1900)$ there are no predictions with which to compare our $K\Lambda$ results. For $G_{17}(2190)$ our $K\Lambda$ amplitude (+0.3) disagrees both in magnitude and sign with the model prediction of $-1.3$ by Capstick and Roberts \cite{capstick98}.
 
 The excellent agreement of our $K\Lambda$ results with model predictions can be attributed to the extensive and non-problematic $K\Lambda$ database. The less extensive and more problematic $\eta N$ data, especially by Brown {\it et al.,} could be the reason for the poorer agreement of our $\eta N$ results with model predictions.
 
 Our results confirm the existence of $P_{11}(1710)$ and are further supported by the excellent agreements of our $K\Lambda$ and $\eta N$ amplitudes with predictions of quark models \cite{capstick94, capstick98}.
 
 In Table X we compare our helicity amplitudes (in units of $10^{-3}~{\rm GeV}^{-1/2}$) with model predictions by Koniuk and Isgur \cite{koniuk80}. For $P_{33}(1232)$, $D_{13}(1520)$, $D_{15}(1675)$, $F_{15}(1680)$, and $F_{37}(1950)$ our results are in excellent agreement both in magnitude and sign with the predictions. For $P_{11}(1440)$, $S_{11}(1535)$, and $F_{35}(1905)$ our results agree with predictions in sign but not in magnitude. For the remaining states our results differ in sign or magnitude with model predictions for one or more helicity amplitudes.

% \clearpage
%\newpage

\begin{table*}[htpb]
\caption{Comparison of decay amplitudes for $I=1/2$ states with predictions of quark models.  The first row gives our results, while the second and third rows list the available $\pi N$, $\eta N$, $K\Lambda$, $\pi\Delta_l$, $\pi\Delta_h$, $\rho_1N$, $\rho_3N_l$, and $\rho_3N_h$ amplitudes predicted by Koniuk and Isgur \cite{koniuk80} and Capstick and Roberts \cite{capstick94, capstick98}, respectively. Here, the $\pi \Delta$ and $\rho N$ amplitudes by Koniuk and Isgur \cite{koniuk80} have been multiplied by $-1$.}
\begin{center}
\begin{ruledtabular}
\begin{tabular}{ccccccccc}
State   & $\pi N$   & $\eta N$  &   $K\Lambda$    & $\pi\Delta_l$    & $\pi\Delta_h$ & $\rho_1 N$ &   $\rho_3N_l$ &   $\rho_3N_h$\\ \hline
$P_{11}(1440)$   &$12.7(1)$  & --   &  --           & $+4.0(2)$     & --   & $+1.8(3)$  &-- &-- \\
    $\ast \ast$$ \ast \ast$                           &$6.8$   &--   &--   &$+2.4$   &-- &$+0.27$ &-- &$+0.09$\\
                               &$20.3_{-0.9}^{+0.8}$ &$+0.0^{+1.0}_{-0.0}$ &--  & $+3.3^{+2.3}_{-1.8} $  &-- &$-0.3^{+0.2}_{-0.3}$ &--&$-0.5^{+0.3}_{-0.5}$\\ \hline
                               
$D_{13}(1520)$  & $8.55(4)$   &-- &--& $-3.3(1)$ &$-2.7(1)$ &--&$-4.9(1)$&--\\
   $\ast \ast$$ \ast \ast$                            &$9.2$   &$+0.4$  &--&$-6.7$      &$-2.5$ &$+0.73$&$-4.98$&$-1.1$\\
                               &$8.6(3)$   &$+0.4^{+2.9}_{-0.4}$  &$0.0_{-0.9}^{+0.0}$ &$-5.7^{+3.6}_{-1.6}$ &$-1.5^{+1.3}_{-3.0}$ &$-0.1^{+0.1}_{-0.3}$&$-2.4^{+1.9}_{-6.4}$&$-0.3^{+0.2}_{-1.0}$\\ \hline

$S_{11}(1535)$  &$7.2(1)$     &$+7.6(2)$  &--  &--&$+1.6(3)$ &$-3.8(2)$&--&$-3.5(2)$\\
   $\ast \ast $$\ast \ast$                            &$5.3$         &$+5.2$  &--&--& $+1.7$  &$-6.1$&--&$+1.6$ \\
                               &$14.7(5)$      &$+14.6^{+0.7}_{-1.3}$ &--&--&$+1.4(3)$ &$-0.7(1)$&--&$+0.4(1)$ \\ \hline

$S_{11}(1650)$  &$8.4(2)$     &$-5.1(2)$  &$-3.3(2)$  &--&$+3.0(3)$ &$-2.8(3)$&--&$-0.6(3)$ \\
     $\ast \ast $$\ast \ast$                          &$8.7$          &$-1.5$       &$-3.0$       &-- &$+8.2$ &$-9.7$&--&$+2.7$ \\
                               &$12.2(8)$       &$-7.8^{+0.1}_{-0.0}$       &$-5.2^{+1.4}_{-0.5}$        & --   &$+3.6^{+0.8}_{-0.6}$   &$+0.9^{+0.3}_{-0.2}$&--&$+0.4(1)$ \\ \hline

$D_{15}(1675)$  &$7.5(1)$     &$+0.6(3)$  &$-0.5(1)$  &$+8.1(2)$&-- &$+0.4(1)$&$-0.6(2)$&--\\
     $\ast \ast$$ \ast \ast$                          &$5.5$      &$-2.8$    &$+0.1$              &$+9.3$&-- &$-1.1$&$-2.0$&$0$\\
                               &$5.3(1)$       &$-2.5(2)$    &$0.0(0)$      &  $+5.7(4)$      &$0.0$&$+0.2(0)$&$-0.4(0)$&$0.0$\\ \hline

$F_{15}(1680)$  &$9.25(4)$     &$+1.1(2)$  &$-0.15(3)$  &$-3.6(2)$&$+1.1(1)$ &--&$-3.1(1)$&$-1.7(1)$\\
     $\ast \ast$$ \ast \ast$                          &$7.1$            &$+0.7$          &$-0.1$        &$-2.0$  & $+0.7$ &$+1.6$&$-3.96$&$-1.3$\\
                               &$6.6(2)$            &$+0.6(1)$          & $-0.1(0)$       &$+1.6(1)$ &$+0.5(1)$ &$-0.2(0)$&$-3.0^{+0.4}_{-0.5}$&$-0.3(1)$\\ \hline

$D_{13}(1700)$  & $1.2(1)$   &-- &--& $-4.2(8)$ &$+1.3(5)$ &--&$-4.6(4)$&-- \\
        $\ast \ast$$\ast$                       & $3.6$         &$-0.7$  &$-0.2$       &$-16$&$+7.7$ &$-0.11$&$-4.3$&$-2.74$\\
                               & $5.8(6)$         &$-0.2(1)$  &$-0.4(2)$       &$-27.5(16)$&$+4.6^{+1.6}_{-1.3}$ &$0.0$&$\pm 0.1(0)$&$-0.9^{+0.3}_{-0.6}$\\ \hline

$P_{11}(1710)$   &$4.1(6)$  & $-3(1)$   &  $-3.1(9)$ & $-2.7(7)$     & --   &$-4.4(7)$&--&-- \\
       $\ast \ast$$ \ast$                         &$6.7$       &$+2.9$          &  $-2.1$    & $-3.6$          &-- &$+5.5$&--&$+2.5$\\
                                &$4.2(1)$       &$+5.7(3)$          &  $-2.8(6)$    & $-13.9(15)$                      &-- &$+0.3(1)$&--& $-3.7^{+0.9}_{-1.2}$\\ \hline
$P_{13}(1720)$   &$5.2(3)$  & $0.0(7)$   &  $-2.4(2)$ & --     & --   &$+1.7(3)$&--&-- \\
     $\ast \ast$$ \ast \ast$                           &$6.5$       & $+1.9$      &  $-1.7$      &$-1.9$  &$+1.0$ &$+11.7$&$-2.6$&$-3.5$\\
                                &$14.1(1)$     &$+5.7(3)$      &   $-4.3^{+0.8}_{-0.7}$      &$-1.7(2)$   &$-1.0^{+0.2}_{-0.3}$  &$-2.6^{+0.7}_{-0.8}$&$+1.8^{+0.6}_{-0.5}$&$+0.7^{+0.3}_{-0.2}$\\ \hline
 
 $F_{15}(1860)$  &$6.1(3)$     &$-3.0(7)$  &$-0.9(2)$  &$-1(1)$&$+0.1(6)$   &--&$-2.6(10)$&$+8.6(9)$\\ 
       $\ast \ast$                       &$1.3$          &$-0.6$        &$+0.9$      &$+7.0$  &$+4.3$ &$-1.7$&$-6.6$&$-4.4$\\
                              &$0.9(2)$          &$-0.8(2)$       &-0.5(3)& $+7.8^{+0.4}_{-0.6}$&$-5.8^{+2.4}_{-3.9}$ &$-0.4(3)$&$-7.8^{+3.1}_{-0.2}$&$-0.2(1)$\\ \hline
                       
$D_{13}(1875)$  & $6(1)$   &-- &--& $-20.8(9)$ &$-4(2)$   &--&$+3.1(22)$&--\\
     $\ast \ast$                          & --  &  --  &-- &--  &-- &--&--&--\\
                               &$8.2^{+0.7}_{-1.7}$   &$+4.0(2)$ &$-5.6^{+1.7}_{-1.3}$ &$-1.4^{+0.5}_{-1.2}$ &$-5.3^{+0.9}_{-0.8}$  &$-4.4^{+1.9}_{-0.7}$&$-6.2(24)$&$-11.3^{+4.9}_{-1.6}$\\ \hline

 $P_{11}(1880)$  &$9(1)$     &$+9(2)$  &$+13(2)$  &--&$-2(2)$ &$+0.3(22)$&--&--\\
       $\ast \ast$                        &$4.4$         &$-0.8$           &$-1.4$     & --     &-- &$+4.6$&--&$-1.1$\\
                               & $2.7^{+0.6}_{-0.9}$  &$-3.7^{+0.5}_{-0.0}$&$-0.1(1)$     & $-8.7^{+2.1}_{-0.4}$     &-- &$+2.3^{+1.7}_{-1.4}$&--&$\pm 0.3^{+0.0}_{-0.1}$\\ \hline
$S_{11}(1895)$  &$9.2(6)$     &$+14(1)$  &$+3.0(7)$  &--&$+6(1)$ &$+2.0(12)$&--&$-6.6(9)$\\
        $\ast \ast$                       &--          &--           &--     & --     &-- &--&--&--\\
                               & $5.7^{+0.5}_{-1.6}$  &$+2.4^{+1.5}_{-2.3}$     &$+2.3(27)$      & --     &$-6.7^{+1.5}_{-1.3}$ &$+2.3(6)$&--&$-17.9^{+7.3}_{-3.8}$\\ \hline
$P_{13}(1900)$   &$2.6(8)$  & $0.0(8)$   &  $-3.7(6)$ & --     & --   &$+8.0(6)$&--&-- \\ 
       $\ast \ast$                         &$3.2$       &$-2.9$   &--                               &$+4.1$ & $+1.5$ &$-0.43$&$-1.32$&$-0.46$\\    
                                &$6.1^{+0.6}_{-1.2}$       &$-4.6(3)$       &$-0.9^{+0.4}_{-0.1}$ &$+3.8(5)$ &$-2.2^{+1.2}_{-1.5}$ &$-1.4^{+0.9}_{-1.0}$&$-1.0(6)$&$+0.2^{+0.5}_{-0.2}$\\     \hline
$F_{17}(1990)$  &$2.1(6)$    &--&$-1.0(5)$   &--&-- &--&--&--\\
        $\ast \ast$                      &$3.1$         &$-2.3$  &$-0.3$    &$+6.0$ &--&$-0.80$&$+4.2$&$0$\\
                              &$2.4(4)$         &$-2.2^{+0.6}_{-0.7}$   &$0.0(0)$&$+5.0^{+2.0}_{-1.4}$ &$0.0$ &$+0.6(3)$&$-1.0^{+0.6}_{-0.5}$&$0.0$\\ %\hline
%$D_{15}(2060)$  & $5(1)$   & $+0.6(15)$ &$0.3(20)$ &+11(2)&--&$-7.9(3)$&$-3.5(2)$&--\\
%          $\ast \ast$                     & --  &  --  &-- &--  &--&--&--&--\\ 
%                               &$5.2$   &$+0.0$ &$-1.7$ &$+7.8$ &$+0.9$&$-0.8$&$+0.8$&$+2.1$\\                              
 \end{tabular}
\end{ruledtabular}
\end{center}
\end{table*}

\begin{table*}[htpb]
\addtocounter{table}{-1}
\caption{Cont'd.}
\begin{center}
\begin{ruledtabular}
\begin{tabular}{ccccccccc}   
State   & $\pi N$   & $\eta N$  &   $K\Lambda$    & $\pi\Delta_l$    & $\pi\Delta_h$ & $\rho_1 N$ &   $(\rho_3N)_l$ &   $(\rho_3N)_h$\\ \hline                                          
$D_{15}(2060)$  & $5(1)$   & $+0.6(15)$ &$+0.3(20)$ &+11(2)&--&$-7.9(25)$&$-3.5(18)$&--\\
         $\ast \ast$                     & --  &  --  &-- &--  &--&--&--&--\\ 
                               &$5.2^{+0.4}_{-1.0}$   &$+0.0^{+0.4}_{-0.2}$ &$-1.7^{+0.5}_{-0.4}$ &$+7.8^{+1.1}_{-1.3}$ &$+0.9^{+0.8}_{-0.4}$&$-0.8^{+0.3}_{-0.4}$&$+0.8^{+0.6}_{-0.4}$&$+2.1^{+2.4}_{-1.2}$\\  \hline  
$G_{17}(2190)$  &$10.0(9)$    &$-3(1)$&$+0.3(4)$   &--&--&--&$-6.7(24)$&--\\
      $\ast \ast$$ \ast \ast$                         & --  &--   & --    & --    &--&--&--&--\\
                               &$6.9(13)$    & $+2.5(7)$   &$-1.3^{+0.4}_{-0.6}$    &$-1.3(2)$&$-2.6^{+0.9}_{-1.3}$ &$-1.9^{+0.7}_{-1.5}$&$-11.4^{+1.0}_{-3.8}$&$-3.7^{+1.4}_{-3.0}$\\                                                            
\end{tabular}
\end{ruledtabular}
\end{center}
\end{table*}

 \begin{table*}[htpb]
\caption{Comparison of decay amplitudes for $I=3/2$ states with predictions of quark models.  The first row gives our results, while the second and third rows list the available $\pi N$, $\pi\Delta_l$, $\pi\Delta_h$, $\rho_1N$, $\rho_3N_l$, and $\rho_3N_h$ amplitudes predicted by Koniuk and Isgur \cite{koniuk80} and Capstick and Roberts \cite{capstick94, capstick98}, respectively. Here, the $\pi\Delta$ and $\rho N$ amplitudes by Koniuk and Isgur \cite{koniuk80} have been multiplied by $-1$.}
\begin{center}
\begin{ruledtabular}
\begin{tabular}{ccccccc}
State   & $\pi N$      & $\pi\Delta_l$    & $\pi\Delta_h$ & $\rho_1 N$ &   $\rho_3N_l$ &   $\rho_3N_h$\\ \hline
$P_{33}(1232)$   &$10.60(2)$             & $0.00(5)$     & --   & --  &-- &-- \\
    $\ast \ast$$ \ast \ast$                           &$11.0$     &--   &-- &-- &-- &--\\
                               &$10.4(1)$ &-- & --   &-- &-- &--\\ \hline
                               
$P_{33}(1600)$  & $4.2(5)$   & $+12.5(4)$ &-- &--&--&--\\
   $\ast \ast$$ \ast \ast$                            &$5.4$   &$+8.6$  &$+0.1$&$-1.3$      &$-5.5$ &$-0.35$\\
                               &$8.7(2)$   &$+8.4^{+3.6}_{-3.5}$   &$0.0$ &$+0.4^{+0.7}_{-0.3}$&$-0.9^{+0.6}_{-1.4}$&$0.0$\\ \hline

$S_{31}(1620)$  &$6.1(1)$     &--  &$-6.0(2)$ &$+5.4(2)$&--&$-0.6(2)$\\
   $\ast \ast $$\ast \ast$                            &$3.3$         &-- & $-8.0$   &$+7.82$&--&$-1.72$ \\
                               &$5.1(7)$      &-- &$-4.2^{+1.3}_{-1.8}$ &$-3.6^{+1.3}_{-2.5}$&--&$-0.3^{+0.1}_{-0.2}$ \\ \hline

$D_{33}(1700)$  &$6.0(2)$     &$+11.6(3)$  &$+1.9(6)$  &-- &$+8.7(4)$&--\\
     $\ast \ast$$ \ast \ast$                          &$4.9$      &$+10.3$    &$+6.3$              &$+4.2$&$+16.5$ &$+0.89$\\
                               &$4.9(7)$       &$+15.4^{+0.9}_{-1.8}$    &$+5.0^{+2.4}_{-1.8}$      &  $-1.2^{+0.6}_{-1.2}$      &$+3.4^{+2.2}_{-1.7}$&$+0.5^{+0.5}_{-0.2}$\\ \hline

$S_{31}(1900)$  &$4.4(5)$     &--  &$-11.5(8)$  &$-5.4(9)$&-- &$-7.3(10)$\\
     $\ast \ast$$ \ast \ast$                          &--            &--          &--        &--  & -- &--\\
                               &$3.1^{+0.4}_{-1.1}$            &--          &$-4.4^{+0.8}_{-0.7}$ &$-2.2(6)$&--&$+2.3^{+1.0}_{-0.4}$\\ \hline

$F_{35}(1905)$  & $4.0(3)$   & $+8.8(11)$ &$+13.3(10)$ &--&$-2.6(14)$&-- \\
        $\ast \ast$$\ast$                       & $4.0$         &$+3.2$  &$+5.5$       &$-0.049$&$-2.1$ &$-6.4$\\
                               & $3.4(3)$               &$-1.5(0)$&$+4.7(6)$ &$-0.7(2)$&$+6.3^{+0.8}_{-0.4}$&$-0.7^{+0.1}_{-0.2}$\\ \hline

$P_{31}(1910)$   &$6.0(2)$  & --   &  -- & --    & --   &-- \\
       $\ast \ast$$ \ast$                         &$5.3$       &$+5.9$          &  --    & $-3.7$          &-- &$-4.9$\\
                                &$9.4(4)$       &$-8.4^{+0.2}_{-0.1}$                                &-- &$+5.6^{+0.9}_{-0.4}$&--&$+2.6^{+0.4}_{-0.2}$\\ \hline
$P_{33}(1920)$   &$7.9(12)$  & $-5.2(21)$   &  -- & --     & --   &-- \\
     $\ast \ast$$ \ast \ast$                           &$5.2$       & $-3.2$      &  $-1.4$      &$-8.1$  &$+6.2$ &$+5.5$\\
                                &$4.2(3)$           &$-8.9^{+0.3}_{-0.2}$   &$+4.4^{+0.8}_{-0.7}$  &$+5.3^{+1.3}_{-0.5}$&$+6.6^{+1.6}_{-0.7}$&$-0.7^{+0.2}_{-0.4}$\\ \hline
 
 $D_{35}(1930)$  &$4.3(4)$     &--  &--  &--&--   &--\\ 
       $\ast \ast$                       &--          &--        &--      &--  &-- &--\\
                              &$5.2(1)$          & $+3.9(2)$&$-0.7(1)$ &$+0.1(0)$&$-2.9^{+0.5}_{-0.8}$&$-0.1^{+0.0}_{-0.1}$\\ \hline
                       
$F_{37}(1950)$  & $10.9(1)$   &+4.7(3) & -- &--  &--&--\\
     $\ast \ast$                          & $7.5$  &  $+5.5$  &$0.0$ &$-4.69$ &$-8.2$&$0$\\
                               &$7.1(1)$   &$+4.8(2)$ &$0.0$ &$+1.3(1)$ &$-2.3(2)$  &$0.0$\\ \hline

 $F_{35}(2000)$  &$5.8(5)$     &$-3.7(17)$  &$+1.7(29)$  &--&$+21.2(12)$ &--\\
       $\ast \ast$                        &$1.0$          &$-6.2$          &$+1.4$     & $+7.2$    &$+17.8$ &$+4.6$\\
                               & $1.2(3)$  &$-14.0^{+1.6}_{-0.1}$     & $+1.5^{+1.5}_{-0.8}$     &$+2.6^{+2.8}_{-2.1}$&$+3.1(12)$&$-3.1^{+2.4}_{-3.2}$\\

 \end{tabular}
\end{ruledtabular}
\end{center}
\end{table*}

%\newpage
\begin{table*}[htpb]
\caption{Comparison of helicity amplitudes with predictions of quark models. The first row lists our results, while the second and third rows list the helicity amplitudes predicted by Koniuk and Isgur \cite{koniuk80} and Capstick \cite{capstick92}, respectively. The first column identifies the states.}
\begin{center}
\begin{ruledtabular}
\begin{tabular}{cccccccccc}
State   & $A^p_{1/2}$   & $A^n_{1/2}$  &   $A^p_{3/2}$    & $A^n_{3/2}$     &State   & $A^p_{1/2}$   & $A^n_{1/2}$  &   $A^p_{3/2}$    & $A^n_{3/2}$     \\ \hline                               
$P_{11}(1440)$   & $-84(3)$            & $40(5)$             &--      &--                       &       $P_{13}(1720)$   &  $57(3)$      & $-2(1)$         & $-19(2)$            & $-1(2)$    \\
   $\ast \ast $$\ast \ast$                            & $-24$                   & $16$                &--      &-- &       $\ast \ast$$ \ast \ast$                          &$-133$       & $57$      &  $46$      &$-10$   \\
                                                                     & $4$             & $-6$          &--     &--  &                                            & $-11$       &$4$           &$-31$          &$11$\\
\hline                               
$D_{13}(1520)$  &  $-34(1)$   & $-38(3)$          & $127(3)$           & $-101(4)$   &   $P_{33}(1232)$   & $-137(1)$    &--         &    $-251(1)$      &--   \\
     $\ast \ast$$ \ast \ast$                          &$-23$   &$-45$  &$128$      &$-122$    &  $\ast \ast $$\ast \ast$                         &$-103$           &--         &$-179$         &--  \\ 
                                                      &$-15$  &$-38$  &$134$  &$-114$                        &                                                                 &$-108$           &--         &$-108$         &--  \\ \hline
$S_{11}(1535)$  & $59(3)$      & $-49(3)$                       &--              &--                &      $P_{33} (1600)$ & $6(5)$            &--        &  $52(8)$           &--\\
   $\ast \ast $$\ast \ast$                            &$147$         &$-119$  & --   &--              &      $\ast\ast$$\ast$       & $-16$              &--       &   $-46$              &--\\
                                          &$76$               &$-63$          &--              &--                      &                                    &$30$                 &--      &    $51$               &--\\
  \hline             

$S_{11}(1650)$  & $30(3)$              & $11(2)$           &  --  &    --                                                 &       $S_{31}(1620)$   & $-3(3)$       &--               &--          &-- \\
    $\ast \ast$$ \ast \ast$                           &$88$          &$-35$       &--       &--                         &       $\ast \ast$$\ast\ast$                           & $59$           &--          &--          &--  \\ 
                                                                & $54$              &$-35$       &--       &--                         &                                                                      &$81$           &--          &--          &--\\ \hline

$D_{15}(1675)$  & $11(1)$            & $-40(4)$    & $20(1)$        &$-68(4)$                            &         $D_{33}$(1700) &  $58(10)$      &--      &  $97(8)$                          &--  \\
   $\ast\ast$$\ast\ast$                            &$12$      &$-37$    &$16$              &$-53$             &             $\ast \ast$$\ast\ast$                          &  $100$     &--            &$105$             &--\\ 
                                                                 &$2$          &$-35$   &$3$                 &$-51$            &                                                                           &$82$          &--           &$68$                &--\\ \hline
$F_{15}(1680)$  & $-17(1)$     & $29(2)$   & $136(1)$       &$-59(2)$                                     &  $F_{35}$(1905) &  $68(18)$      &--      &  $-223(29)$                          &--  \\
      $\ast \ast$$ \ast \ast$                         &$\sim0$            &$26$          &$91$        &$-25$  &   $\ast \ast$$\ast\ast$                          &  $8$     &--            &$-33$             &--\\ 
                               &$-38$            &$19$          & $56$       & $-23$                                           &                                                                 &$26$    &--            &$-1$                &--    \\  \hline
$D_{13}(1700)$  &   $21(5)$    &  $-49(8)$ & $50(9)$     & $-92(14)$                                      & $P_{31}(1910)$   &  $30(2)$      & --         &--            &--    \\
       $\ast \ast$$\ast$                        & $-7$         &$-15$  &$11$       &$-76$                       &    $\ast \ast$$\ast \ast$                         &$\sim 0$       &--          &  --    & --         \\ 
                                                             &$-33$       &$18$   &$-3$         &$-30$                      &                                                                  &$-8$              &--          &--      &--          \\ \hline
                                                             
   $P_{11}(1710)$   &  $-8(3)$      & $17(3)$         &--            &--                               & $F_{37}$(1950)  &  $-65(1)$  & --   & $-83(1)$    & --\\
    $\ast \ast$$ \ast$                           &$-47$       &$-21$          &  --    & --  &  $\ast \ast$$ \ast \ast$                               &  $-50$&--&$-69$&--\\
                                                             &$13$         &$-11$          &--      &--   &                                                 &$-33$         &-- & $-42$&--\\
                      
\end{tabular}
\end{ruledtabular}
\end{center}
\end{table*}

\begin{figure*}[H]
%\caption{Argand diagrams for two body or quasi-two body amplitudes}
\vspace{-10mm}
\scalebox{0.35}{\includegraphics*{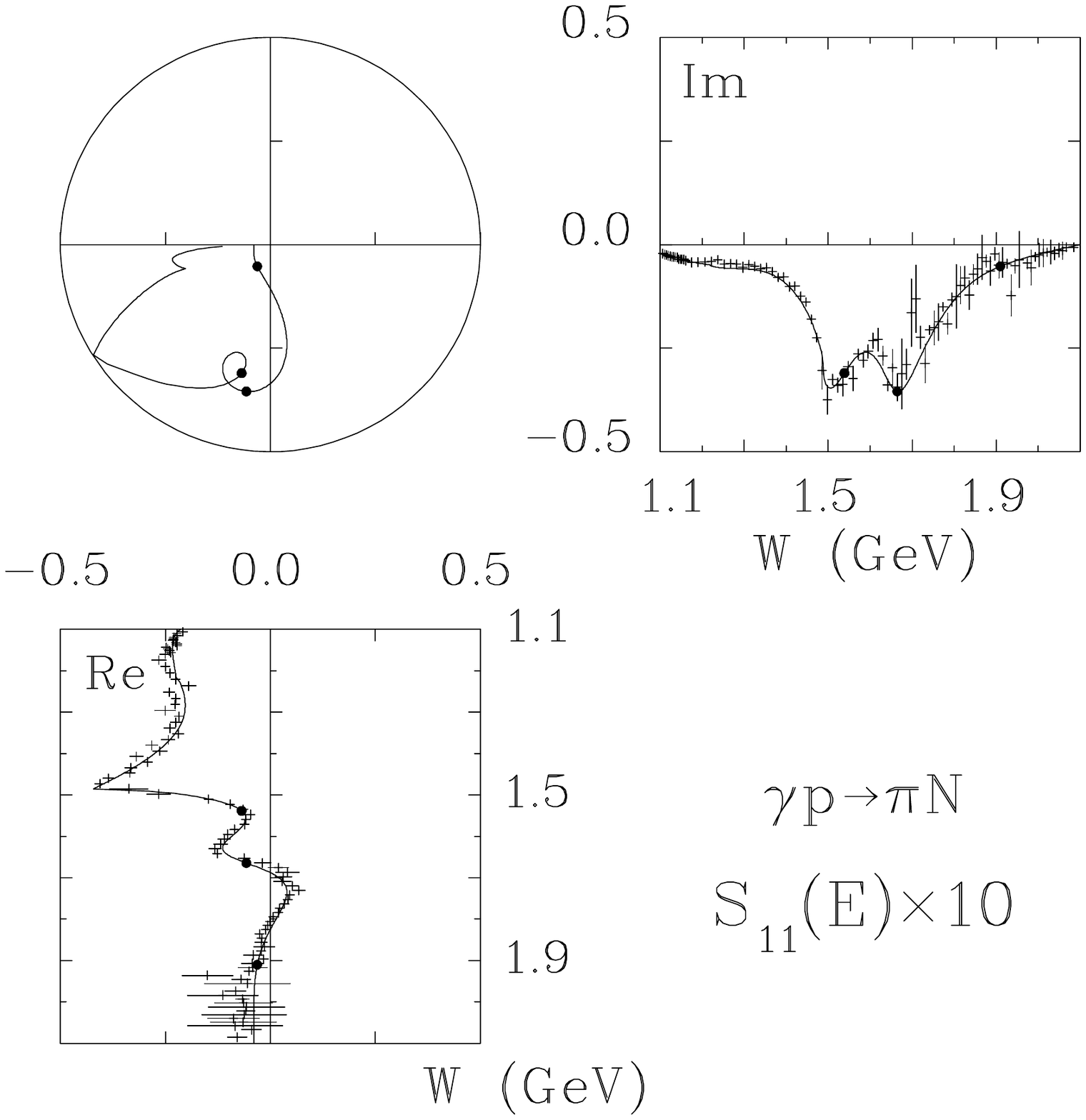}}
\vspace{-25mm}
\scalebox{0.35}{\includegraphics*{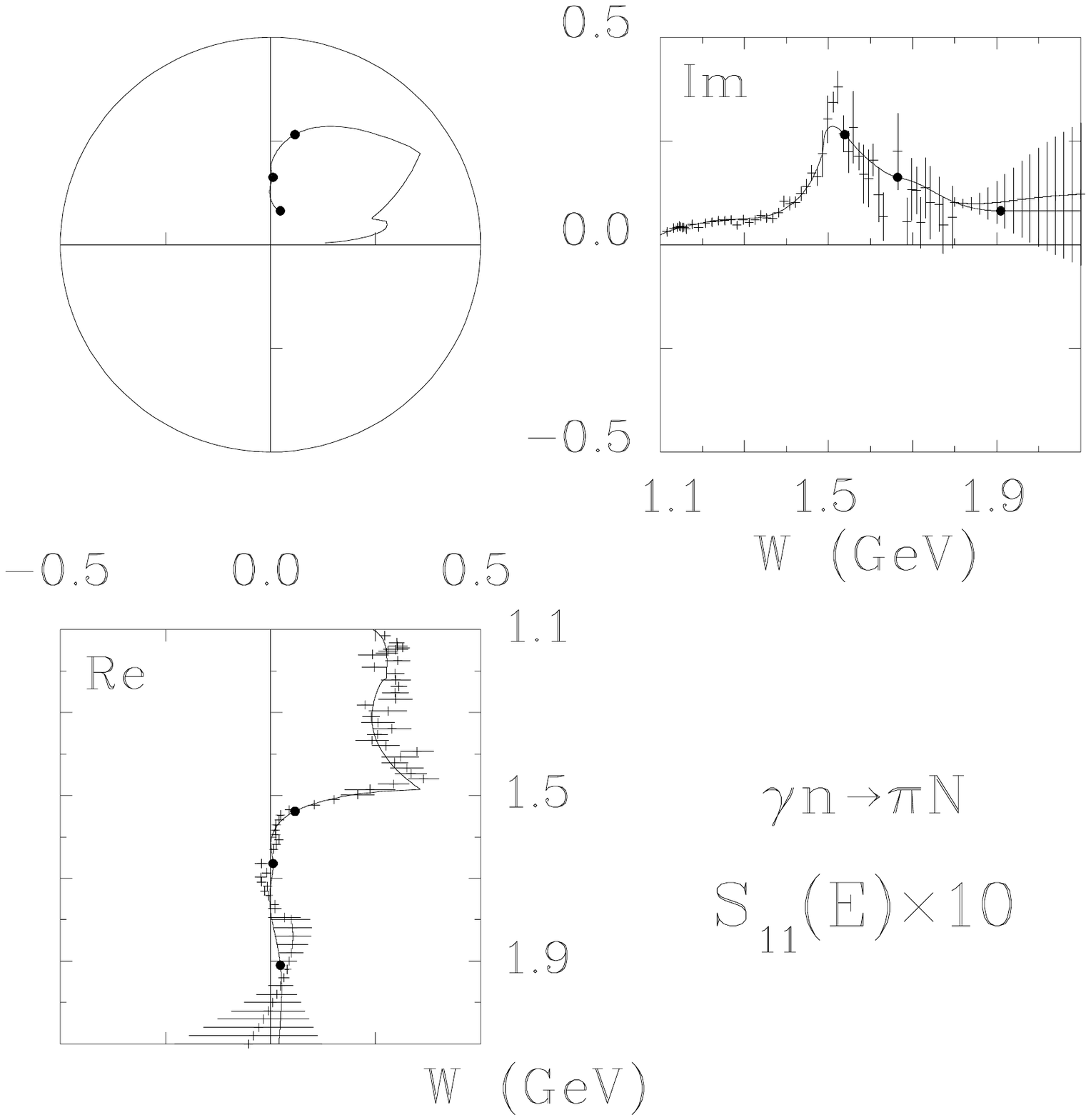}}
\scalebox{0.35}{\includegraphics*{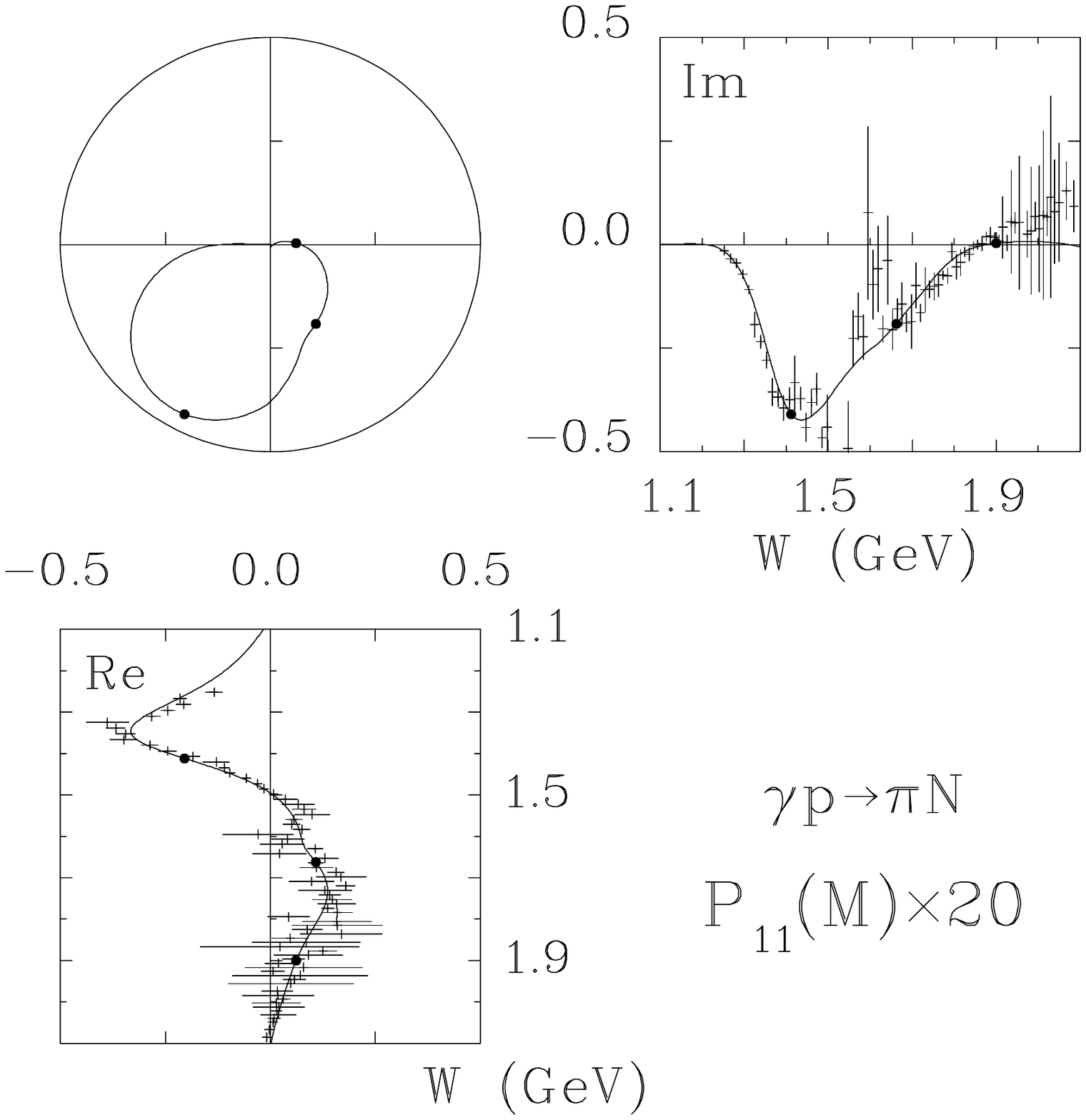}}
\vspace{-25mm}
\scalebox{0.35}{\includegraphics*{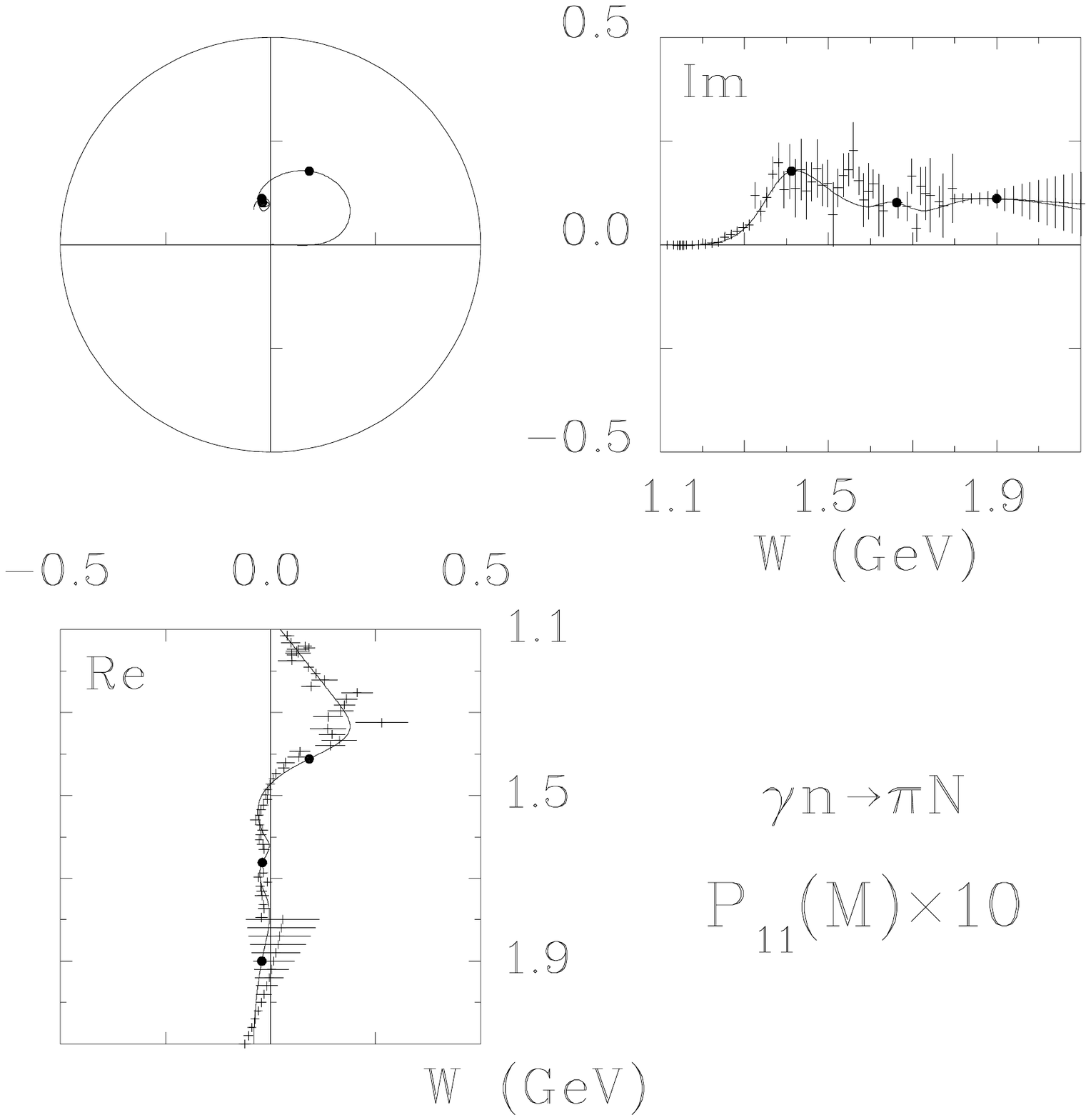}}
\vspace{-15mm}
\scalebox{0.35}{\includegraphics*{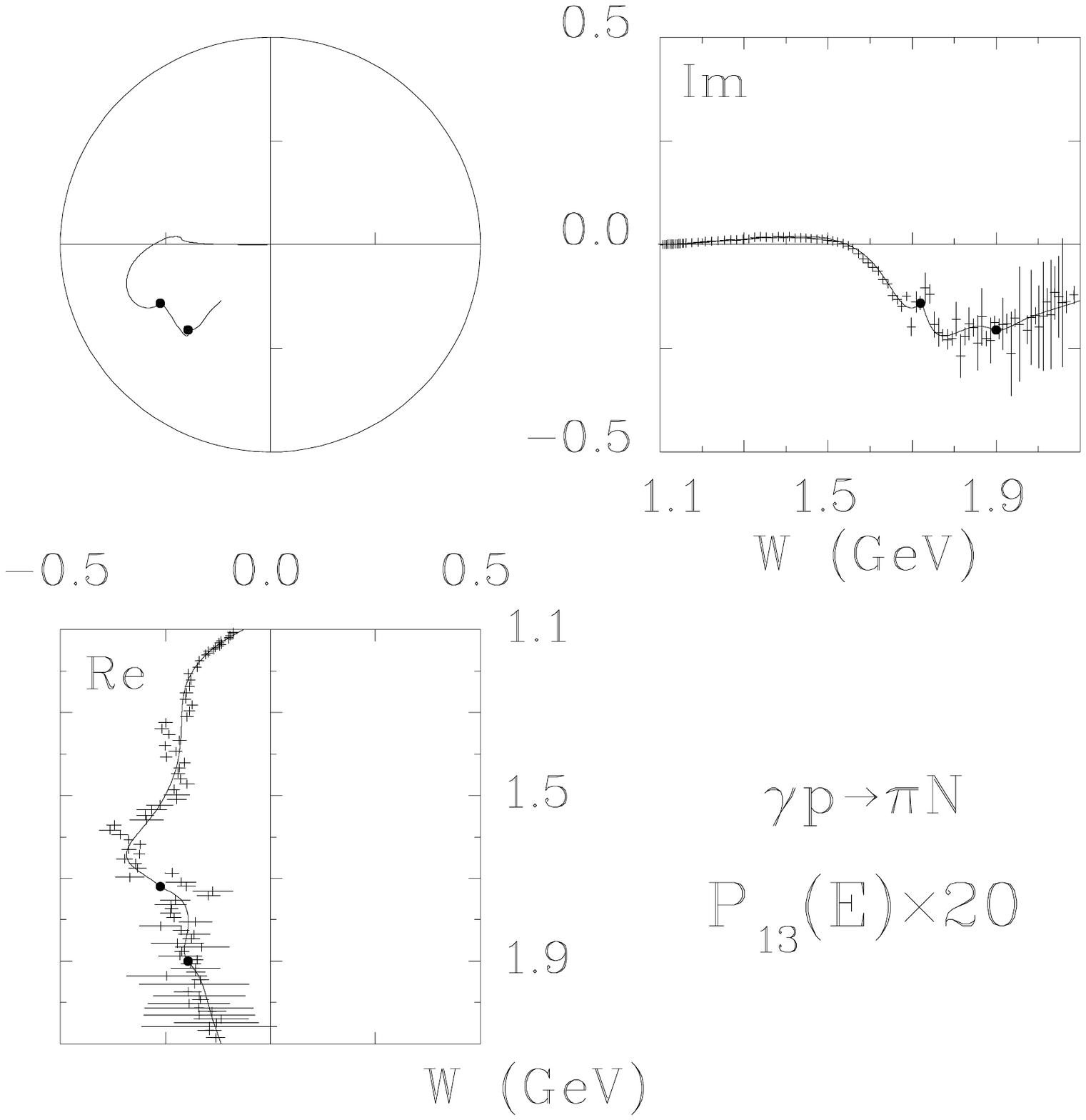}}
\scalebox{0.35}{\includegraphics*{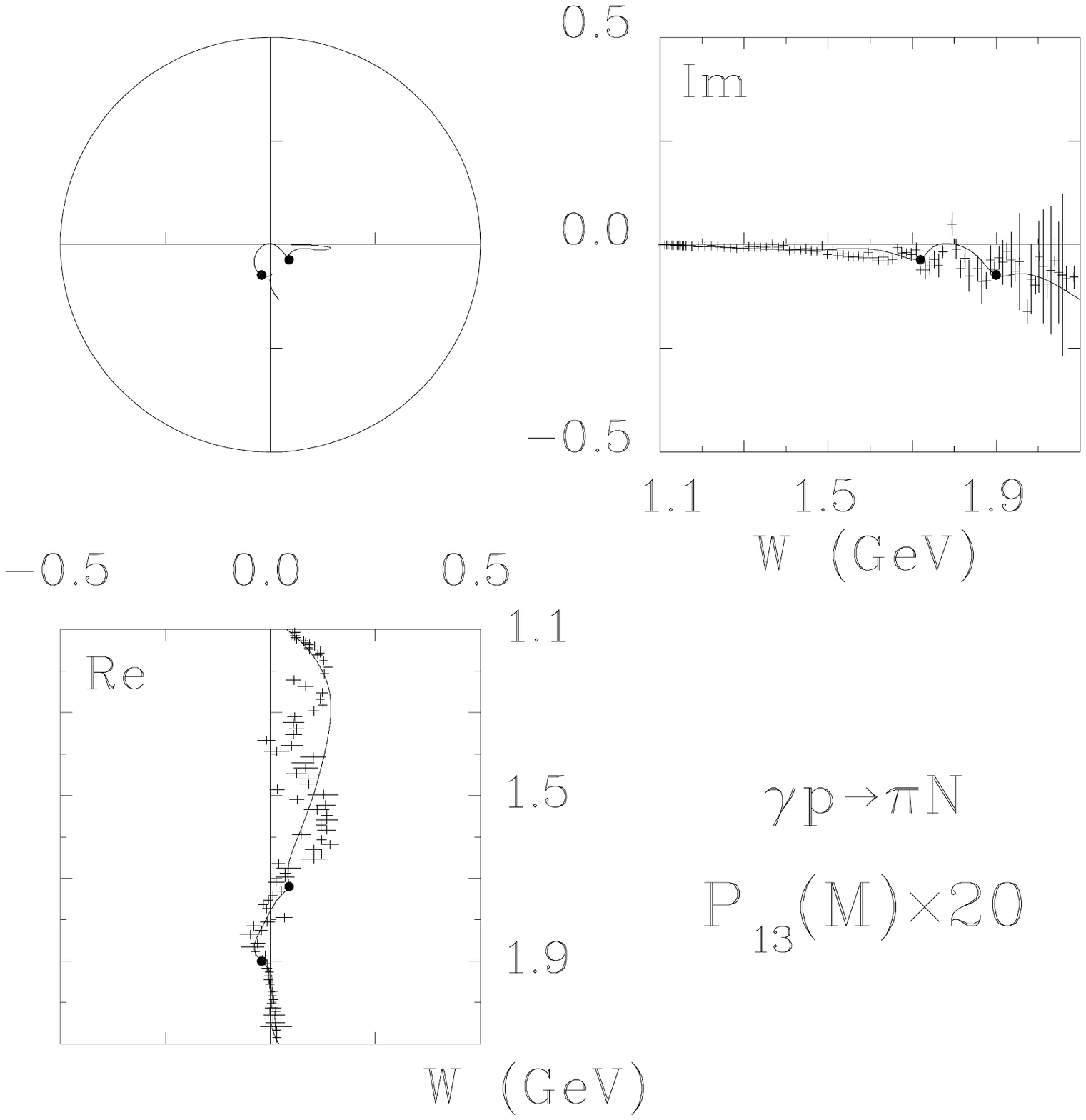}}
\caption{Argand diagrams for pion-photoproduction amplitudes.}
\end{figure*}

\begin{figure*}[htpb]
\addtocounter{figure}{-1}
%\caption{Cont'd.}
\vspace{-10mm}
\scalebox{0.35}{\includegraphics*{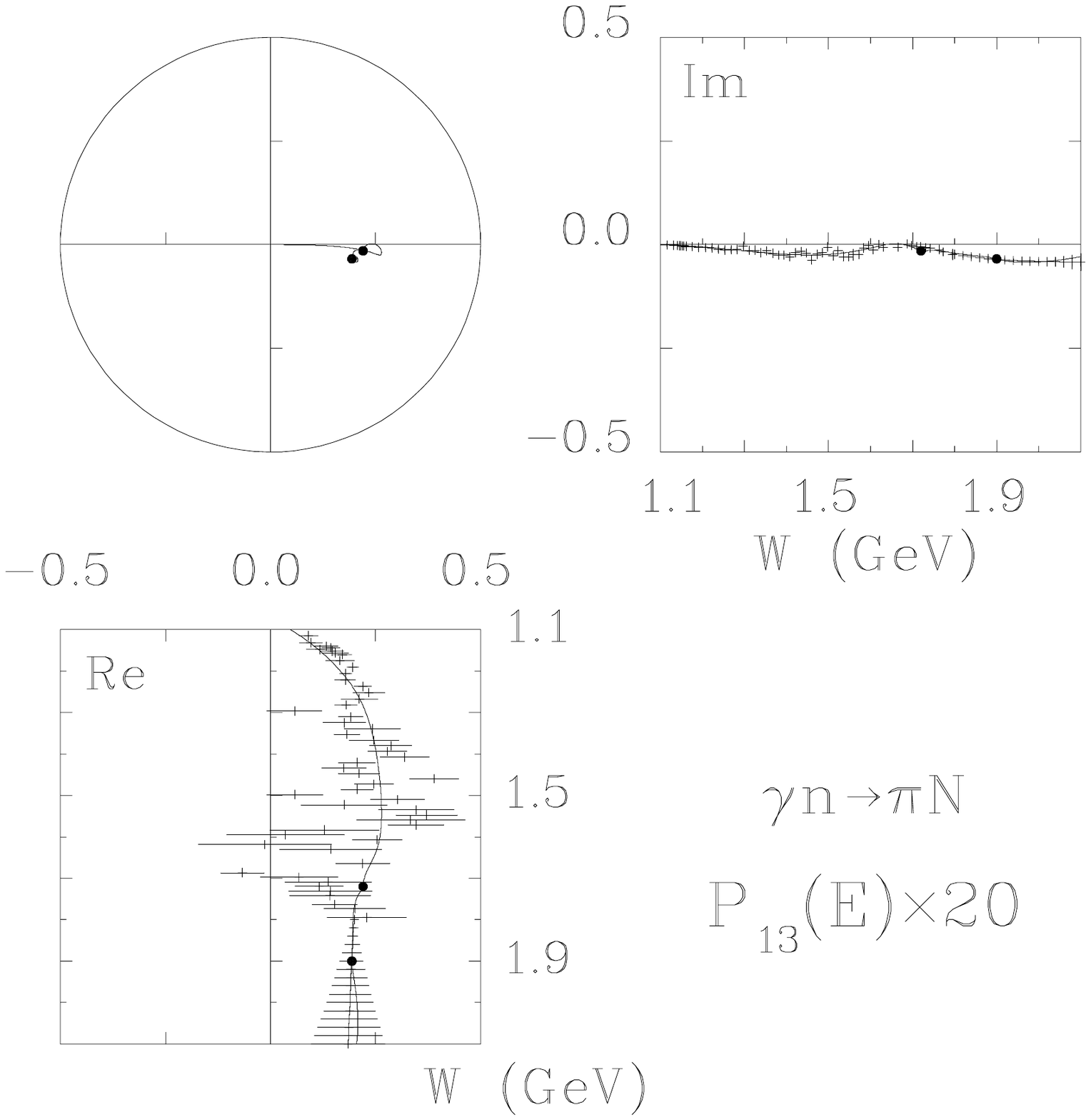}}
\vspace{-25mm}
\scalebox{0.35}{\includegraphics*{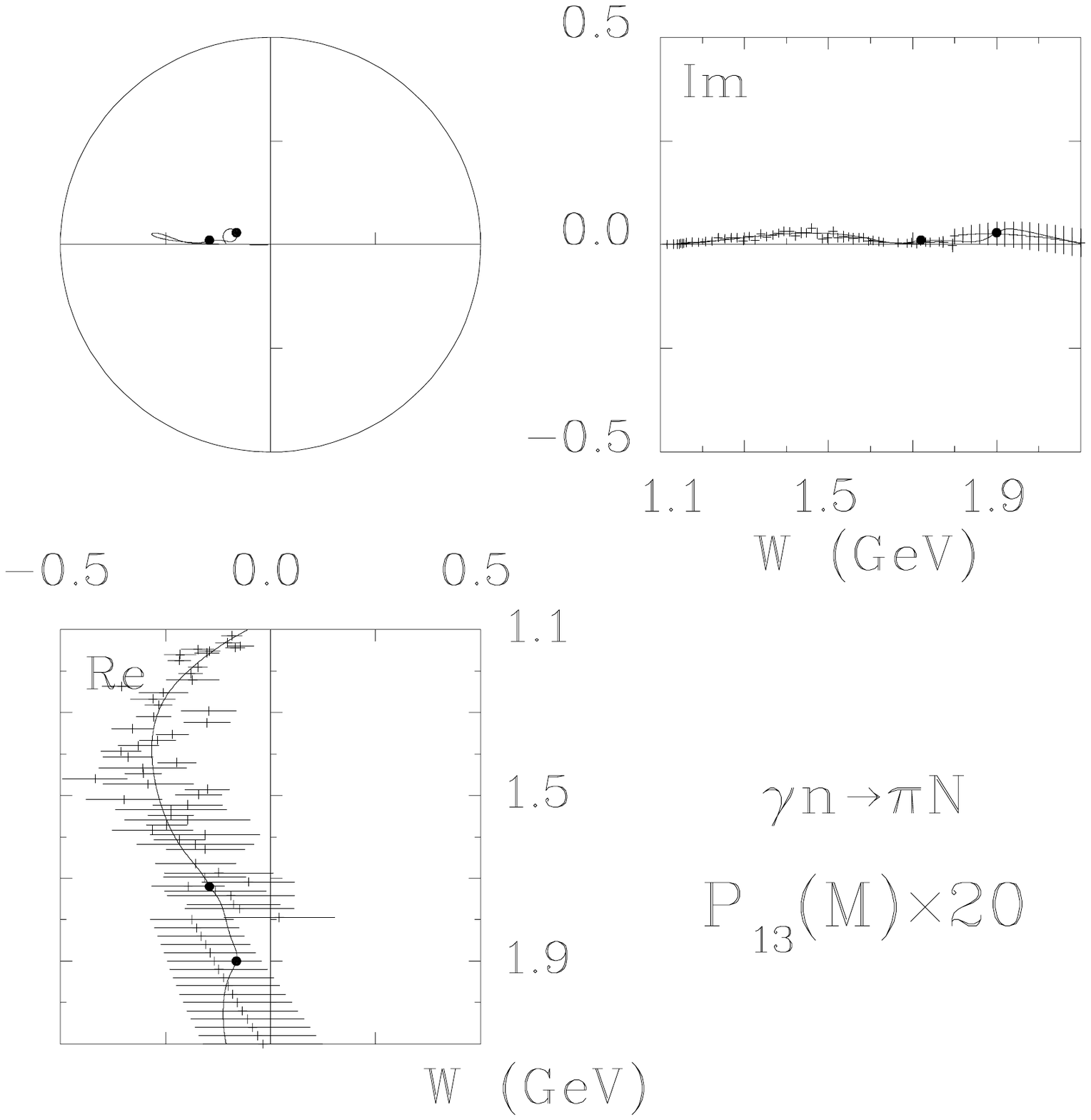}}
%\vspace{15mm}
\scalebox{0.35}{\includegraphics*{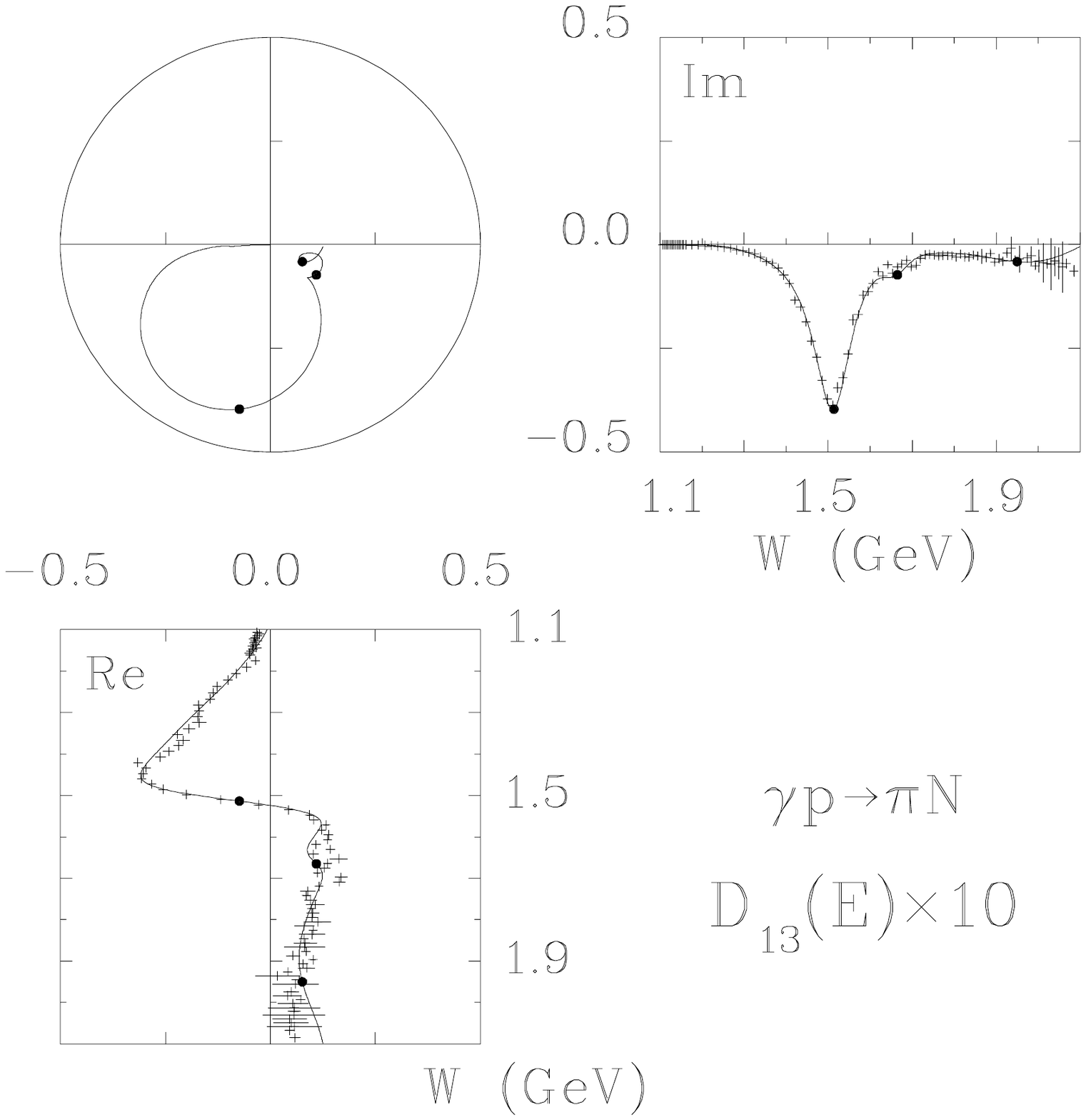}}
\vspace{-25mm}
\scalebox{0.35}{\includegraphics*{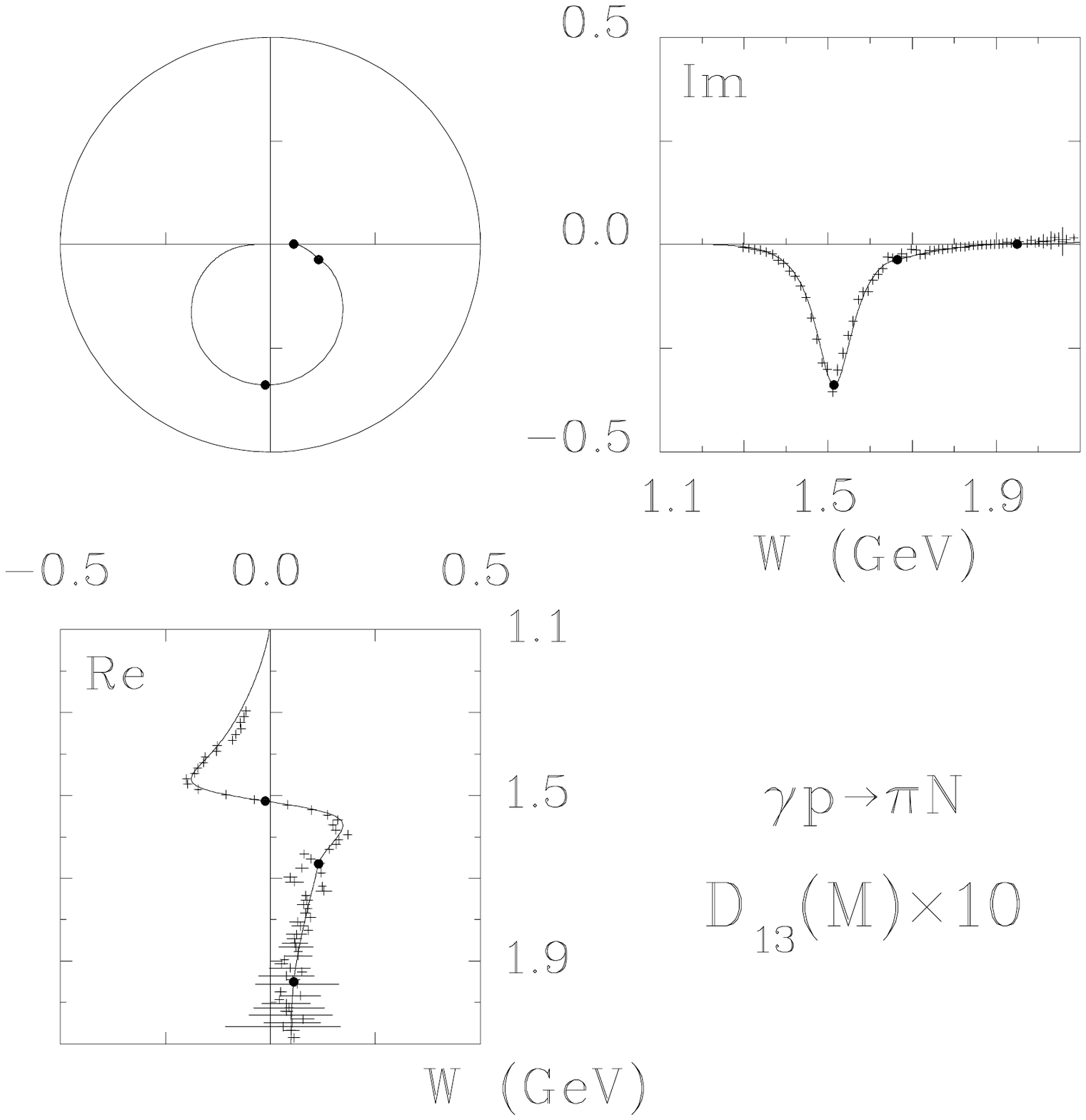}}
\vspace{-15mm}
\scalebox{0.35}{\includegraphics*{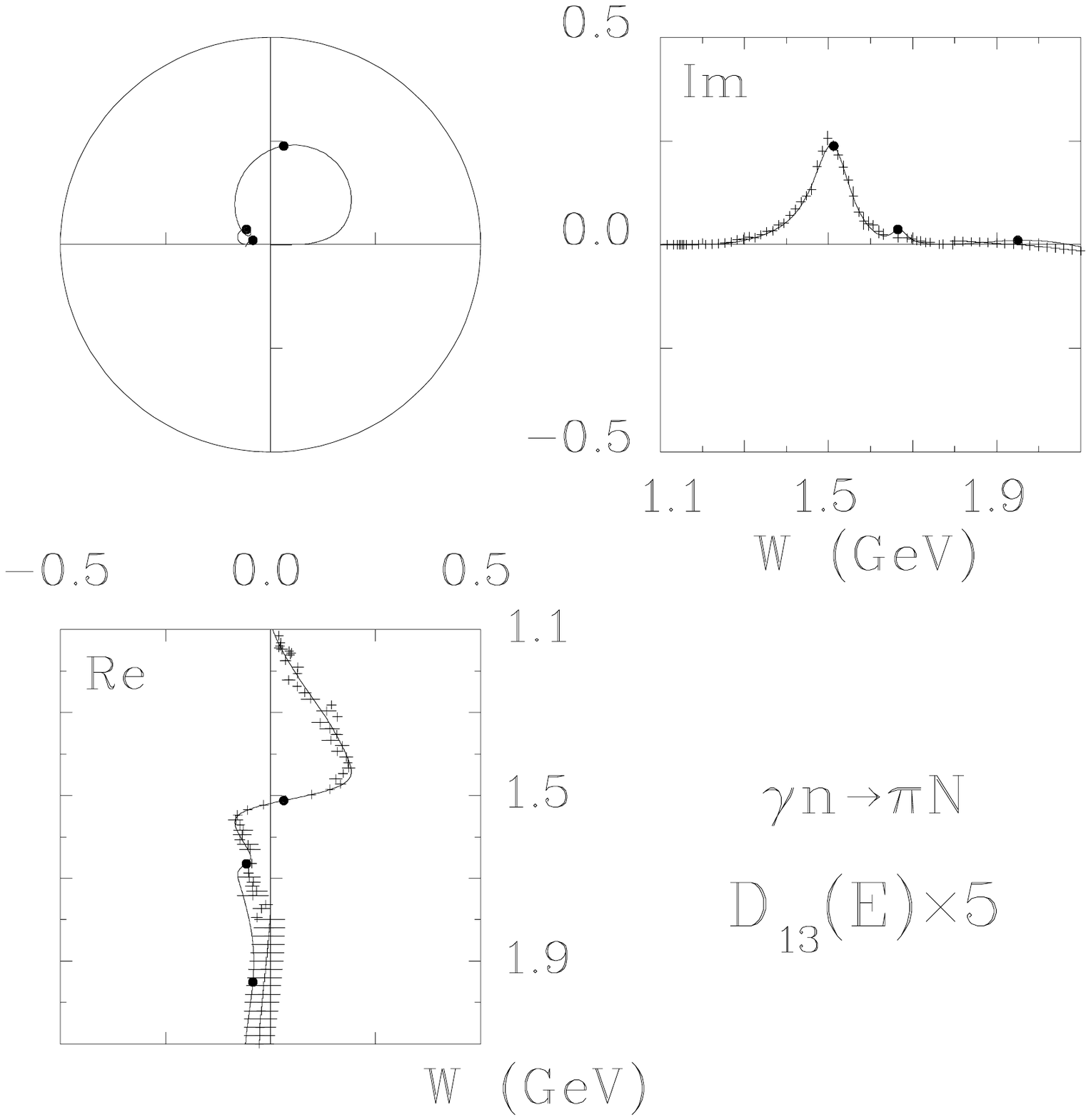}}
\scalebox{0.35}{\includegraphics*{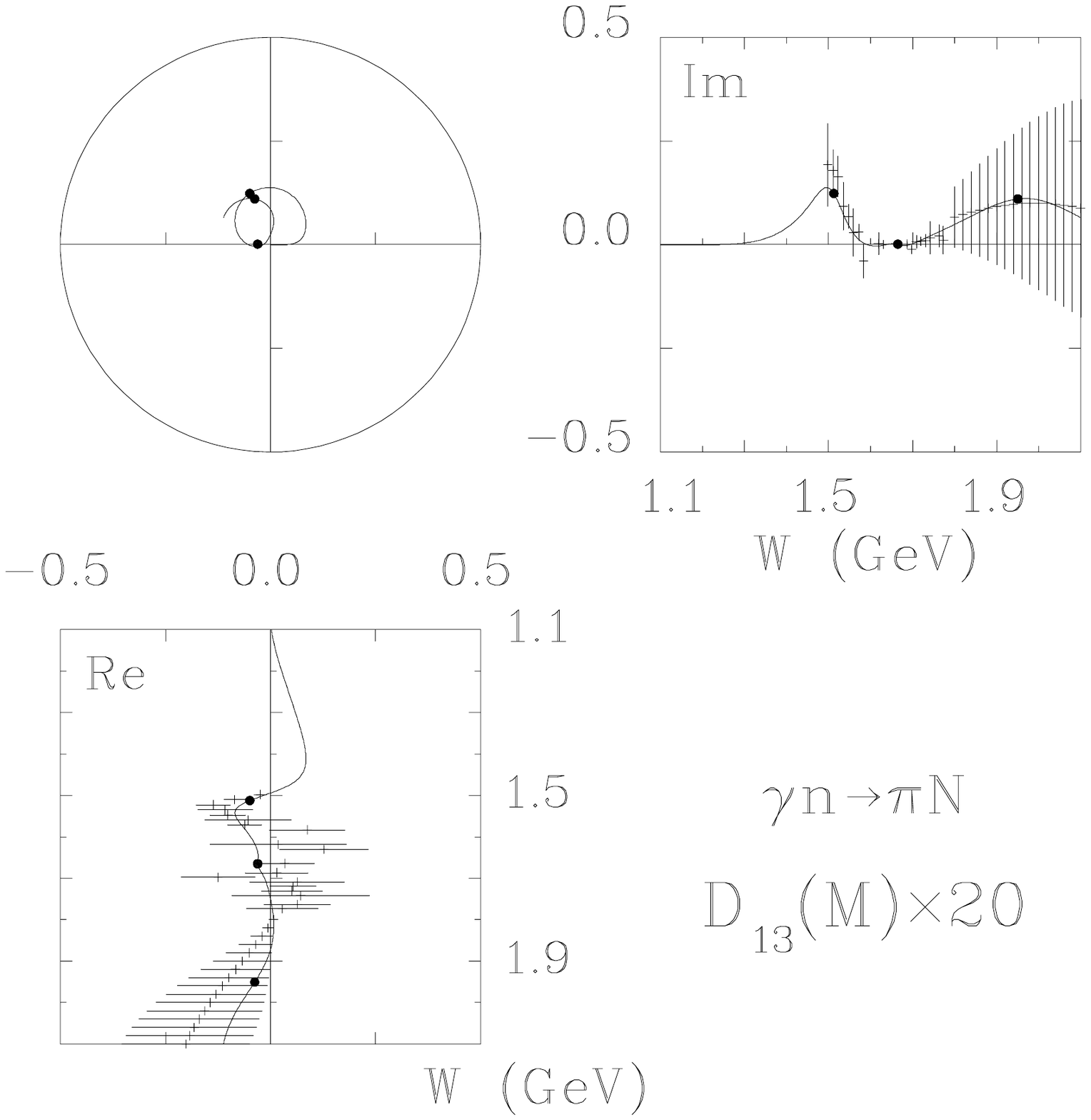}}
\caption{Cont'd.}
\end{figure*}

% %\newpage 
\begin{figure*}[htpb]
\addtocounter{figure}{-1}
%\caption{Cont'd.}
\vspace{-10mm}
\scalebox{0.35}{\includegraphics*{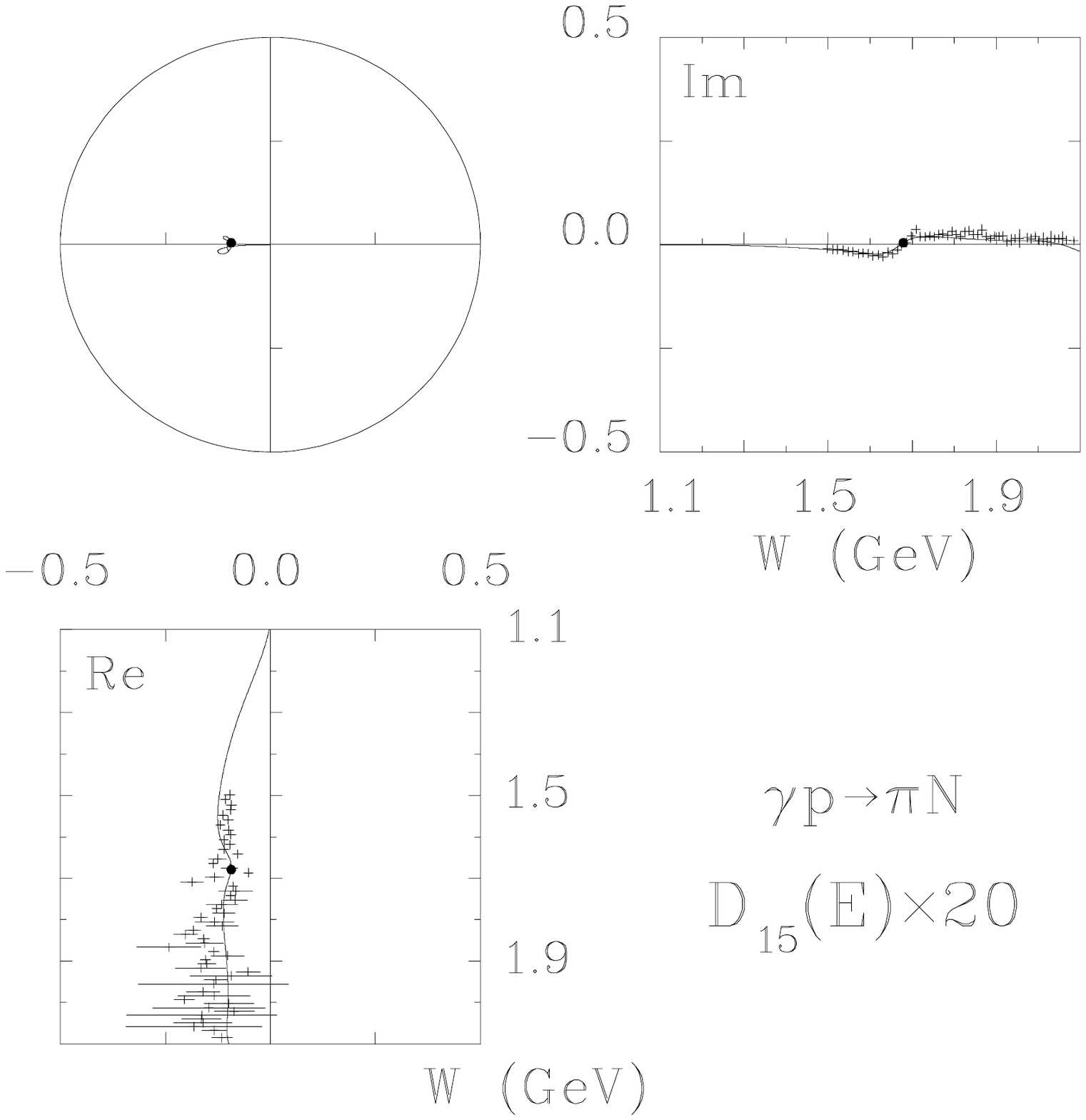}}
\vspace{-25mm}
\scalebox{0.35}{\includegraphics*{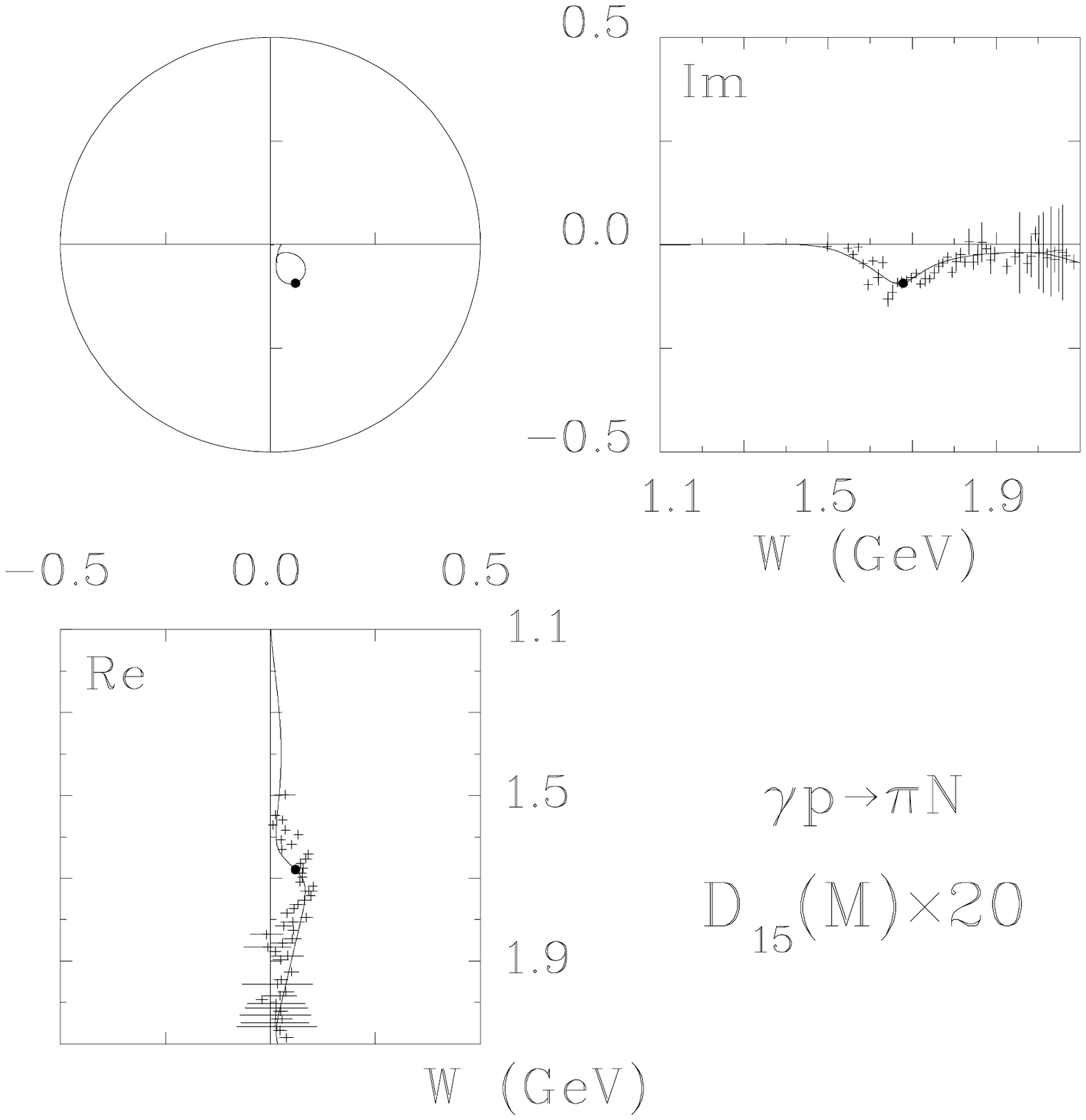}}
%\vspace{15mm}
\scalebox{0.35}{\includegraphics*{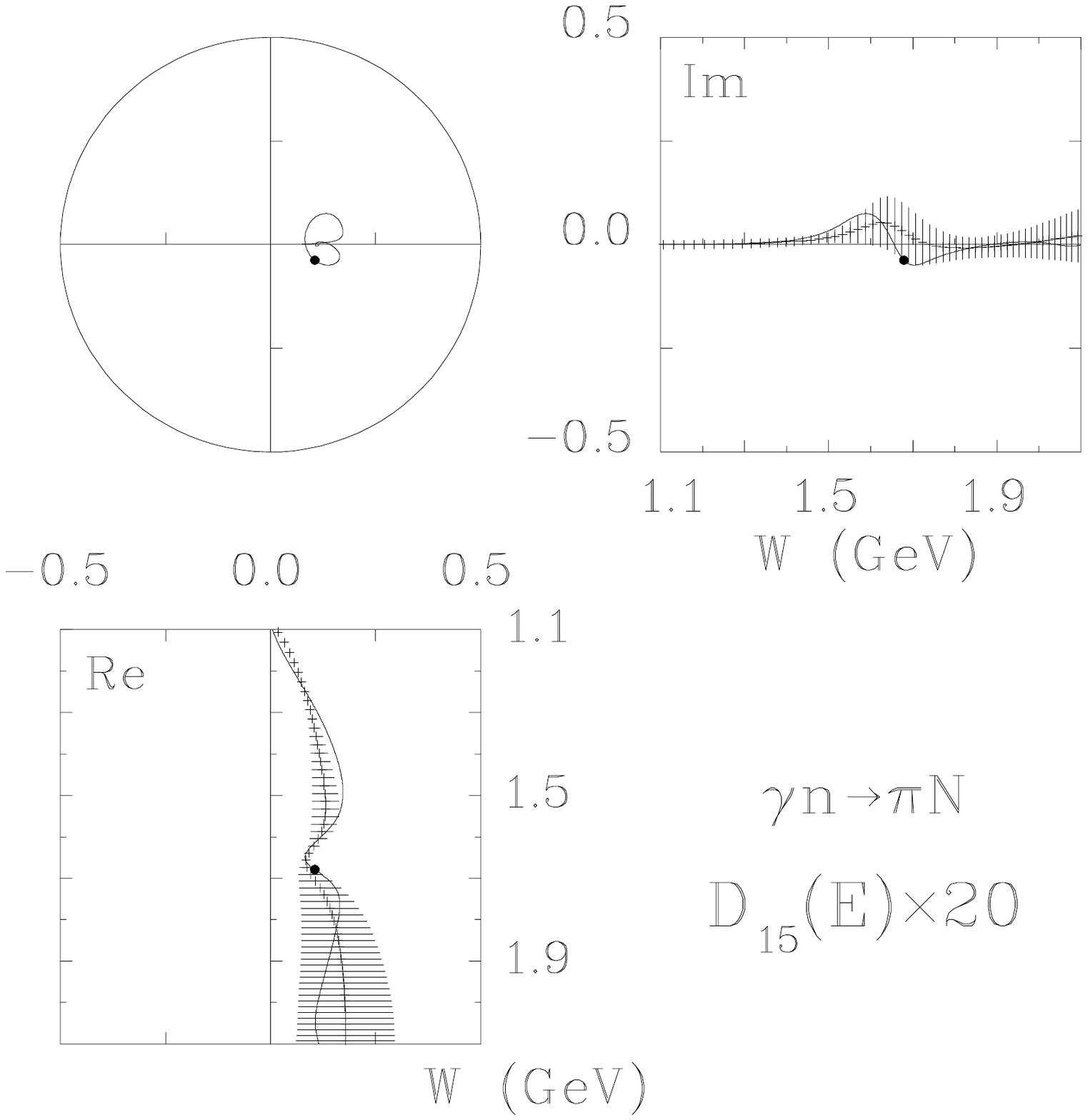}}
\vspace{-25mm}
\scalebox{0.35}{\includegraphics*{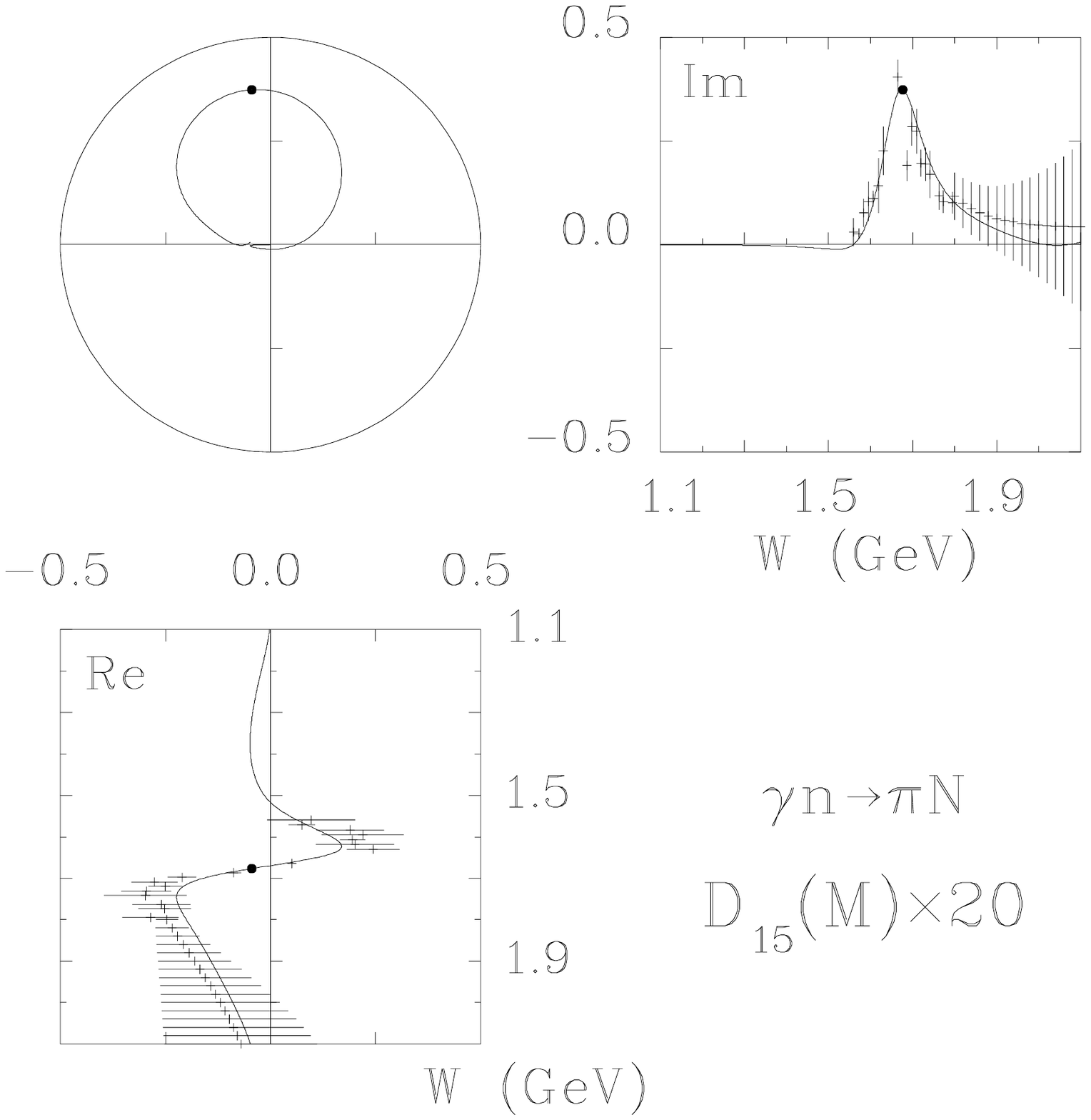}}
\vspace{-15mm}
\scalebox{0.35}{\includegraphics*{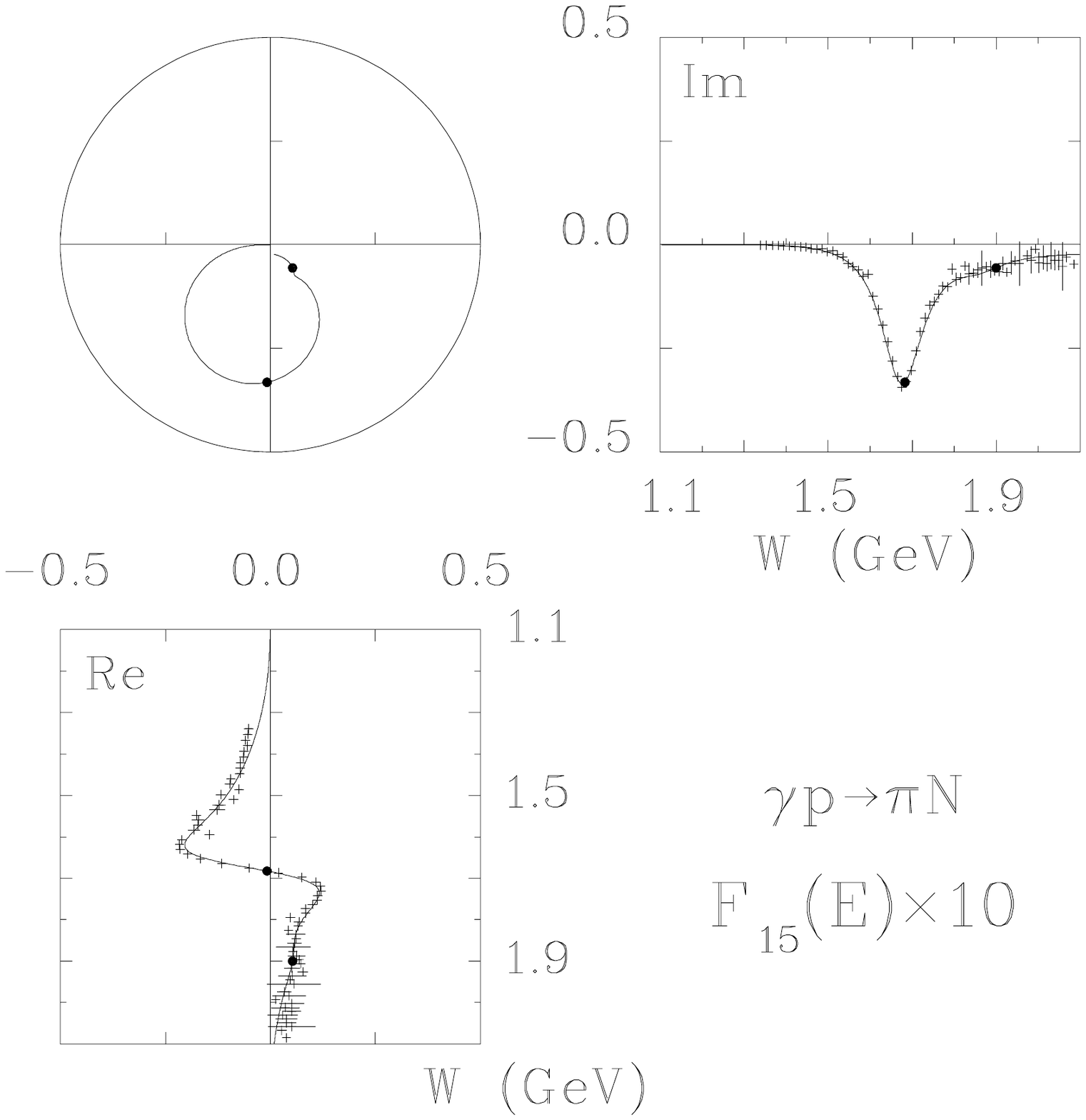}}
\scalebox{0.35}{\includegraphics*{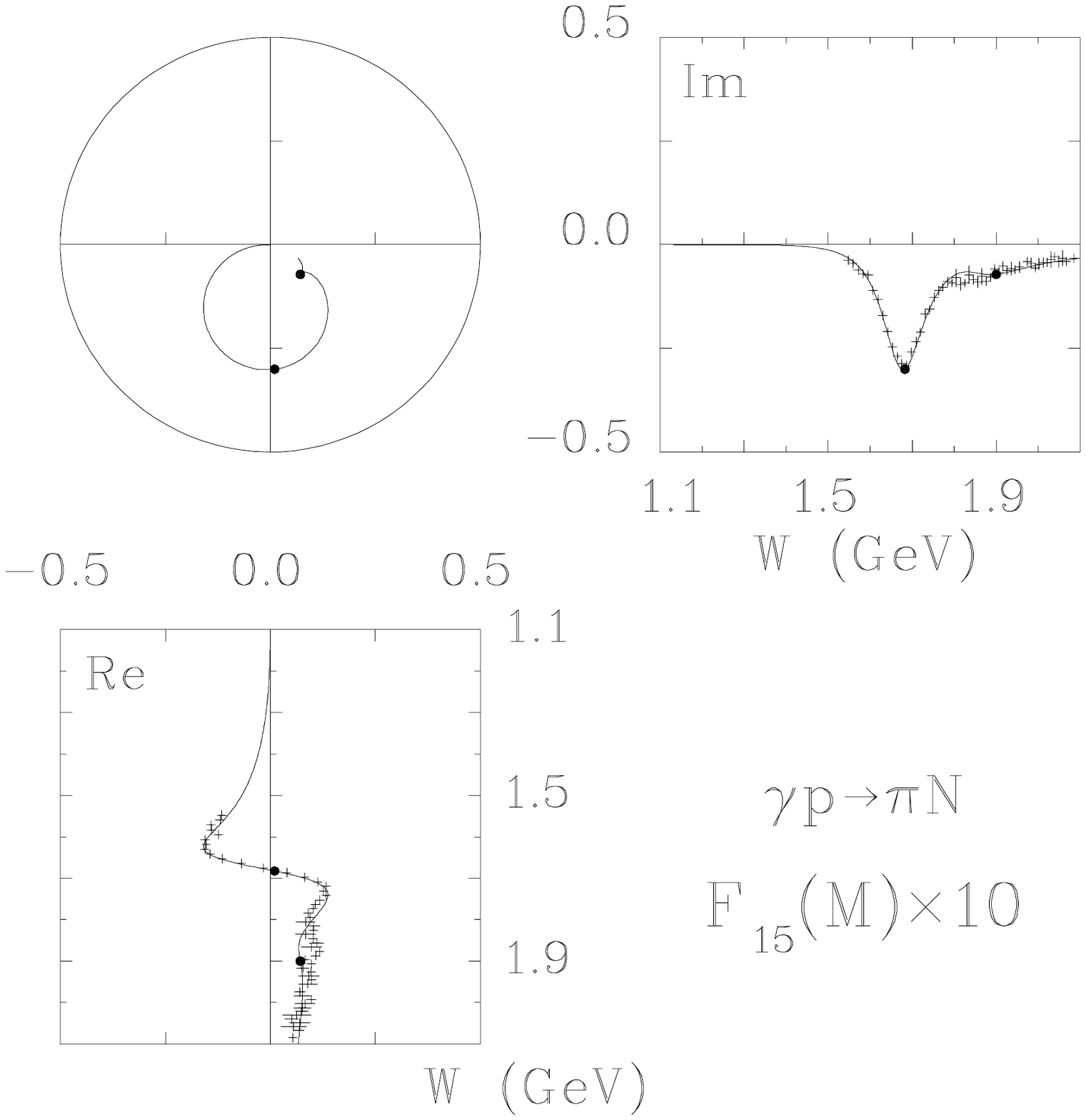}}
\caption{Cont'd.}
\end{figure*}

\begin{figure*}[htpb]
\addtocounter{figure}{-1}
%\caption{Cont'd.}
\vspace{-10mm}
\scalebox{0.35}{\includegraphics*{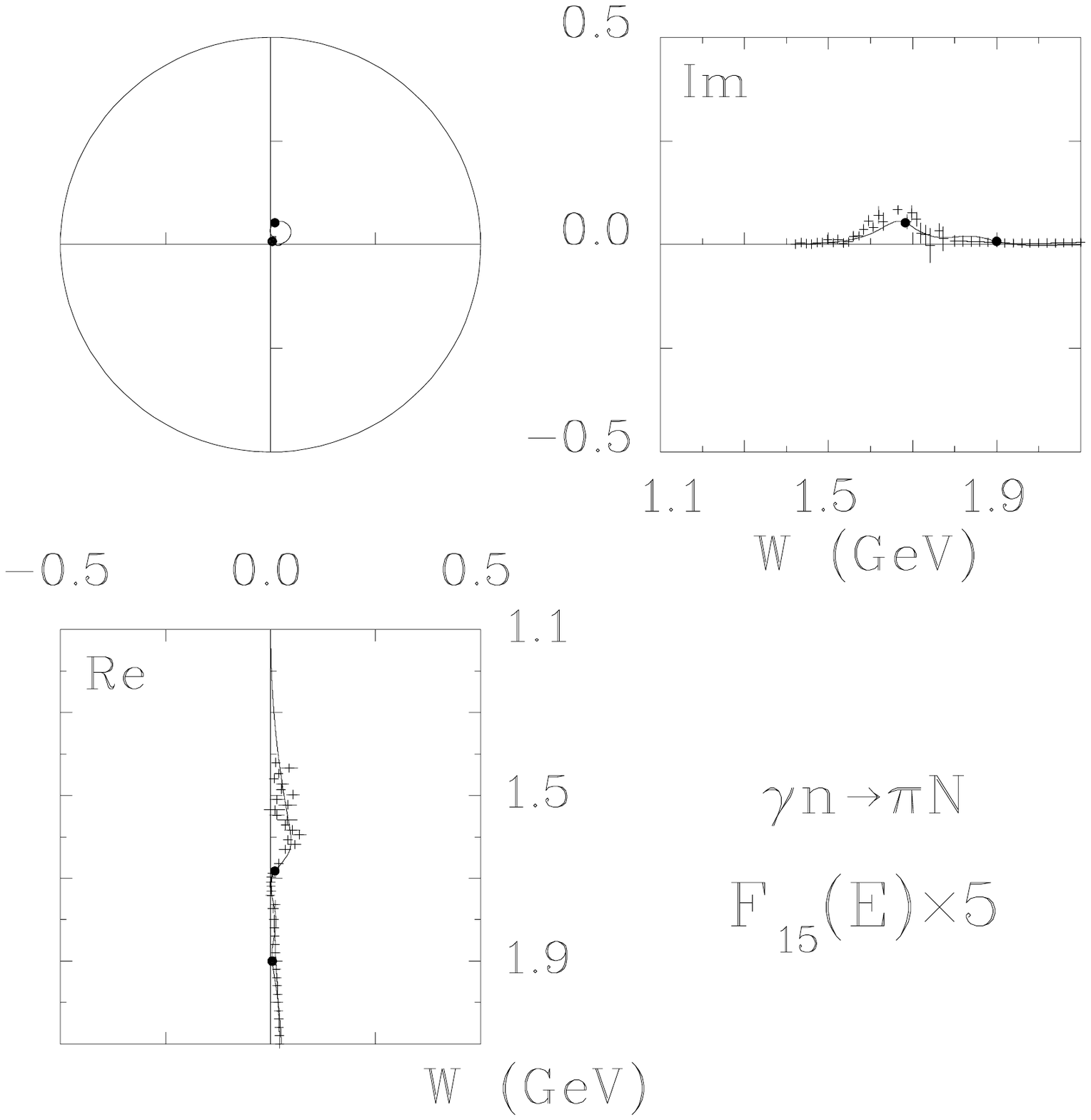}}
\vspace{-25mm}
\scalebox{0.35}{\includegraphics*{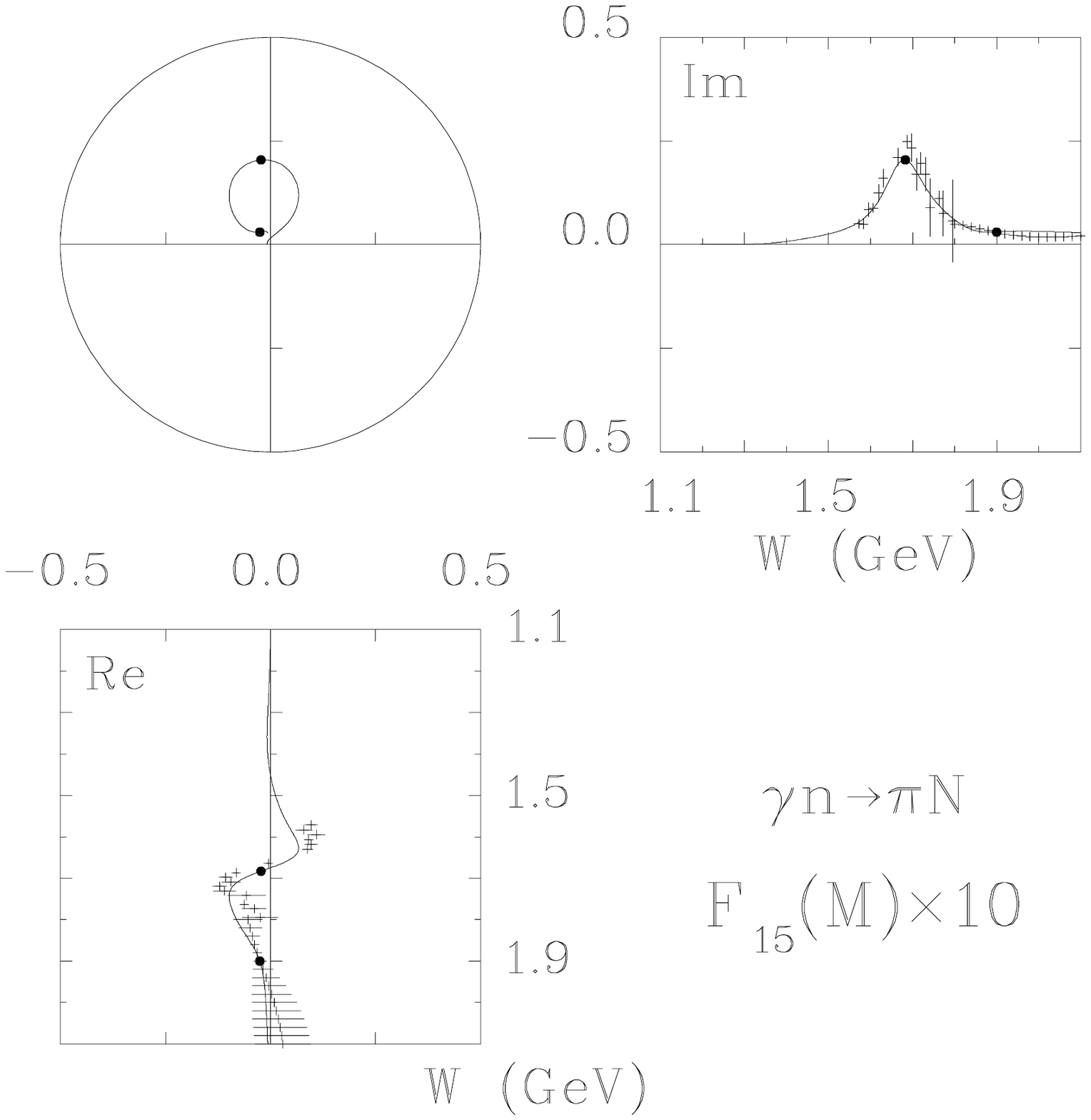}}
%\vspace{15mm}
\scalebox{0.35}{\includegraphics*{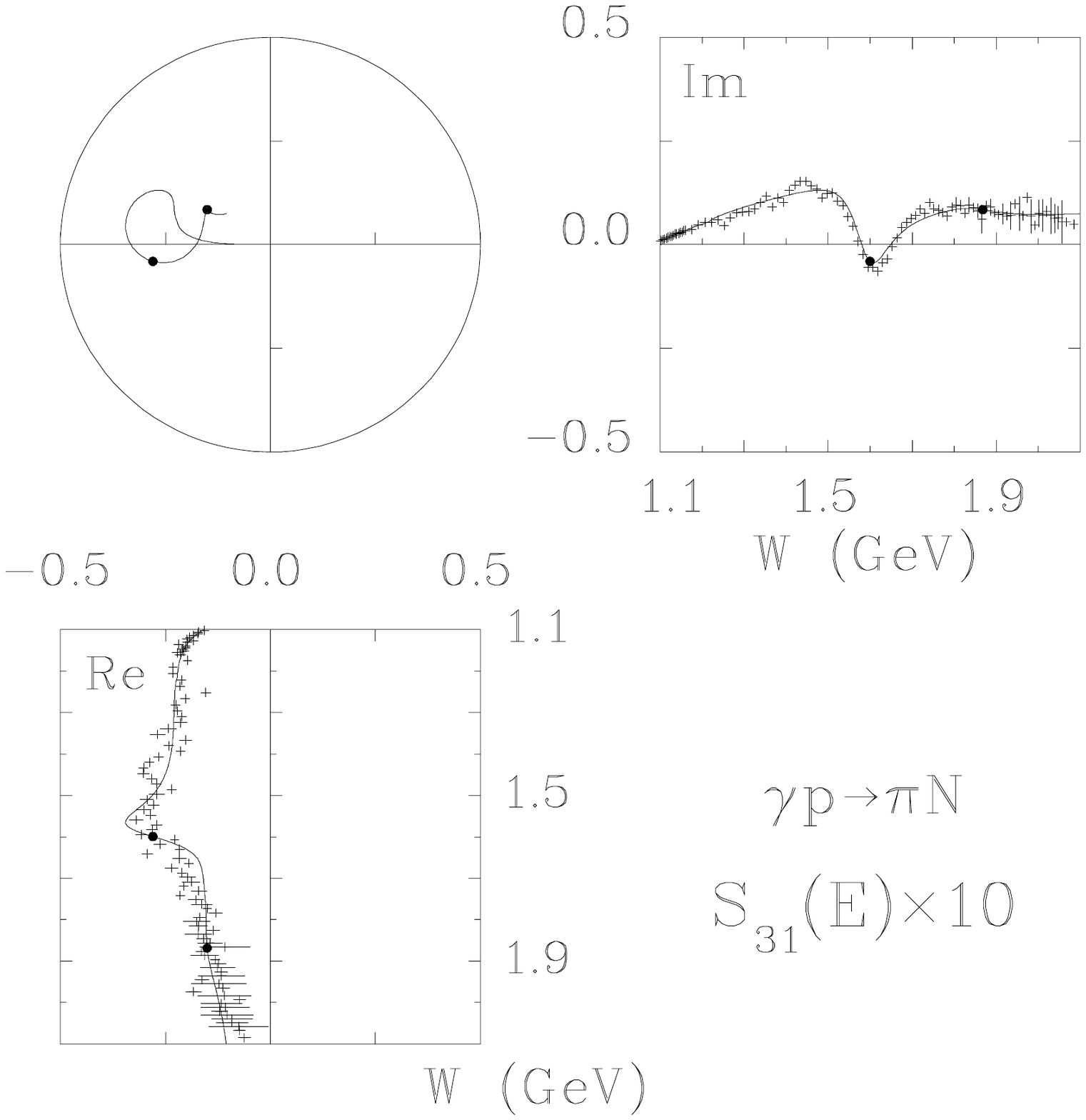}}
\vspace{-25mm}
\scalebox{0.35}{\includegraphics*{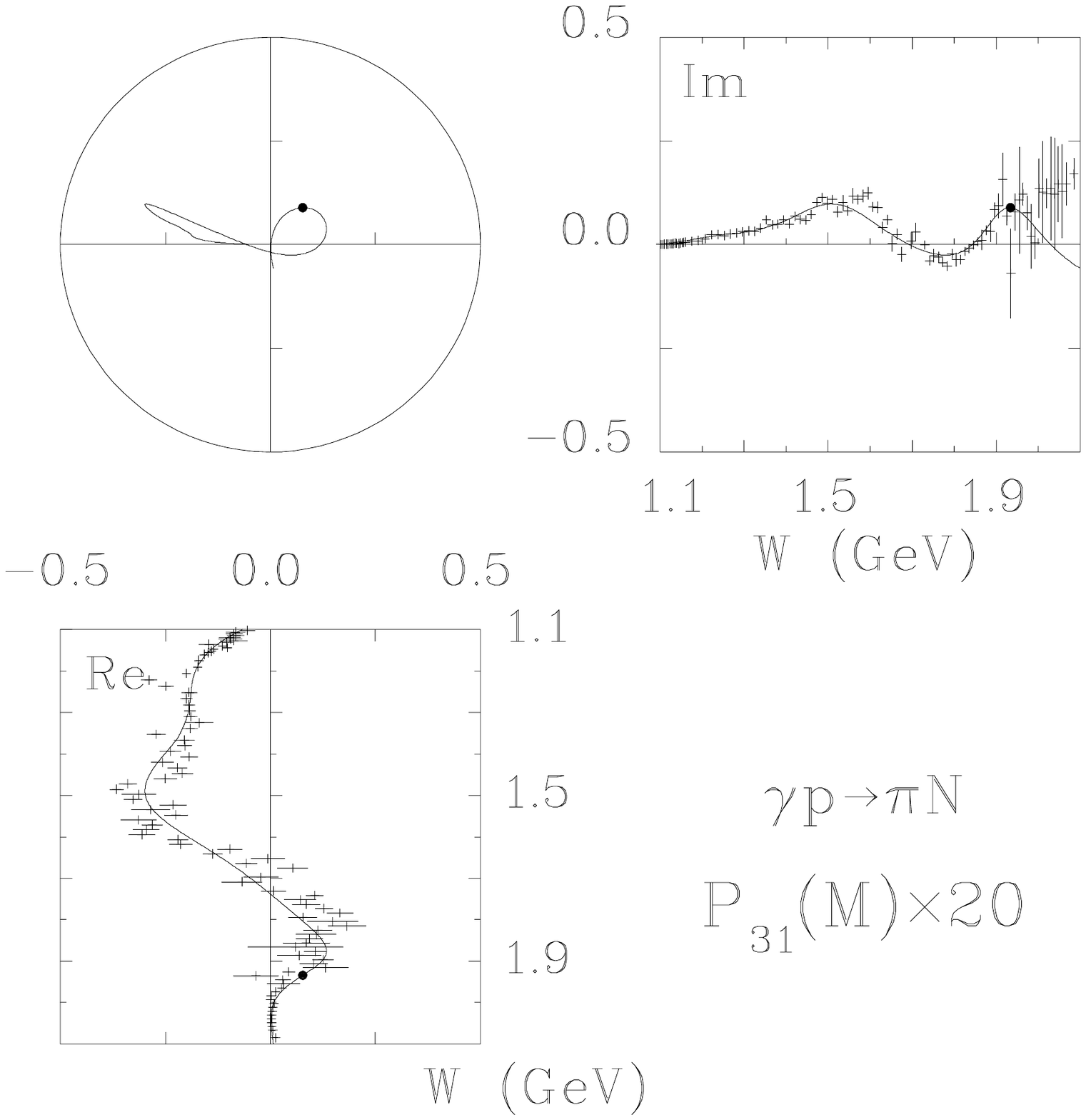}}
\vspace{-15mm}
\scalebox{0.35}{\includegraphics*{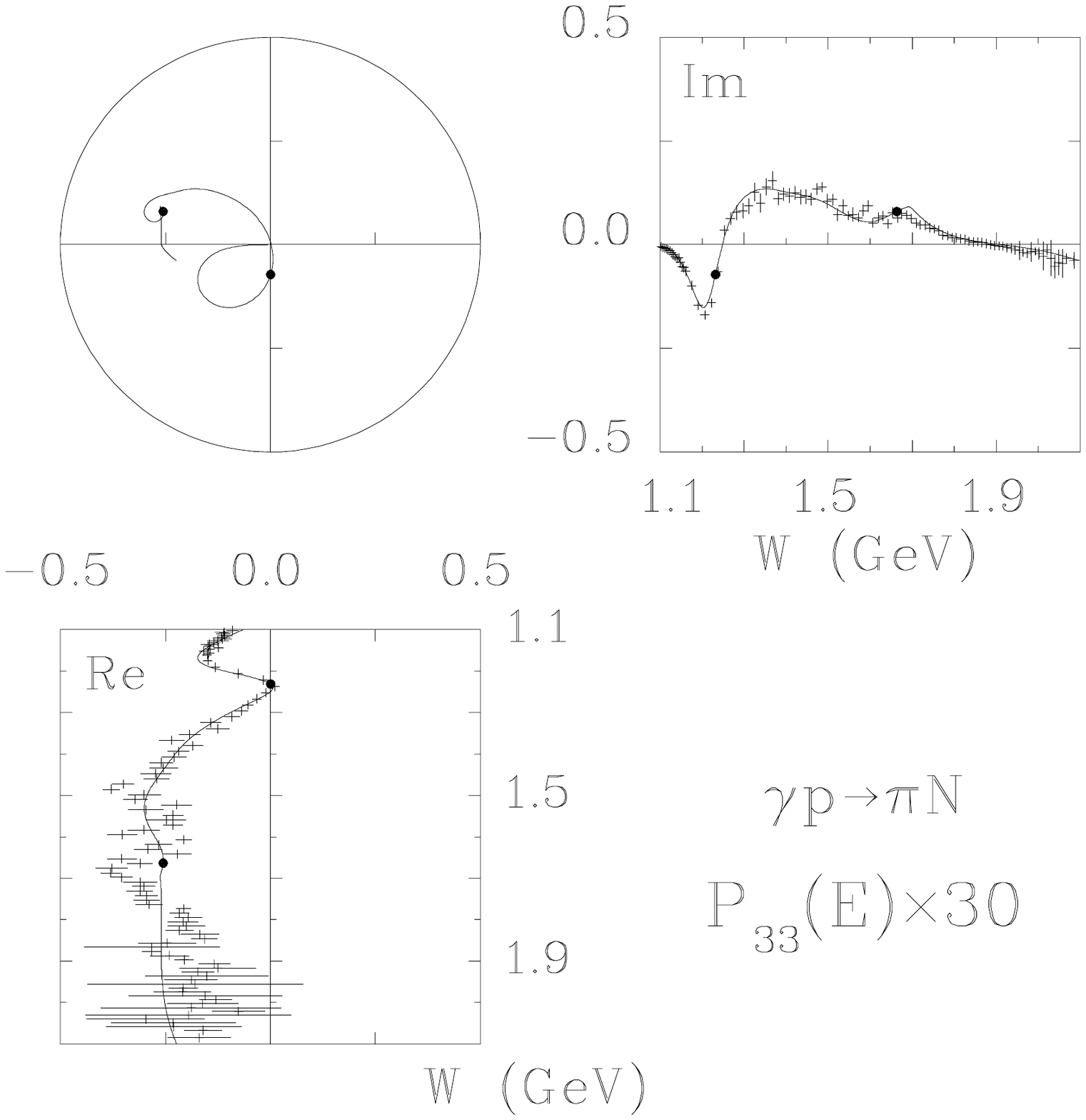}}
\scalebox{0.35}{\includegraphics*{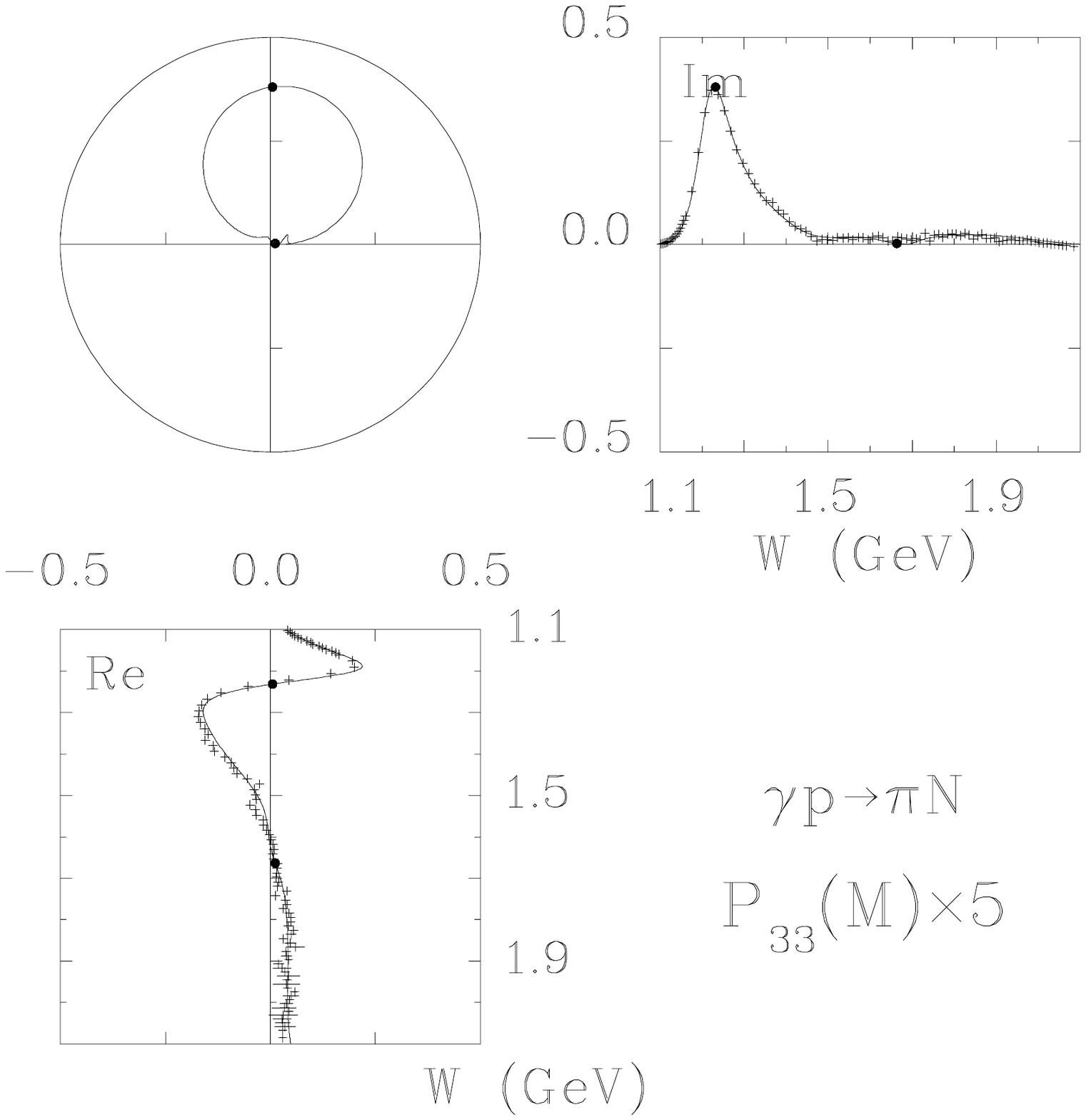}}
\caption{Cont'd.}
\end{figure*}

\begin{figure*}[htpb]
\addtocounter{figure}{-1}
%\caption{Cont'd.}
\vspace{-10mm}
\scalebox{0.35}{\includegraphics*{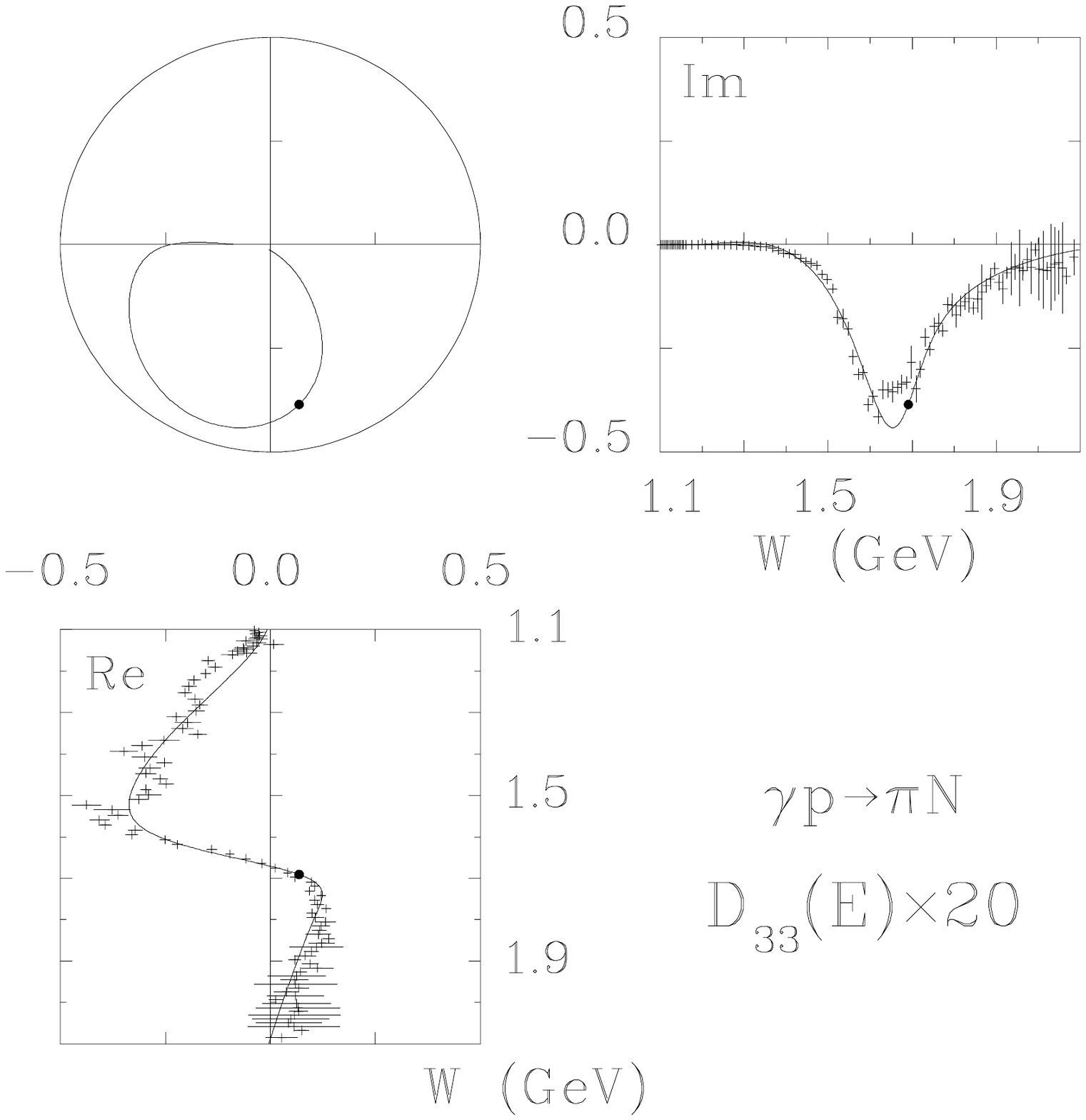}}
\vspace{-25mm}
\scalebox{0.35}{\includegraphics*{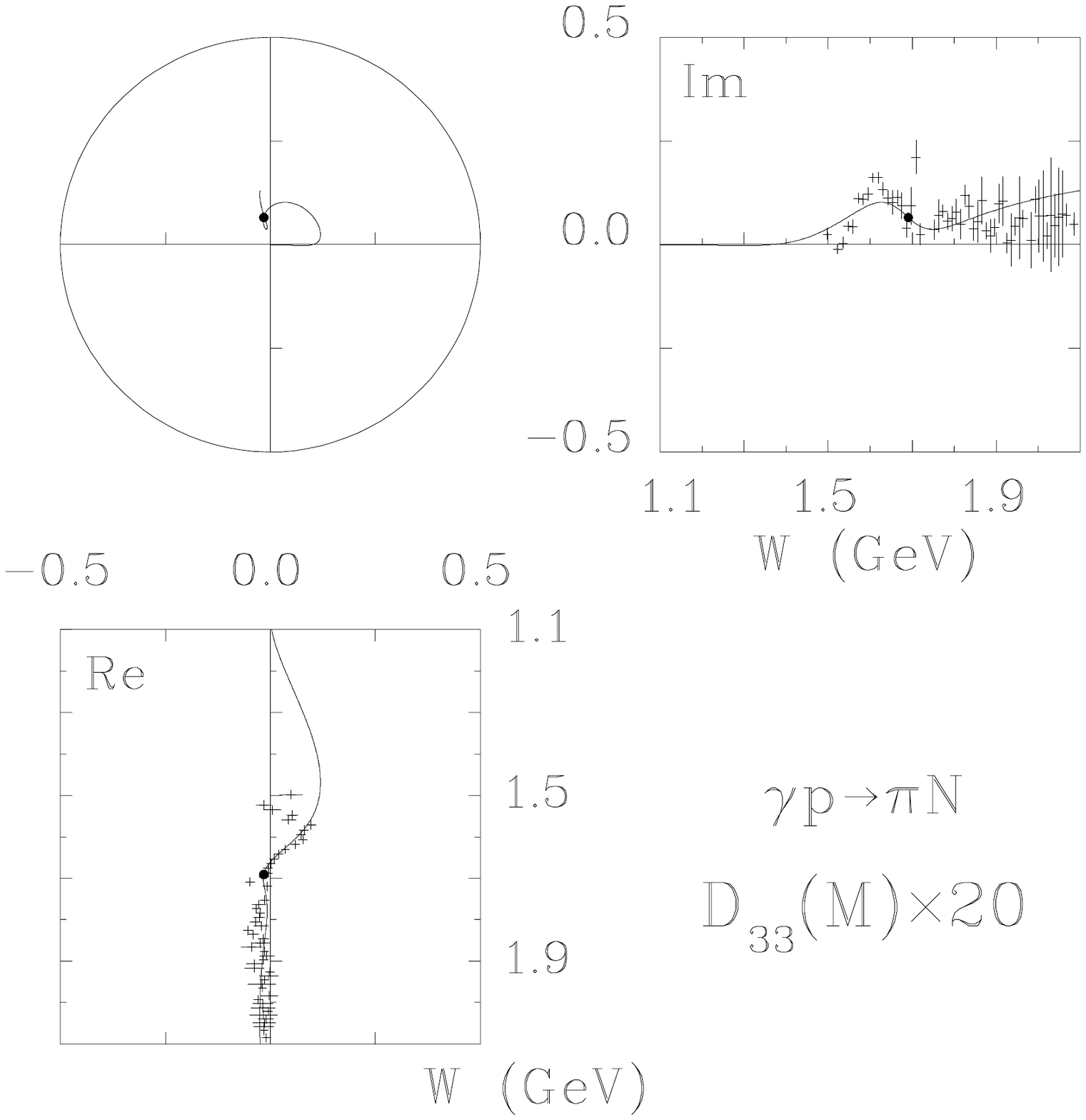}}
%\vspace{15mm}
\scalebox{0.35}{\includegraphics*{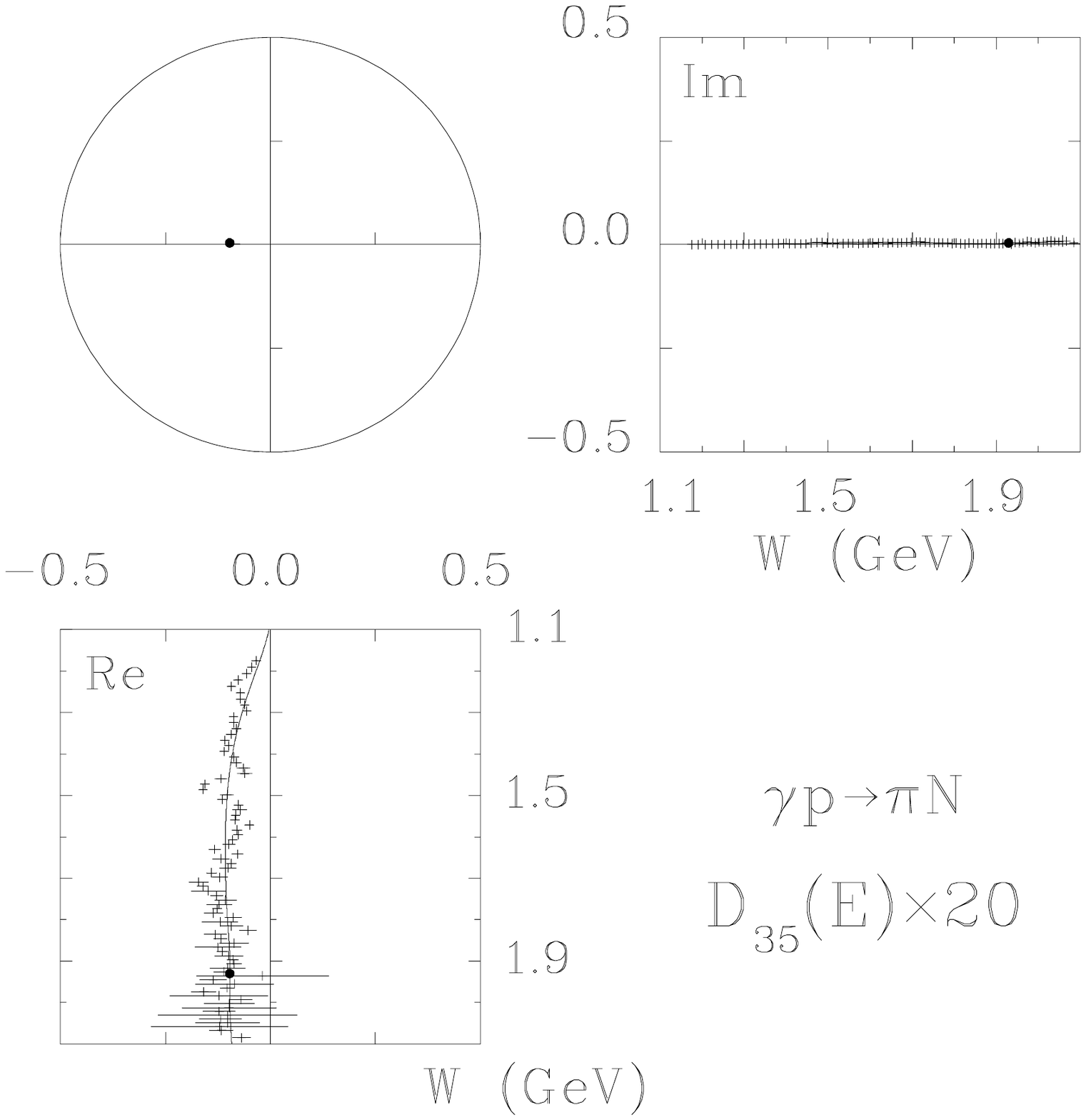}}
\vspace{-25mm}
\scalebox{0.35}{\includegraphics*{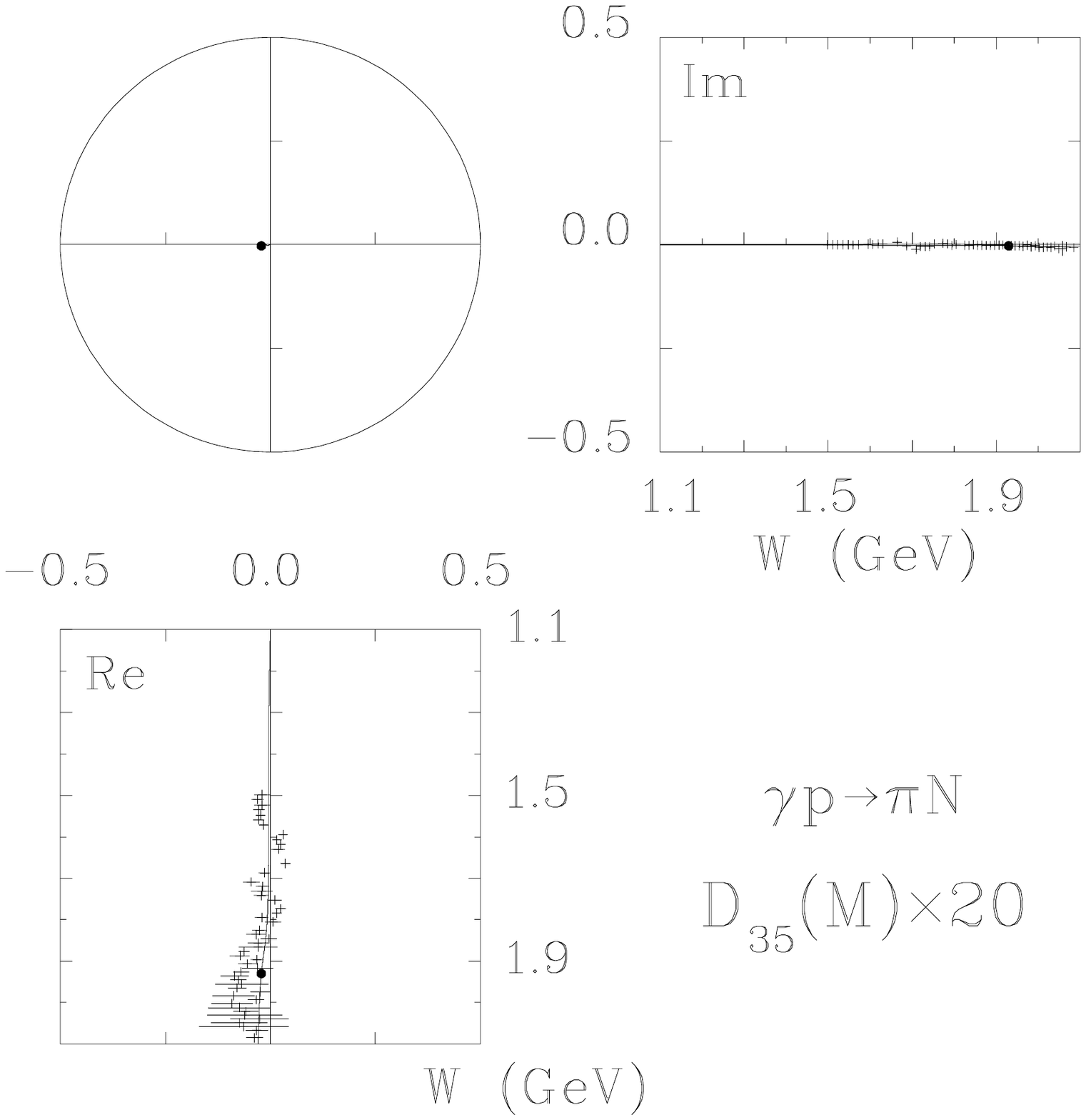}}
\vspace{-15mm}
\scalebox{0.35}{\includegraphics*{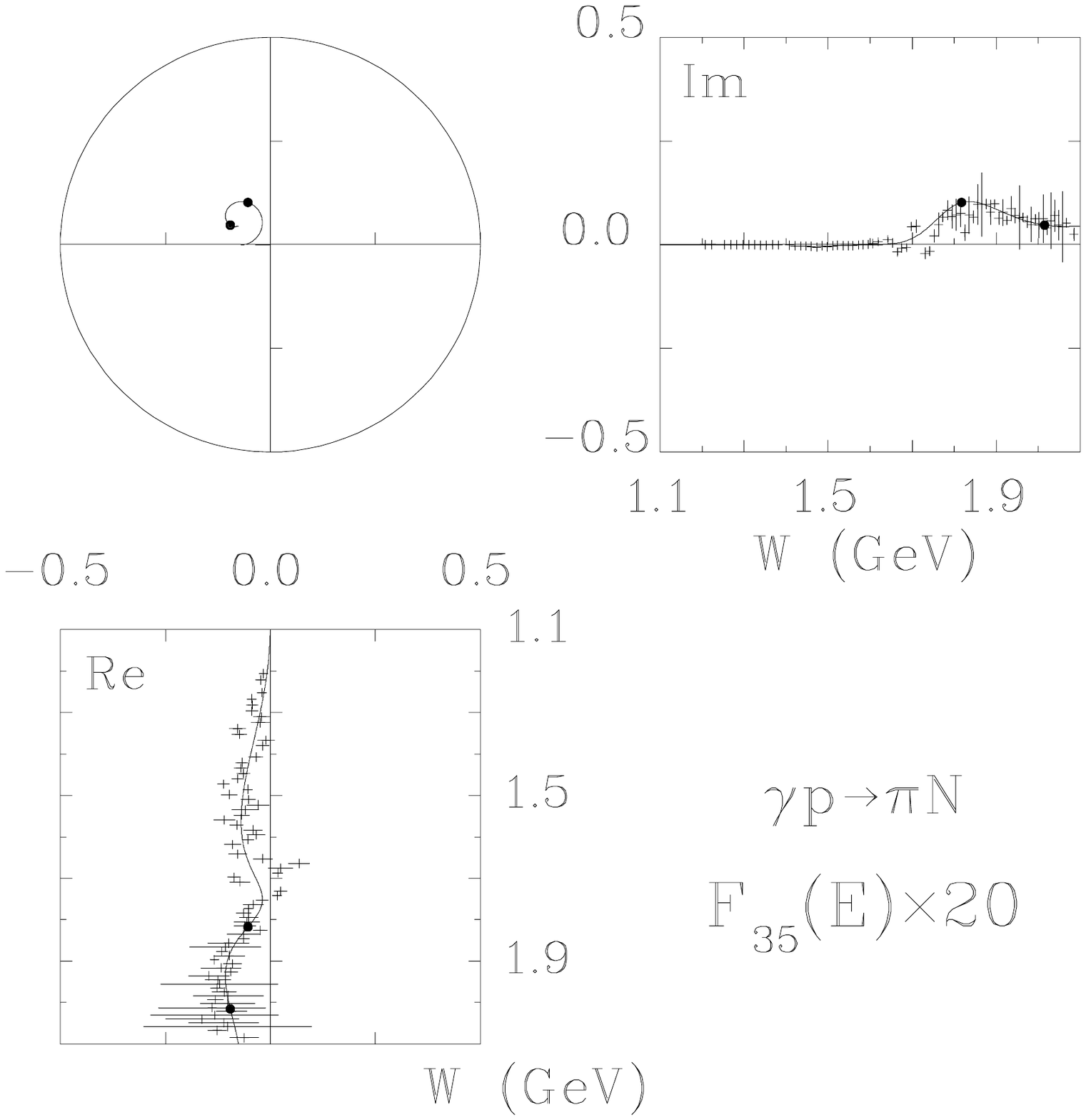}}
\scalebox{0.35}{\includegraphics*{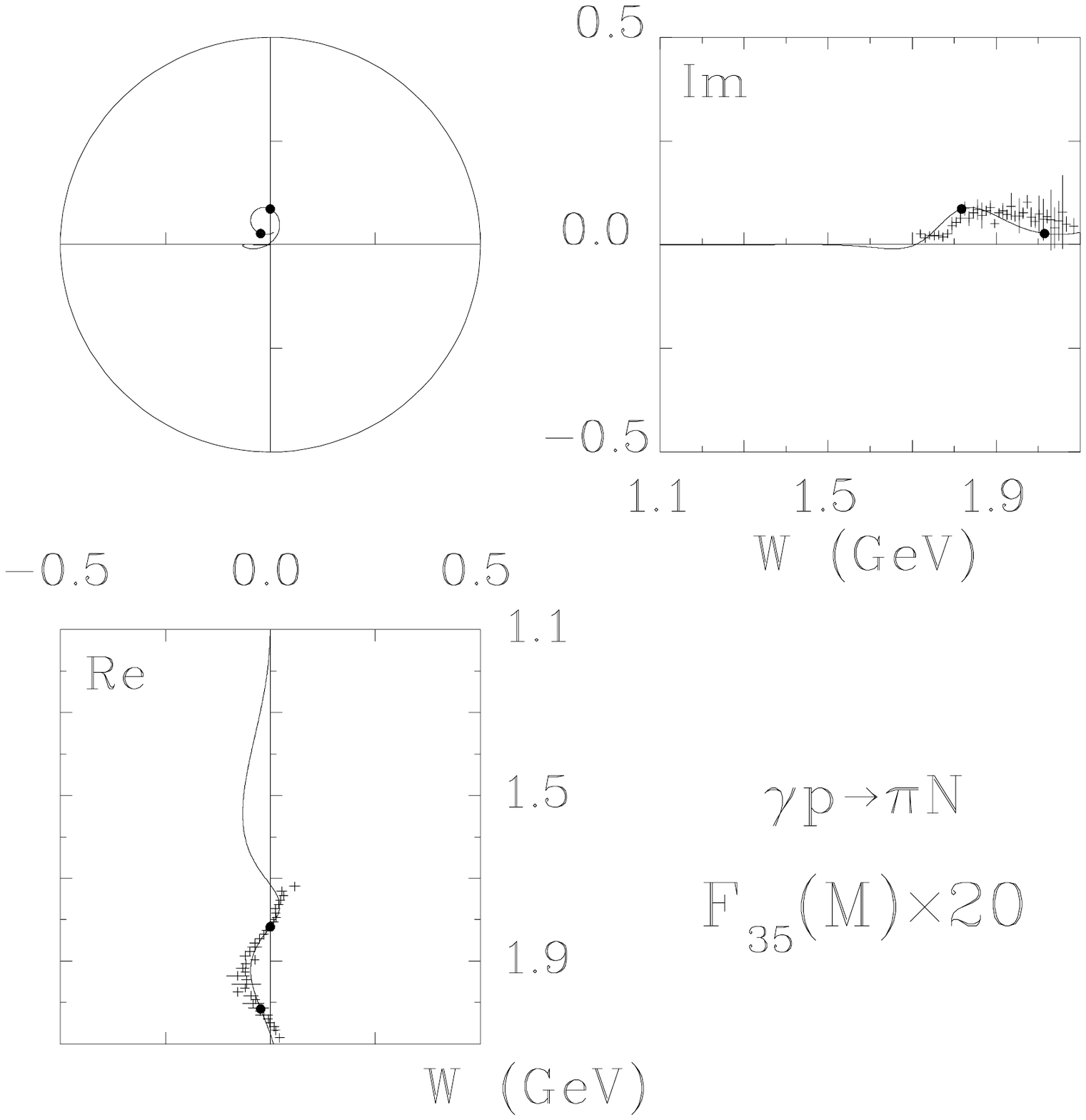}}
\caption{Cont'd.}
\end{figure*}

%%\newpage  NEED TO REMOVE COMMENT for the followings: OK 
\begin{figure*}[htpb]
\addtocounter{figure}{-1}
%\caption{Cont'd.}
\vspace{-10mm}
%\scalebox{0.35}{\includegraphics*{../f35/plot6.pdf}}
\vspace{15mm}
\scalebox{0.35}{\includegraphics*{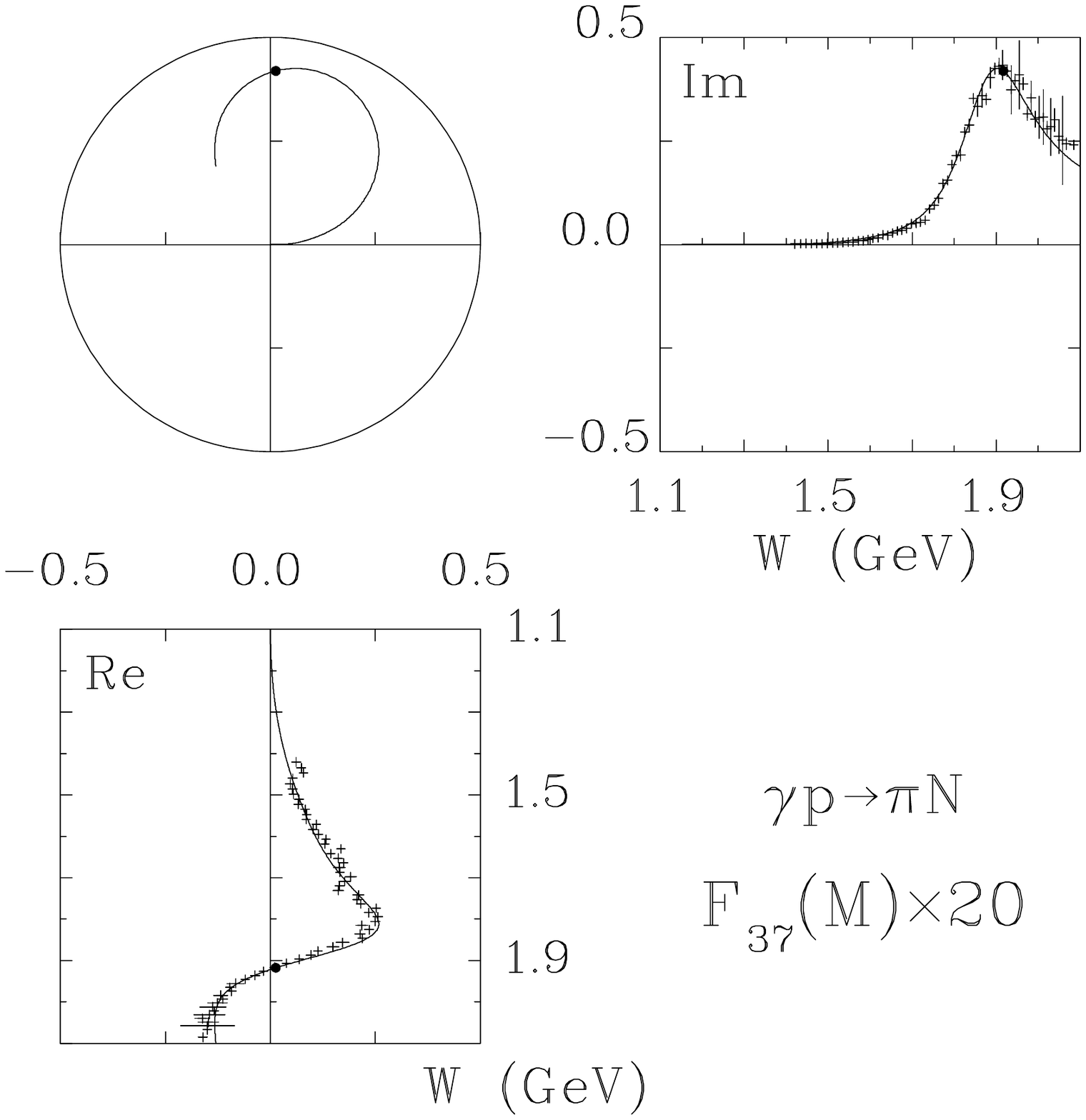}}
%\vspace{15mm}
%\scalebox{0.35}{\includegraphics*{../d33/plot7.pdf}}
%\vspace{15mm}
%\scalebox{0.35}{\includegraphics*{../d35/plot4.pdf}}
%\vspace{15mm}
%\scalebox{0.35}{\includegraphics*{../d35/plot5.pdf}}
%\scalebox{0.35}{\includegraphics*{../f35/plot5.pdf}}
\caption{Cont'd.}
\end{figure*}

%   Our results on helicity amplitudes (Tables VII and VIII), in most cases, are comparable  both
  \section{Summary and Conclusions}
  This work was undertaken to determine the parameters of $N^*$ and $\Delta^*$ resonances with masses up to about 2.1 GeV using a global multichannel fit. For the first time, we explicitly include amplitudes for $\pi N\rightarrow \eta N$ and $\pi N\rightarrow K\Lambda$ in addition to those for $\pi N\rightarrow \pi N$, $\pi N\rightarrow \pi\pi N$, and $\gamma N\rightarrow \pi N$. Most resonance parameters determined from this work agree satisfactorily with previous analyses \cite{arndt06, manley92, cutkosky80, hoehler79}. We find significant couplings of $S_{11}(1650)$ and $P_{11}(1710)$ to both $\eta N$ and $K\Lambda$. These results confirm the existence of $P_{11}(1710)$, for which no evidence was found in the analysis by Arndt {\it et al.\ }\cite{arndt06}. Also our work finds considerable couplings of $P_{13}(1900)$ to $\pi N$ and $K\Lambda$. Our results, on the whole, agree well with the predictions of quark models \cite{capstick94, capstick98, koniuk80}. 
  
  It is worthwhile to compare our results with the recent multichannel analysis by Anisovich {\it et al.\ }\cite{anisovich12}. Their analysis claims the existence of a number of new states. Interestingly we find all resonances listed in their analysis with masses below about 2100 MeV. Moreover, we find two additional resonances $D_{35} (1930)$ and $F_{35} (2000)$. The other difference is we obtain both $\gamma p$ and $\gamma n$ helicity couplings while their analysis gives only $\gamma p$ couplings. We have good agreement with their results for $P_{11}(1880)$, $F_{15}(1860)$, $P_{13}(1900)$, and $D_{15}(2060)$, which strengthens the evidence for these newly proposed states. 
  
\acknowledgements{This work was supported by the U.S. Department of Energy Grant No. DE-FG02-01ER41194. The authors thank the GWU group and especially Igor Strakovsky for providing part of the database for $\pi^-p\rightarrow \eta n$}.     
%biblilograghy    


\begin{thebibliography}{99}    
\bibitem{isgur-karl} N. Isgur and G. Karl, Phys. Rev. D {\textbf{19}}, 2653 (1979); erratum: Phys. Rev. D {\textbf{23}}, 817 (1981).
\bibitem{capstick-isgur} S. Capstick and N. Isgur, Phys. Rev. D. {\textbf{34}}, 2809 (1986).
\bibitem{glozman88} L. Y. Glozman, W. Plessas, K. Varga, and R. F. Wagenbrunn, Phys. Rev. D. {\textbf{58}}, 094030 (1998).
\bibitem{loring01} U. L{\"o}ring, B. Ch. Metsch, and H. R. Petry, Eur. Phys. J. A {\textbf{10}}, 395 (2001).
\bibitem{capstick-roberts} S. Capstick and W. Roberts, Prog. Part. Nucl. Phys. {\textbf{45}}, S241 (2000).
\bibitem{pdg12} J. Beringer {\textit{et al.} (Particle Data Group), Phys. Rev. D {\textbf{86}}, 010001 (2012).
\bibitem{arndt91} R. A. Arndt, Z. Li, L. D. Roper, R. L. Workman, and J. M. Ford, Phys. Rev. D {\textbf{43}}, 2131 (1991).
\bibitem{cutkosky80} R. E. Cutkosky {\textit{et al.}}, Pion-Nucleon Partial Wave Analysis, Toronto Conf. {\textbf{19}} (1980).%Phys. Rev. D {\textbf{20}}, 2839 (1979).
\bibitem{capstick94} S. Capstick and W. Roberts, Phys. Rev. D {\textbf{49}}, 4570 (1994).
\bibitem{julia-diaz07} B. Julia-Diaz, T. S. H. Lee, A. Matsuyama, and T. Sato, Phys. Rev. C {\textbf{76}}, 065201 (2007).
\bibitem{sarantsev09} A. Sarantsev, Chinese Phys. C {\bf 33}, 1085 (2009).
\bibitem{arndt06} R. A. Arndt, W. J. Briscoe, I. I. Strakovsky, and R. L. Workman, Phys. Rev. C {\textbf{74}}, 045205 (2006).
\bibitem{huang12} F. Huang {\textit{et al.}}, Phys. Rev. C {\textbf{85}}, 054003 (2012).
\bibitem{manoj12} M. Shrestha and D. M. Manley, submitted to Phys. Rev. C (hep-ph 1205.5294).
\bibitem{manoj012} M. Shrestha, Ph.D. dissertation, Kent State University (2012).
\bibitem{manley03} D. M. Manley, Int. J. Mod. Phys. A {\textbf{18}}, 441 (2003).
\bibitem{lee06} T.-S. H. Lee, in {\textit{NSTAR 2005}}, Proceedings of the Workshop on the Physics of Excited Nucleons, edited by S. Capstick, V. Crede, and P. Eugenio (World Scientific, 2006), p. 1.
\bibitem{manley92} D. M. Manley and E. M. Saleski, Phys. Rev. D {\textbf{45}}, 4002 (1992).
\bibitem{niboh97} M. M. Niboh, Ph.D. dissertation, Kent State University (1997).
\bibitem{john07} J. Tulpan, Ph.D. dissertation, Kent State University (2007).
\bibitem{hongyu08} H. Zhang, Ph.D. dissertation, Kent State University (2008).
\bibitem{arndt07} R. A. Arndt, private communication, (2007).
\bibitem{manley84} D. M. Manley, R. A. Arndt, Y. Goradia, and V. L. Teplitz, Phys. Rev. D {\textbf{30}}, 904 (1984).
\bibitem{gwu95}  GWU Center for Nuclear Studies Data Analysis Center http://gwdac.phys.gwu.edu.
\bibitem{abaev96} V. V Abaev and B. M. K. Nefkens, Phys. Rev. C. {\textbf{53}}, 385(1996).
\bibitem{vrana00} T. P. Vrana {\textit{et al.}}, Phys. Lett. C. {\bf 328}, 181 (2000).
\bibitem{saxon80} D. H. Saxon {\textit{et al.}}, Nucl. Phys. B. {\bf 162}, 522 (1980).
\bibitem{hoehler79} G. H\"ohler {\textit{et al.}}, Physik Daten 12-1(1979).
\bibitem{batanic95} M. Batini\'c, I. \v{S}laus, A. \v{S}varc, and B. M. K. Nefkens, Phys. Rev. C. {\bf 51}, 2310 (1995).
\bibitem{cutkosky90} R. E. Cutkosky and S. Wang, Phys. Rev. D {\bf 42}, 235 (1990).
 \bibitem{hoehler93} G. H\"ohler, $\pi N$ Newsletter {\bf 9}, 1 (1993). 
\bibitem{anisovich12} A. V. Anisovich {\textit{et al.}}, Euro. Phys. J.  A {\bf 48}, 15 (2012).
\bibitem{longacre78} R. S. Longacre {\textit{et al.}}, Phys. Rev. D {\bf 17}, 1795 (1978).
\bibitem{anisovich10} A. V. Anisovich, Euro. Phys. J.  A {\bf 44}, 203 (2010).
\bibitem{anisovich09} A. V. Anisovich, Euro. Phys. J.  A {\bf 41}, 13 (2009).
\bibitem{piN98} D. M. Manley, $\pi N$ Newsletter {\bf 14}, 158 (1998).
\bibitem{dugger07} M. Dugger {\textit{et al.}} (CLAS Collaboration), Phys. Rev. C {\textbf{76}}, 025211 (2007); Phys. Rev. C {\textbf{79}}, 065206 (2009). 
\bibitem{arndt96} R. A. Arndt, I. I. Strakovsky, and R. L. Workman, Phys. Rev. C {\bf 53}, 430 (1996).
\bibitem{ahrehs02} J. Ahrens {\textit{et al.}} Phys. Rev. Lett.  {\bf 88}, 232002 (2002).
\bibitem{crawford83} R. L. Crawford and W. T. Morton, Nucl. Phys. B{\bf 211}, 01 (1983).
\bibitem{awaji81} N. Awaji {\textit{et al.}} Bonn Conf. {\bf 352} (1981).
\bibitem{krusche97} B. Krusche, N. C. Mukhopadhyay, and J. F. Zhang, Phys. Lett. B {\bf 397}, 171 (1997).
\bibitem{bemner95} M. Benmerrouche, N. C. Mukhopadhyay, and J. F. Zhang, Phys. Rev. D {\bf 51}, 3237 (1995). 
\bibitem{fuji81} K. Fujii {\textit{et al.}}, Nucl. Phys. B {\bf 187}, 53 (1981).
\bibitem{arndt02} R. A. Arndt, W. J. Briscoe, I. I. Strakovsky, and R. L. Workman, Phys. Rev. C {\textbf{66}}, 055213 (2002). 
 \bibitem{koniuk80} R. Koniuk and N. Isgur, Phys. Rev. D. {\textbf{21}}, 1868 (1980); R. Koniuk, Nucl. Phys. B {\bf 195}, 452 (1982).
\bibitem{capstick98} S. Capstick and W. Roberts, Phys. Rev. D {\textbf{58}}, 074011 (1998).
\bibitem{capstick92} S. Capstick, Phys. Rev. D {\textbf{46}}, 2864 (1992).
%\bibitem{koniuk82} R. Koniuk, Nucl. Phys. B. {\bf 195}, 452 (1982)
%\bibitem{julich11} M. D\"oring {\textit{et al.}}, Nucl. Phys. A {\bf 851}, 58 (2011).
%\bibitem{matsuyama07} A. Matsuyama, T. Sato, and T. -S. Lee, Phys. Rep. {\textbf{439}}, 193 (2007).
%\bibitem{schutz98} C. Schutz {\textit{et al.}}, Phys. Rev. C {\textbf{57}}, 1464 (1998).
%\bibitem{krehl00} O. Krehl {\textit{et al.}}, Phys. Rev. C {\textbf{62}}, 025207 (2000).
%\bibitem{gasparyan03} A. M. Gasparyan {\textit{et al.}}, Phys. Rev. C {\textbf{68}}, 045207 (2003).
%\bibitem{schweber05} S. S. Schweber, {\textit {An introductin to Relativistic Quantum Field Theory}} (Dover, Mineola, NY, 2005).

%
%
}
%\bibitem{shkylar05} V. Shkylar {\textit{et al.}} Phys. Rev. C. {\bf 72}, 015210 (2005).
\end{thebibliography}
\end{document}